\documentclass[useAMS,usenatbib]{mnras}

\usepackage{amssymb}
\usepackage{hyperref}
\usepackage{mathtools}
\usepackage{multirow}

\title[Kinematic Modelling of Disc Galaxies using GPUs]
{
Kinematic Modelling of Disc Galaxies using Graphics Processing Units
}

\author[G. Bekiaris et al.]
{
G.~Bekiaris,$^{1}$\thanks{E-mail: gbekiaris@swin.edu.au}
K.~Glazebrook,$^{1}$
C.J.~Fluke,$^{1}$
R.~Abraham$^{2}$ 
\\
$^{1}$Centre for Astrophysics and Supercomputing, Swinburne University of Technology, P.O. Box 218, Hawthorn, VIC 3122, Australia\\
$^{2}$Department of Astronomy and Astrophysics, University of Toronto, 50 St. George Street, Toronto, ON M5S 3H8, Canada
}

\begin{document}


\pagerange{\pageref{firstpage}--\pageref{lastpage}} \pubyear{2015}

\maketitle

\label{firstpage}

%
%

\begin{abstract}
With large-scale Integral Field Spectroscopy (IFS) surveys of thousands of galaxies currently under-way or planned, the astronomical community is in need of methods, techniques and tools that will allow the analysis of huge amounts of data. We focus on the kinematic modelling of disc galaxies and investigate the potential use of massively parallel architectures, such as the Graphics Processing Unit (GPU), as an accelerator for the computationally expensive model-fitting procedure. We review the algorithms involved in model-fitting and evaluate their suitability for GPU implementation. We employ different optimization techniques, including the Levenberg--Marquardt and Nested Sampling algorithms, but also a naive brute-force approach based on Nested Grids. We find that the GPU can accelerate the model-fitting procedure up to a factor of $\sim100$ when compared to a single-threaded CPU, and up to a factor of $\sim10$ when compared to a multi-threaded dual CPU configuration. Our method's accuracy, precision and robustness are assessed by successfully recovering the kinematic properties of simulated data, and also by verifying the kinematic modelling results of galaxies from the GHASP and DYNAMO surveys as found in the literature. The resulting \textsc{gbkfit} code is available for download from: \mbox{\url{http://supercomputing.swin.edu.au/gbkfit}}.
\end{abstract}


%
%

\section{Introduction}

Galaxy kinematics play a crucial role in answering many fundamental questions about the structure, formation and evolution of galaxies. The missing mass problem which was first raised by \citet{1933AcHPh...6..110Z} and later confirmed by the flat rotation curves found in the majority of observed disc galaxies (\citealt{1972ApJ...176..315R}; \citealt{1973A&A....26..483R}; \citealt{1978PhDT.......195B}; \citealt{1978ApJ...225L.107R}), eventually led to the theory of dark matter (\citealt{1987ARA&A..25..425T}; \citealt{1992PASP..104.1109A}) which revolutionised our understanding of galaxies. Modern theories of galaxy formation are based on the $\Lambda$CDM theory (\citealt{1985ApJ...292..371D}; \citealt{1990Natur.348..705E}; \citealt{2015arXiv150201589P}), where galaxies form and grow inside massive dark matter haloes that dominate the baryons dynamically and whose merging history drives galaxy growth (\citealt{1978MNRAS.183..341W}; \citealt{2002PhR...372....1C}; \citealt{2010gfe..book.....M}). Galaxy kinematics play a vital role in studying this dynamical evolution (\citealt{2001PhDT.........3K}; \citealt{2005A&A...438..491D}; \citealt{2006ApJ...646..107E}; \citealt{2006ApJ...653.1049W}).

When studying the kinematics of disc galaxies, one of the most important measurements is the rotation curve \citep{2001ARA&A..39..137S}. By studying the shape of the rotation curve we can infer or constrain various dynamical quantities such as the baryonic and dark matter mass distribution (\citealt{1997ASPC..117...74B}; \citealt{2008MNRAS.383..297S}). Rotation curves allow us to derive the maximum circular velocity and thereby construct the Tully-Fisher relation \citep{1977A&A....54..661T} which is a correlation between the intrinsic luminosity of a galaxy and its rate of rotation. The Tully-Fisher relation can be used to measure distances and peculiar velocities \citep{1992ApJS...81..413M}, constrain the structural properties of dark matter haloes \citep{2007ApJ...671.1115G} or indicate merger activity \citep{2002AJ....123.2358K}.

Integral Field Spectroscopy (IFS) allows us to obtain spatially-resolved spectral information across the sky, enabling us to investigate galaxy kinematics in greater detail. IFS removes many of the problems and limitations of long-slit spectroscopy \citep{1995A&AS..113..347B} and reveals all the non-circular and peculiar velocity structure across the observed galaxy while allowing us to derive orientation parameters, including the major kinematic axis, from the observed data. 

Galaxy kinematics based on IFS data can be used to study a wide variety of topics including: galaxy formation and evolution scenarios (\citealt{2004MNRAS.350...35F}; \citealt{2008A&A...477..789Y}; \citealt{2008ApJ...687...59G}; \citealt{2011MNRAS.413..813C}; \citealt{2011MNRAS.417.2601W}), the supermassive black holes at the centre of galaxies \citep{2002MNRAS.335..517V} and galaxies with exotic kinematic signatures \citep{2001ApJ...548L..33D}.

To extract the rotation curve from a galaxy we create kinematic models and we fit them to observations. For high resolution observations the tilted-ring modelling approach \citep{1974ApJ...193..309R}, which extracts the rotation curve in a piecewise manner, has become a standard. This method uses a series of parametrized rings to describe the circular velocities at different radii of the galactic disc. For low resolution observations, where high-parameter models cannot be used due to low signal-to-noise ratio and beam smearing effects, different model are employed. Usually these models combine a parametric function for the rotation curve model, the inclination and orientation of the galaxy, and a photometric profile.

Parametric rotation curve models can have a theoretical basis or be purely empirical. Models such as the exponential disc \citep{1970ApJ...160..811F} and the pseudo-isothermal sphere \citep{1987PhDT.......199B} are derived from the stellar and dark matter mass distribution respectively. The linear ramp model \citep{2007ApJ...658...78W}, although it has no theoretical basis, seems to describe high-redshift data very well \citep{2010MNRAS.401.2113E}. The more elaborate empirical Multi-Parameter Function of \citet{1997AJ....114.2402C} can be used to describe rotation curves with an increasing or decreasing velocity plateau. 

In order to compare models to observations, one has to take the spatial resolution into account, by convolving the model with the Point Spread Function (PSF) of the observation. While this is not very important for high resolution local galaxy data, it is essential for high redshift galaxies and low resolution observations where the size of the PSF is comparable to the angular size of the galaxy (\citealt{2006ApJ...645.1062F}; \citealt{2011A&A...528A..88G}; \citealt{2012A&A...539A..92E}). 

Several kinematic model-fitting codes are publicly available. \textsc{rotcur} \citep{1987PhDT.......199B} derives the kinematic parameters from the observed velocity field by fitting a tilted-ring model to it. \textsc{diskfit} \citep{2007ApJ...664..204S} fits a physically motivated model that handles non-axis-symmetric rotation to a velocity map. \textsc{kinemetry} \citep{2006MNRAS.366..787K} extracts the kinematic properties of a galaxy by using a generalisation of photometry to higher-order moments of the line-of-sight velocity distribution.

Although most of the methods discussed in the literature (including the above codes) fit their models to 2D projections, such as the velocity map, there are cases where the fit is done directly to the observed intensity datacube. By doing so, one can take into account the effects of beam smearing, including the asymmetries on the spectral lines and the increase of the velocity dispersion towards the centre of the galaxy \citep{2011ApJ...741...69D}. \textsc{tirific} \citep{2007A&A...468..731J} and \textsc{$^{\mathrm{3D}}$barolo} \citep{2015MNRAS.451.3021D} use a tilted-ring approach to generate a parametrized model datacube and fit it directly to the observation. \textsc{galpak$^{\mathrm{3D}}$} \citep{2015AJ....150...92B} extracts the morphological and kinematic parameters of a galaxy by fitting a ten-parameter disc model to three-dimensional data. These, and other model-fitting approaches, employ an optimization algorithm to fit the kinematic model to the observed data.

While the Levenberg--Marquardt algorithm (\citealt{levenberg44}; \citealt{marquardt:1963}) has been the most popular choice for optimization of parameters in kinematic modelling (e.g., \citealt{2006MNRAS.371..170B}; \citealt{2008A&A...490..589C}; \citealt{2014MNRAS.437.1070G}), other optimization methods have also been used successfully. Some of the most notable include the Nelder-Mead minimizer (e.g., \citealt{2004MNRAS.349..225G}), Markov Chain Monte Carlo samplers (e.g., \citealt{2011MNRAS.417.2601W}; \citealt{2015AJ....150...92B}) and genetic algorithms (e.g., \citealt{2009ApJ...697..115C}; \citealt{2014NewA...26...40W}).

The fitting procedure can be a time-demanding process, especially when it involves complex many-parameter models, high resolution data, and optimization algorithms that require a substantial number of iterations before converging to a solution. Unfortunately most existing kinematic fitting software does not fully utilise all the available computer hardware efficiently in order to decrease the computational time requirements. 

The lack of high-performance software for galaxy kinematic modelling has not been a major problem so far because of the limited amount of 3D spectroscopic data available for analysis. Some of the largest spatially resolved spectroscopic surveys completed to date include: GHASP \citep{2008MNRAS.388..500E}, SINS \citep{2009ApJ...706.1364F}, DiskMass \citep{2010ApJ...716..198B}, and ATLAS$^{3D}$ \citep{2011MNRAS.413..813C}. These surveys consist of up to a few hundreds of galaxies, with the last being the largest one (260 galaxies). However, the number of observed galaxies will increase substantially in the near future because of the large-scale optical and radio surveys such as CALIFA \citep{2012A&A...538A...8S}, KMOS$^{3D}$\citep{2015ApJ...799..209W}, SAMI \citep{2015MNRAS.447.2857B}, MaNGA \citep{2015ApJ...798....7B}, HECTOR \citep{2015IAUS..309...21B} and WALLABY \citep{2012PASA...29..359K}, which are currently underway or planned. Furthermore, new- and next-generation instruments like MUSE \citep{2015A&A...575A..75B} will be able to capture massive amounts of optical spectra in a single exposure, making the analysis of each individual galaxy computationally challenging.

The vast amounts of data that will be produced by the above surveys will require a substantial amount of computational power for their analysis. Therefore, new high-performance tools and software methods that will accelerate the kinematic modelling process are needed. In addition, high-performance software can also benefit the kinematic analysis of numerical simulations where thousands of line-of-sights and time steps can be extracted (e.g., \citealt{2014A&A...562A...1P}).

Although in some of the above-mentioned codes there has been some effort to accelerate the fitting procedure by utilising all the available cores of the Central Processing Unit (CPU), none is currently able to take advantage of other many-core architectures such as the Graphics Processing Unit (GPU).

Over the past decade the GPU has been transformed from a specialised graphics rendering hardware into a massively parallel architecture with high computational power. Its enormous memory bandwidth and arithmetic capabilities make it highly suitable for scientific computing. GPUs have already been used for different problems in astronomy, including radio astronomy cross-correlation \citep{2009PASP..121..857W}, gravitational microlensing \citep{2010NewA...15...16T}, N-Body simulations \citep{2008NewA...13..103B}, real-time control of adaptive optics \citep{2014SPIE.9148E..6OG}, and astronomical data visualization \citep{2013MNRAS.429.2442H}. In most of the above cases the GPUs provided a substantial performance increase compared to the CPU. 

In this work we investigate how the graphics processing unit can benefit galaxy kinematic modelling by accelerating the model fitting procedure. In Section 2 we describe the modelling method including the choice of kinematic models, the different optimization techniques used and the GPU implementation of the fitting algorithm. In Section 3 we introduce and provide details about the datasets we have selected for this work. In Section 4 we evaluate the consistency and robustness of the presented method by fitting models to the selected data and comparing our results with the literature. In section 5 we evaluate the performance of the proposed approach and we compare it to CPU-based solutions. In section 6 we discuss and interpret our findings, but also suggest future research and developments that can follow this work. Finally in Section 7 we list a summary of our conclusions.

%
%

\section{The Method}
\label{sec:method}

Fitting models to observations of galaxies in order to extract their kinematic and structural parameters is a very common problem in astronomy. The fitting procedure employs a model with a number of free parameters, the observed data and an optimization technique which chooses a `better-fitting' solution based on a previous, less-optimal fit. Although the model and optimization technique used for the fitting procedure may vary, in most cases the following pattern is more or less respected:
\begin{enumerate}
\renewcommand{\theenumi}{(\arabic{enumi})}
\item Evaluate the model on a cube using the current set of parameters.
\item Convolve the model cube with the line and point spread functions of the observation.
\item Compare the convolved model cube or its moments with the observed data and evaluate the goodness of fit.
\item Check termination criteria and, if met, return the optimized model parameters. Otherwise, adjust model parameters and repeat from Step 1.
\end{enumerate}

The above steps describe a typical iteration of the fitting procedure. Steps 1 to 3 usually remain the same regardless of the chosen optimization technique, while step 4 can differ substantially. The number of iterations required to converge to an optimal solution is usually heavily dependent on the chosen optimization technique, the number of free parameters in the selected model, and the choice of initial model parameter values.

The method we present in this paper follows a similar approach to the one described above but employs two different modelling strategies. The first one uses a velocity field model instead of an intensity cube, and resembles the method used in \citet{2008MNRAS.390..466E}. It does not take the line and point spread functions into account and thus it is more appropriate when modelling high-resolution observations. The second one uses an intensity cube model and resembles the method used in \citet{2014MNRAS.437.1070G}. It deals with beam smearing effects by taking into account the line and point spread functions, hence it is a much better option when working with low-resolution data.

While our method is very similar to previous work, its implementation differs substantially from the majority of the available kinematic model-fitting codes since it takes advantage of the massively parallel architecture of the GPU. More detailed information on the fitting procedure and its implementation on the GPU can be found in Section~\ref{sec:method-fitting-procedure}.

%
%

\subsection{The models}
\label{sec:method-models}

Every galaxy model in this work is a combination of a disc and a rotation curve. In addition, when an intensity cube model is used, a photometric profile and an intrinsic velocity dispersion are employed as well. We now describe these model components in detail.

\subsubsection{The disc model}
\label{sec:method-models-disc}

To model the geometry of the galaxy we use a thin and flat disc. To align any given point $(x,y)$ of the observation with the coordinate system defined by the projected major and minor axes of the disc, we use the following equations:
\begin{equation}
\label{eq:model-disc-01}
x_{\mathrm{e}} = - (x-x_{\mathrm{o}})\sin{PA} + (y-y_{\mathrm{o}})\cos{PA},
\end{equation}
\begin{equation}
\label{eq:model-disc-02}
y_{\mathrm{e}} = - (x-x_{\mathrm{o}})\cos{PA} - (y-y_{\mathrm{o}})\sin{PA},
\end{equation}
where $(x_{\mathrm{o}},y_{\mathrm{o}})$ is the centre and $PA$ the position angle of the projected disc. The radius $r$ of the circle on the disc plane that passes through the projected position $(x,y)$ is calculated as follows:
\begin{equation}
\label{eq:model-disc-03}
r = \sqrt{x_{\mathrm{e}}^2 + \left(\frac{y_{\mathrm{e}}}{\cos{i}}\right)^2 },
\end{equation}
where $i$ is the inclination of the disc. Finally, we calculate the position angle, $\theta$, of the projected point $(x,y)$ relative to the disc major axis using:
\begin{equation}
\label{eq:model-disc-04}
\cos{\theta} = \frac{x_{\mathrm{e}}}{r}.
\end{equation}

\subsubsection{The surface brightness model}
\label{sec:method-models-brightness}

The surface brightness of a disc galaxy can be modelled using two components: an inner spheroid component for the bulge and an outer exponential component for the disc \citep{1959HDP....53..311D}. 

In this work we only used an exponential disc profile. While this distribution describes correctly the stellar component, it is less appropriate for describing the ionised gas lines whose emission is dominated by HII regions. However, when dealing with low-resolution data, due to the effects of beam smearing, an exponential disc profile can be a reasonable choice (\citealt{2009ApJ...697..115C}; \citealt{2009ApJ...699..421W}; \citealt{2014MNRAS.437.1070G}). On the contrary, when working with high-resolution data where the HII regions are clearly resolved, more complex models may be used. In many cases, HII regions are organised into ring- or spiral-like structures but they can also be randomly distributed across the galactic disc (\citealt{1995ApJS..101..287T}; \citealt{1996ApJS..105...93E}). This makes their modelling a non-trivial task. Since the focus of this work is not to investigate modelling in detail, the decision to use only an exponential disc model can be considered acceptable. Furthermore, by using only an exponential disc profile we were able to compare our modelling results directly to the results found in the literature (see Section~\ref{sec:fitting-results-dynamo}) However, our code is flexible enough to cope with a wide range of user-specified intensity models.

The exponential disc profile is described by the following equation:
\begin{equation}
\label{eq:model-brightness-exp}
I(r) = I_{\mathrm{0}}\exp{\left(-\frac{r}{r_{\mathrm{0}}}\right)},
\end{equation}
where $I(r)$ is the surface brightness at radius $r$, $I_{\mathrm{0}}$ is the surface brightness at the centre of the galaxy, and $r_{\mathrm{0}}$ is the scale length.

\subsubsection{The rotation curve models}
\label{sec:method-models-rcurves}

In this work we have used two different rotation curve models: an \emph{arctan} profile \citep{1997AJ....114.2402C} and the analytical function of \citet{2008MNRAS.388..500E} which from now on we refer to as the \emph{Epinat function}. 

The two-parameter arctan function is given by:
\begin{equation}
\label{eq:model-rcurve-arctan}
V_{\mathrm{rot}}(r) = \frac{2}{\pi}V_{\mathrm{t}}\arctan{\frac{r}{r_{\mathrm{t}}}},
\end{equation}
where $V_{\mathrm{rot}}(r)$ is the circular velocity at radius $r$, $V_{\mathrm{t}}$ is the asymptotic circular velocity and $r_{\mathrm{t}}$ the turn-over radius. Since $V_{\mathrm{t}}$ is never reached in practice, this model describes rotation curves with an increasing velocity plateau. 

The four-parameter Epinat function is given by:
\begin{equation}
\label{eq:model-rcurve-epinat}
V_{\mathrm{rot}}(r) = V_{\mathrm{t}}\frac{(r/r_{\mathrm{t}})^g}{1+(r/r_{\mathrm{t}})^a},
\end{equation}
where $V_{\mathrm{t}}$ and $r_{\mathrm{t}}$ control the turn-over velocity and radius while $g$ and $a$ control the inner and outer slope of the rotation curve. As a result, this function allows us to describe rotation curves with flat ($a = g$), rising ($a < g$), and declining ($a > g$) velocity plateaus.

\subsubsection{The velocity dispersion model}
\label{sec:method-models-dispersion}

Our galaxy model includes an intrinsic velocity dispersion component, $\sigma_{\mathrm{model}}$, which is assumed to be constant across the galactic disc. Fitting a disc model which includes the velocity dispersion allows us to correctly account for beam smearing \citep{2011ApJ...741...69D}, but it can be problematic when the velocity field is not representative of a disc or if the intrinsic velocity dispersion is not constant \citep{2014MNRAS.437.1070G}. 

An alternative method for extracting the intrinsic velocity dispersion of a galaxy is to use the $\sigma_{\mathrm{m}}$ measure of \citet{2009ApJ...697.2057L}, which is the flux-weighted mean of the velocity dispersions measured in each individual pixel. This quantity provides a more robust way to measure the velocity dispersion when working with data of low signal-to-noise ratio and it does not require assumptions about an underlying kinematic model. However, it can be biased due to beam smearing effects. To correct for this, \citet{2014MNRAS.437.1070G} subtracted a map of unresolved velocity shear present in the best-fitting disc model from the observed velocity dispersion map before calculating $\sigma_{\mathrm{m}}$.

Since the purpose of this work is to evaluate the feasibility of the GPU for the kinematic model-fitting procedure, and not to study the velocity dispersion of galaxies in detail, the assumption of a constant intrinsic velocity dispersion is reasonable. Therefore, all the velocity dispersion results derived by our method in this paper refer to the fitted velocity dispersion $\sigma_{\mathrm{model}}$.

\subsubsection{The complete kinematic model}
\label{sec:method-models-complete}

The observed radial velocity in the sky plane at $(x,y)$ of a purely rotating thin disc with systemic velocity $V_{\mathrm{sys}}$ is described by:
\begin{equation}
\label{eq:model-complete-01}
v_{\mathrm{model}}(x,y) = V_{\mathrm{sys}} + V_{\mathrm{rot}}(r)\sin{i}\cos{\theta}.
\end{equation}
The above radial velocity model has seven or nine free parameters depending on the selected rotation curve profile. 

Our final model can be either a velocity field or an intensity cube. In the case of a velocity field, equation~\ref{eq:model-complete-01} is evaluated for each pixel directly. In the case of an intensity cube, equation~\ref{eq:model-complete-01} along with equation~\ref{eq:model-brightness-exp} and $\sigma_{\mathrm{model}}$ are used to infer the shape of the spectral lines prior to the convolution step. However, as a last step, the velocity and velocity dispersion maps are extracted from the intensity cube model. This is done because our fitting method always expects observations of velocity and velocity fields as input. For more details on the evaluation of the velocity field and intensity cube models see Sections~\ref{sec:method-fitting-procedure-vfieldval}~to~\ref{sec:method-fitting-procedure-moments}.

\subsubsection{The initial model parameter values}
\label{sec:method-models-initialvalues}

In order to start the fitting procedure, an initial set of model parameter values has to be given to the optimization algorithm. Ideally we want the initial parameter values to be as close to the optimal solution as possible in order to decrease our chances of getting trapped into a local optimum. These values are calculated by analysing the continuum, velocity and velocity dispersion maps of the observation. 

The coordinates of the kinematic centre ($x_{\mathrm{o}}$,$y_{\mathrm{o}}$) are estimated by calculating the central moments of the continuum map. $V_{\mathrm{sys}}$ is calculated by extracting the velocity mean around the previously estimated kinematic centre from the velocity map. To estimate $PA$, we draw a line that passes through the estimated kinematic centre and splits the velocity field into two parts. The orientation of the line is selected in a way that the difference of the means of the two parts is maximized. The initial values of $i$ and $r_{\mathrm{t}}$ are set to the photometric inclination and scale radius found in online databases such as HyperLeda (\citealt{2003A&A...412...45P}, for the GHASP survey) and SDSS (\citealt{2006ApJS..162...38A}, for the DYNAMO survey). For the $V_{\mathrm{t}}$ and $\sigma_{\mathrm{model}}$ parameters we use the values of 150~km~s$^{-1}$ and 30~km~s$^{-1}$ respectively. The $a$ and $g$ parameters of the Epinat rotation curve function are both set to 1.

The parameter values of the surface brightness profile are assumed to be known and they remain fixed during the fitting procedure. The value of $I_{\mathrm{0}}$ is technically a scaling factor, and since our fit is done to the velocity and velocity dispersion maps, its value is not expected to affect the fitting results. The scale length ($r_{\mathrm{0}}$) is fixed to its actual value (for the mock data) or to the value found in the SDSS database (for the DYNAMO survey). This is a reasonable decision since the scale length can be determined from the photometry.

More information about the simulated data and the two surveys studied in this work (i.e., GHASP and DYNAMO) can be found in Section~\ref{sec:data}.

\subsubsection{The point and line spread functions}
\label{sec:method-models-psf}

Finally, we need to know how the imaging system responds to a point source. This is provided by the point and line spread functions. The point spread function (PSF) is modelled using a two-dimensional Gaussian function and its standard deviation, $\sigma_{\mathrm{PSF}}$, depends primarily on the atmospheric seeing. The line spread function (LSF) is modelled using a one-dimensional Gaussian function and its standard deviation, $\sigma_{\mathrm{LSF}}$, depends on the instrument's spectral resolution. These two functions can be combined to form a three-dimensional Gaussian function:
\begin{equation}
\label{eq:gauss3d}
f(x,y,z)=\frac{1}{\sigma_{\mathrm{PSF}}^2\sigma_{\mathrm{LSF}}^{}\sqrt{(2\pi)^3}}\exp{\left(-\frac{x^2 + y^2}{2\sigma_{\mathrm{PSF}}^2}-\frac{z^2}{2\sigma_{\mathrm{LSF}}^2}\right).}
\end{equation}

%
%

\subsection{The fitting procedure and implementation}
\label{sec:method-fitting-procedure}

In this section we provide details about our fitting method and how it is implemented on the graphics processing unit. 

GPU-programming requires a very different mental model from that required for programming general purpose processors because of their substantially different architectures. CPUs feature a small number of processing cores, large caches, and units for managing and scheduling tasks such as branch prediction, instruction ordering and speculative execution. On the other hand, GPUs sacrifice the large caches and the advanced management and scheduling units in order to maximize the number of processing cores on the chip \citep{Thomas:2009:CCG:1508128.1508139}. 

Previous research on achieving high performance on a GPU has already been done (\citealt{Ryoo:2008:OPA:1345206.1345220}; \citealt{2010ASPC..434..209B}; \citealt{Brodtkorb20134}), and although each application might require manual tuning, the following are considered rules of thumb:

\begin{itemize}
\item \textbf{Maximize parallelism:} In order to fully utilise the GPU one has to make sure that all its processing cores are occupied at any given time. One way to achieve this is by using data-parallel algorithms \citep{hillis1986data} which perform simultaneous operations across large sets of data using the parallel processors of the underlying platform. However, the performance of such algorithms may suffer if the size of the data is small relative to the number of processing cores available on the GPU.
\item \textbf{Maximize arithmetic intensity:} Arithmetic intensity is defined as the number of mathematical operations performed per memory access. Although GPUs are coupled with high-bandwidth main memory (also known as \emph{global memory}) and a small but very fast on-chip memory, their strongest point is to execute arithmetic instructions. As a result, if a GPU program performs only a few arithmetic operations per memory access it becomes \emph{memory-bound} and its performance is limited by the rate its memory can be accessed. 
\item \textbf{Minimize time spent on memory accesses:} Instead of (or in addition to) increasing the arithmetic intensity of the program by increasing the number of instructions per memory access, one can minimize the time spent on global memory transactions. One way to achieve this is to move frequently-accessed data and data that remain constant during the code execution to the on-chip shared memory and the constant memory respectively. The advantage of shared and constant memory over global memory is the superior bandwidth and the much lower latency. Another strategy to reduce the time spent on memory transactions is to use coalesced global memory access by making sure that successive GPU threads access adjacent locations in a contiguous stretch of memory. This strategy takes advantage of the fact that whenever a GPU thread reads or writes the global memory it always accesses a large chunk of memory at once. If other threads make similar accesses at the same time then the GPU can reuse the same chunk for those threads without attempting a new memory transaction.
\end{itemize}

Due to the image nature of our data we expect our method to benefit from GPU parallelization. 3D spectroscopy data involves 2D and 3D images of thousands or millions of pixels that can be processed independently and thus we expect a high GPU utilisation. Furthermore, due to the well behaved and regular structure of image data  it is relatively easy to write algorithms that follow particular memory access patterns and achieve coalesced global memory access.

To investigate whether our method is suitable for GPU parallelization we follow a similar approach to the one discussed in \citet{2010ASPC..434..209B}. We outline each step involved in the fitting iteration and then identify which ones resemble known algorithms. For each identified algorithm we refer to its pre-existing analysis, while for the steps that do not appear to resemble any known algorithms we perform the analysis ourselves.

Our method fits a model to the velocity and velocity dispersion maps of the observation. Because of this, when we are using an intensity cube model the method extracts the velocity and velocity dispersion maps from the convolved datacube in order to compare them to the input data and calculate the goodness-of-fit. In addition, one can also supply measurement error and mask maps if they are available. In Sections~\ref{sec:method-fitting-procedure-vfieldval} to \ref{sec:method-fitting-procedure-optimization} we describe all the steps involved in the two modelling strategies in detail.

\subsubsection{Velocity field evaluation}
\label{sec:method-fitting-procedure-vfieldval}

For the velocity field modelling, the first step of the fitting procedure evaluates the radial velocity model (equation~\ref{eq:model-complete-01}) on a two-dimensional grid using the current set of model parameters. The evaluation of the velocity field is a highly parallel transform operation because each pixel evaluation is completely independent of the others. To implement the velocity field evaluation on the GPU we assign each thread to evaluate a different pixel. When working with low-resolution data the number of pixels in our velocity field might be less than the number of available GPU cores and thus the GPU may remain underutilised. Therefore the velocity field modelling strategy will perform best with high-resolution data. In order to achieve high performance we also make sure that memory coalescing is achieved by assigning consecutive pixels to consecutive GPU threads. When using the velocity field model strategy the Steps~\ref{sec:method-fitting-procedure-cubeeval} to \ref{sec:method-fitting-procedure-moments} are skipped and the iteration continues at Step~\ref{sec:method-fitting-procedure-residual}.

\subsubsection{Intensity cube evaluation}
\label{sec:method-fitting-procedure-cubeeval}

For the intensity cube modelling strategy, the first step of the fitting procedure generates an intensity cube defined by two spatial and one velocity coordinate. Using the current set of model parameter values, for each spatial coordinate of the cube we calculate a line-of-sight velocity and velocity dispersion by evaluating the full kinematic model described by equation~\ref{eq:model-complete-01}. The spatial intensities of the cube are calculated with an exponential disc profile (equation~\ref{eq:model-brightness-exp}) for which the scale length is assumed to be known in advance. The intensities along the velocity dimension of the cube are modelled using a Gaussian profile with its mean and standard deviation corresponding to the calculated line-of-sight velocity and velocity dispersion respectively. 

\begin{figure}
	\centering
	\includegraphics[width=0.47\textwidth]{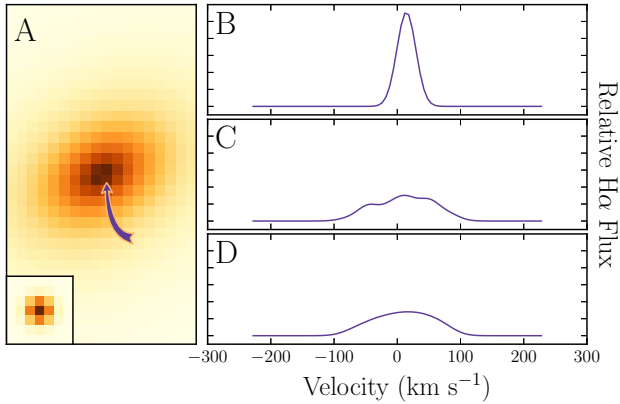}
	\caption{Artefacts may appear on the spectral lines of low resolution model datacubes after the PSF convolution. (A)~Integrated flux of a low resolution model datacube before the convolution with the PSF. (B,C,D)~All three plots refer to the same spectral line on datacube A. (B)~The spectral line before the PSF convolution. (C)~The spectral line after the PSF convolution. (D)~The spectral line after the PSF convolution with a high resolution version of datacube A. The spatial and spectral resolutions of the cubes are: 0.7~arcsec~pixel$^{-1}$ and 30~km~s$^{-1}$ for the original cube, and 0.35~arcsec~pixel$^{-1}$ and 15~km~s$^{-1}$ for the high-resolution cube.}
	\label{fig:upsampling}
\end{figure}

We define the number of pixels on the spatial plane of the observed data as $N_{\mathrm{DATA,x}} \times N_{\mathrm{DATA,y}}$. Furthermore, we define the number of pixels required to cover the cube's modelled spectral lines and their velocity shift as $N_{\mathrm{DATA,z}}$. The number of pixels along the spatial and velocity dimensions of the intensity cube must be an integer multiple of the number of pixels in the observed data. From now on we will refer to this number as the \emph{upsampling factor} or $u$. The higher the upsampling factor, the higher the resolution of the intensity cube. When dealing with low resolution observations, evaluating the intensity cube at a higher resolution allows us to avoid artefacts that might appear in its spectral lines after the PSF convolution (Fig.~\ref{fig:upsampling}).

Padding must be applied to the model cube prior to PSF convolution. First, the cube is padded by $N_{\mathrm{PSF,n}}-1$ across all its dimensions, where $N_{\mathrm{PSF,n}}$ is the size of dimension $n$ of the PSF image. This is required in order to perform the convolution in the frequency domain through the Fast Fourier Transform (FFT, \citealt{Duhamel1990259}). Then, each dimension of the cube is padded to the next power of two because the majority of the FFT algorithms perform optimally when working with that size \citep{4607357}. The length of each dimension of the cube is:
\begin{equation}
\label{eq:cube-dim-size}
N_{\mathrm{CUBE,n}} = 2^{ \textstyle{\big\lceil \log_{2} \left[u\times (N_{\mathrm{DATA,n}}+N_{\mathrm{PSF,n}}-1)\right] \big\rceil } }
\end{equation}

The evaluation of the cube is a highly parallel transform operation because each pixel evaluation is completely independent of the others. To implement the cube evaluation on the GPU we assign each thread to evaluate a different pixel. Since the number of pixels in a cube is usually more than the available GPU cores, each GPU thread evaluates more than one pixel. In order to achieve high performance we also make sure that memory coalescing is achieved by assigning consecutive pixels to consecutive GPU threads.

\subsubsection{Point spread function convolution}
\label{sec:method-fitting-procedure-psfconv}

The PSF of the observation is a three-dimensional image, which from now on we will refer to as the \emph{PSF kernel}, and it is created by combining the two-dimensional image of the point spread function and the one-dimensional image of the line spread function. Alternatively, since in this work both the point and the line spread functions are modelled using Gaussian profiles, the PSF kernel can be generated by evaluating equation~(\ref{eq:gauss3d}) directly on a three-dimensional grid.

In this step of the fitting iteration we convolve the generated intensity cube with the PSF kernel of the observation. The naive 3D convolution algorithm, although simple, has a high computational complexity of $O(N^6)$ and hence it is preferred to perform the convolution in the frequency domain where the computational complexity is reduced to $O(N^3log_{2}^3N)$. According to the circular convolution theorem, in order to perform a convolution via FFT, the two input sequences must have the same size. Hence, we need to expand the PSF kernel to the intensity cube size. 

The intensity cube and PSF kernel are transformed to frequency space using the FFT which is a highly parallel divide-and-conquer algorithm and suitable for GPU parallelization. Next, a point-wise complex multiplication between the two transformed images, which is essentially a transform operation, is performed on the GPU. Because the data of both the cube and the PSF kernel are located in contiguous memory locations coalesced access is achieved once again. Finally an inverse FFT is applied to the results of the multiplication in order to get the final result of the convolution back in geometric space.

\subsubsection{Down-sampling to the resolution of the observation}
\label{sec:method-fitting-procedure-downsampling}

Before proceeding further we have to down-sample the intensity cube in order to create a new cube that matches the resolution of the observed data. This is achieved by a transform and reduce operation on the GPU: We assign each GPU thread to calculate the value of one or more pixels of the down-sampled cube by averaging the corresponding pixels of the original cube. Here we achieve coalesced access to the down-sampled cube but not for the original cube. As a result we do not expect a high efficiency from this step.

\subsubsection{Velocity and velocity dispersion map extraction}
\label{sec:method-fitting-procedure-moments}

In this step we extract the velocity and velocity dispersion maps from the down-sampled cube, in order to allow for comparison with the input data. This can be done by fitting an analytic profile (e.g., Gaussian) to each of the cube's spectral lines. Alternatively, one can calculate the first- and second-order moments of each spectrum, which correspond to the velocity and velocity dispersion respectively.

\citet{1989AJ.....97.1227S} and \citet{anthony2009image} investigated the two methods thoroughly and assessed their accuracy, computational requirements and robustness. The first method can achieve superior accuracy compared to the second but it is computationally more expensive and may occasionally fail to converge, returning infinite or undefined results for some parameters. The moment-based method, although computationally faster, it can be heavily biased if the underlying continuum is asymmetric, contains multiple spectral lines, or the data have a low signal-to-noise ratio.

Both methods have successfully been used for galaxy kinematics in the literature (\citealt{2006ApJ...645.1062F}; \citealt{2008MNRAS.388..500E}). However, the moment-based approach is usually favoured when computation time is an issue \citep{2014MNRAS.437.1070G}. Furthermore, in cases where a continuum is clearly present in the data, more sophisticated methods and techniques are employed, such as template fitting to remove the continuum and then Gaussian line fitting to extract the emission lines, e.g. using \textsc{ppxf} \citep{2004PASP..116..138C} and \textsc{gandalf} \citep{2006MNRAS.366.1151S}.

In this work we followed the moment-based approach which is a transform and reduce operation. Each GPU thread is assigned to calculate the first- and second-order moments of one or more spectra. Because consecutive GPU threads process consecutive spectra and since the cube data have a slice-major alignment, memory coalesced access is achieved once again. For additional information as to why the moment-based approach was preferred over Gaussian fitting, see Section~\ref{sec:performance-results-analysis}.

\subsubsection{Residual and goodness-of-fit evaluation}
\label{sec:method-fitting-procedure-residual}

After obtaining the velocity and velocity dispersion model maps from the previous step we compare them to the input data and we evaluate the goodness-of-fit using a statistical test. The majority of the statistical tests require the residual between the observed and the modelled data, which is the pixel-wise difference between the two datasets weighted by the error associated with each measurement. This is a trivial transform and reduce operation which is performed in parallel on the GPU, producing a residual velocity map and a residual velocity dispersion map. Although coalesced memory access is achieved, this step is expected to be memory bound since it has a very low arithmetic intensity and it accesses the memory four times for every pixel in the output residual maps: three reads from the model, input and error maps, and one write to the residual map.

The goodness-of-fit calculation takes place on the CPU instead of the GPU. This was a software design decision we had to take in order to make the implementation of our method optimizer-agnostic. To calculate the goodness-of-fit we first have to transfer the residual maps from the GPU to the CPU memory. Although the size of the residual maps can be considered small, the transfer is done via the low-bandwidth PCI Express bus (PCIe, \citealt{budruk2004pci}). Therefore, this transfer might take up a big portion of the total iteration time if the fitted model involves a small amount of operations.

To evaluate the goodness of fit we use the following quantity \citep{2014MNRAS.437.1070G}:
\begin{equation}
\label{eq:goodness-of-fit}
\chi^2 = \sum_{\mathrm{i=1}}^{\mathrm{N}} \left[ \frac{(v_{\mathrm{i,model}}-v_{\mathrm{i,obs}})^{2}}{E(v_{\mathrm{i,obs}})^{2}} + W\frac{(\sigma_{\mathrm{i,model}}-\sigma_{\mathrm{i,obs}})^{2}}{E(\sigma_{\mathrm{i,obs}})^{2}} \right],
\end{equation}
where $\mathrm{N}$ is the number of pixels in the data, $v$ and $\sigma$ are the velocity and velocity dispersion, $E$ is the measurement error and $W$ is the importance of the velocity dispersion in the goodness-of-fit. Since $W$ is a constant, we incorporate it in the measurement error of the velocity dispersion prior the fitting procedure. In case a velocity dispersion field is not present in the fitting procedure the second term of equation~\ref{eq:goodness-of-fit} is ignored.

The value of $W$ is decided based on how well we expect the selected velocity dispersion model to describe the observed data. By using a small value, $W < 1$, we decrease the contribution of the velocity dispersion residual to the $\chi^{2}$ statistic, and we can still obtain high-quality velocity field fits, even if our velocity dispersion model does not describe the observed data very well. \citet{2014MNRAS.437.1070G} used $W = 0.2$ to ensure a good fit to the velocity fit, while they used the intensity weighted mean of the velocity dispersion ($\sigma_{\mathrm{m}}$, \citealt{2009ApJ...697.2057L}) to estimate the velocity dispersion of the galaxy.

\subsubsection{Optimization step}
\label{sec:method-fitting-procedure-optimization}

In this step the selected optimization method decides if the fitting procedure should continue or terminate. The termination criteria depend on the optimization algorithm, but they are usually related to the goodness of fit. If the termination criteria are not met the optimization technique will adjust the model parameter values and a new fitting iteration will take place.

For our implementation we used a modular software design which enabled us to use third party optimization libraries and switch between them during runtime. This allowed us to test our approach using different optimization methods without having to modify the source code each time. While this feature is not common in existing kinematic modelling software, it can be very useful since it enables us to cross-check our fitting results using the same software.

All the optimization libraries we have chosen for our software are CPU-based, thus their logic does not take advantage of the GPU parallelization. This limitation is not an important issue since the computational requirements of the optimization logic are expected to be negligible compared to the computational requirements of the rest of the steps involved in the fitting procedure. However, the above statement might be false if the fitted model is simple and has very small runtime requirements. For more details on how the runtime requirements of the optimizer's logic compare to the total runtime of the fitting procedure, see Section~\ref{sec:performance-results}.

%
%

\subsection{The optimization algorithms}
\label{sec:method-optimization-algorithms}

The optimization technique is one of the most important factors in the fitting procedure. It has the main impact on both the goodness-of-fit selected and the computational runtime. In this work we use three different optimization methods: the Levenberg--Marquardt Algorithm (LMA), Nested Sampling (NS) and brute-force optimization based on Nested Grids (NG). The three optimizers were chosen based on the approach they follow to explore the parameter space. This also translates to the number of model evaluations required before converging to a solution, and subsequently to the overall runtime requirements of the fitting procedure. For a brief discussion as to why Markov Chain Monte Carlo (MCMC) methods were not considered as practical solution in this work, see Appendix~\ref{app:mcmc}.

\subsubsection{The Levenberg--Marquardt Algorithm}
\label{sec:method-optimization-algorithms-lm}

The Levenberg--Marquardt Algorithm is one of the most commonly used optimization methods for galaxy kinematic modelling in astronomy (e.g., \citealt{2006MNRAS.371..170B}; \citealt{2008A&A...490..589C}; \citealt{2014MNRAS.437.1070G}). Starting from an initial set of parameter values, LMA uses gradient and curvature information to explore the parameter space. Compared to the other techniques, it requires a small number of model evaluations before converging to a solution. LMA is a local optimizer and will converge to the closest minimum to the initial guess. Thus, in order to find the global minimum and avoid getting trapped into a local one, the starting point of the optimization must be as close to the global minimum as possible.

In this work we have used the C version of the \textsc{mpfit}\footnote{\url{http://purl.com/net/mpfit}} library \citep{2009ASPC..411..251M} which implements both the `vanilla' LMA algorithm and a modified version that supports constrained optimization through box constraints. Although LMA provides a few opportunities for parallelization \citep{Cao:2009:PLA:1542275.1542338}, the \textsc{mpfit} library is not parallelized.

\subsubsection{Nested Sampling}
\label{sec:method-optimization-algorithms-ns}

The Nested Sampling algorithm is a Monte Carlo technique for efficient Bayesian evidence calculation and parameter estimation \citep{2004AIPC..735..395S}. The algorithm starts with a population of $N$ \emph{active} points drawn from a prior distribution. Then, the point with the lowest likelihood $\mathcal{L}_{\mathrm{i}}$ is recorded and replaced by a point drawn from the prior subject to the constraint that the new point has a likelihood $\mathcal{L} > \mathcal{L}_{\mathrm{i}}$. This procedure is repeated and the population of active points moves progressively to areas of higher likelihood. 

Several extensions to the original algorithm have been introduced to increase its efficiency and robustness. To improve the acceptance rate when drawing samples from the prior with the hard constraint $\mathcal{L} > \mathcal{L}_{\mathrm{i}}$, \citet{2006ApJ...638L..51M} introduced Ellipsoidal Nested Sampling. Their algorithm uses an elliptical bound around the current set of live points in order to restrict the region from which new samples are drawn. To improve the sampling efficiency for multimodal posteriors, \citet{2007MNRAS.378.1365S} introduced Clustered Nested Sampling which allows the formation of multiple ellipsoids centered on the modes of the posterior, and hence reducing the region of the prior sampled. This algorithm was further improved by \citet{2008MNRAS.384..449F} and \citet{2009MNRAS.398.1601F} to efficiently sample from posteriors with pronounced degeneracies between parameters. Finally, \citet{cameron2014} introduced Importance Nested Sampling which uses the discarded draws from the NS process (i.e., the sampled points failing to satisfy the constraint $\mathcal{L} > \mathcal{L}_{\mathrm{i}}$) and calculates the Bayesian evidence at up to an order of magnitude higher accuracy than the `vanilla' NS.

In this work we used the \textsc{multinest}\footnote{\url{http://ccpforge.cse.rl.ac.uk/gf/project/multinest}} library which implements the NS algorithm with all the extensions mentioned above. \textsc{multinest} is relatively easy to use since NS does not need an initial set of parameter values but rather the user has to define a prior, which in many problems is just a uniform distribution. In addition, its behaviour is controlled by only three main parameters: the number of active points $N$, the maximum efficiency $e$, and the tolerance $tol$.

\subsubsection{The Nested Grids Optimizer}
\label{sec:method-optimization-algorithms-ng}

The Nested Grid optimizer follows a brute-force approach in order to find the optimal solution to the problem. Initially a multi-dimensional regular grid representing the parameter space is defined. For each node of the grid a model is generated and a goodness-of-fit statistic is derived from it. Then, the node with the best goodness-of-fit is selected and a new grid is built around that area. The newly created grid has the same number of nodes and dimensions as the previous one, but it covers a smaller volume on the parameter space. This procedure is repeated until the area of the parameter space under the grid is smaller than a user-defined accuracy threshold $a$. The resolution of the grid has to be high enough to detect the features in the parameter space and not miss the global optimum. 

Although the Nested Grid optimizer is a brute-force method and it is expected to be computationally inefficient, previous research \citep{2011PASA...28...15F} has shown that, under certain circumstances, brute-force approaches can be acceptable when taking advantage of the enormous computational power of the GPU. Furthermore they are also conceptually easier to implement on GPUs as a first solution.

%
%

\section {The Data}
\label{sec:data}

In order to assess our method we tested our software using mock data. This allowed us to specify the exact configuration of the observations, including the spatial and spectral resolution (i.e., the point and line spread functions), the galaxy's kinematic and morphological properties, and the amount of noise in the data. Since the simulated galaxies are generated with known model parameter values, we can use mock data to evaluate the correctness, accuracy and robustness of our fitting method.

Furthermore, we also investigated how our method behaves under real science scenarios by using data from the GHASP \citep{2008MNRAS.388..500E} and DYNAMO \citep{2014MNRAS.437.1070G} surveys. The two surveys were chosen based on their spatial and spectral resolution. GHASP was selected as a representative of high-resolution survey, while DYNAMO is a good example of a survey with low spatial resolution where correct PSF treatment is essential. Other examples of low-spatial resolution surveys are SINS \citep{2006ApJ...645.1062F} and MASSIV \citep{2012A&A...539A..91C}. Summary information about all the data used in this work can be found in Table~\ref{tab:data-info}.

\begin{table*}
	\centering
	\begin{minipage}{165mm}
	\caption{Information about the different datasets described in Section~\ref{sec:data}. We used four versions of the same mock data, each version has a different level of noise: 0, 1, 5, and 20~km~s$^{-1}$ on the velocity maps, and 0, 1.4, 7, and 28~km~s$^{-1}$ on the velocity dispersion maps.}
	\begin{tabular}{ccccccc}
	\hline	
	Survey name & Number of galaxies & Redshift	  & Number of pixels	   & Spatial sampling      & Spectral resolution & Seeing       \\
				&                    &            &                        & (arcsec pixel$^{-1}$) & (km~s$^{-1}$)       & (arcsec)	    \\
	\hline
	Mock data   & 1000               & --         &  $32^{2}$ -- $96^{2}$  & $1.0$                 & $10$ -- $60$        & $4.0$        \\
	GHASP       & 173                & $\sim0$    & $256^{2}$ or $512^{2}$ & $0.96$ or $0.68$      & $\sim30$            & $3.0$ (mean) \\
	DYNAMO      & 44                 & $\sim0.1$  & $16\times32$           & $0.7$                 & $\sim30$            & $1.4$ (mean) \\
	\hline
	\end{tabular}
	\label{tab:data-info}
	\end{minipage}
\end{table*}

%
%

\subsection{The mock data}
\label{sec:data-mock}

Each mock observation consists of an intensity cube convolved with a 3D Gaussian PSF kernel (equation~\ref{eq:gauss3d}). Galaxies are modelled using the arctan rotation curve profile (equation~\ref{eq:model-rcurve-arctan}). A velocity map and a velocity dispersion map are then extracted from the convolved intensity cube by calculating the first- and second-order moments of the spectral lines respectively. In addition, an intensity map is extracted by integrating the intensity across each spectral line. Finally, a mask image is defined in order to simulate pixels with low signal-to-noise ratio and instrument faults (e.g., dead detector pixels or IFU fibres).

For each velocity and velocity dispersion map we created four distinct versions, each with a different level of noise. For the velocity map we used Gaussian noise of 0, 1, 5, and 20~km~s$^{-1}$. The noise on the velocity dispersion map, since it is a second-order moment, was chosen to be $\sim\sqrt{2}$ times larger\footnote{This mathematically motivated choice was validated by fitting Gaussian profiles to simulated spectra with different signal-to-noise ratios, showing that it is a good approximation in most circumstances.}, resulting noise levels of 0, 1.4, 7, and 28~km~s$^{-1}$. The zero-noise maps (which we will also refer to as the \emph{clean dataset}) can be used to verify the accuracy of our method, while the last three can be used to evaluate how our method behaves under low- and high-noise level scenarios. Gaussian noise is not always realistic since real data can contain correlated noise between adjacent pixels (e.g., due to non-circular motions or sky line residual artefacts). However, the inclusion of such noise would not affect our timing results significantly as it does not change the conceptual structure of the problem, and hence it was not explored further.

We generated 1000 mock datacubes by randomly sampling the parameter space defined by the resolution of the observation, and the galaxy's morphological and kinematic parameters (Table~\ref{tab:data-info-moc}). To reduce the degrees of freedom of this parameter space we used a fixed point spread function size, while the turn-over radius $r_{\mathrm{t}}$ and scale length $r_{\mathrm{0}}$ were calculated as a function of the point spread function's FWHM. In addition, we assumed that the systemic velocity $V_{\mathrm{sys}}$ has already been subtracted from the moment maps and thus it is equal to zero for all galaxies. This is a reasonable decision since the value of the systemic velocity is not expected to affect our fitting results.

\begin{table}
	\centering
	\begin{minipage}{84mm}
	\caption{Parameter space used to generate the mock dataset.}
	\begin{tabular}{cccc}
	\hline	
	Parameter & Units & Range (min) & Range (max)                                                                    \\
	\hline
	$N_{\mathrm{data,xy}}$    & pixels      & 32                                 & 96                                \\
	$PSF_{\textsc{FWHM}}$     & pixels      & 4.0 (fixed)                        & 4.0 (fixed)                       \\
	$LSF_{\textsc{FWHM}}$     & km~s$^{-1}$ & 20                                 & 60                                \\
	$V_{\mathrm{sys}}$        & km~s$^{-1}$ & 0 (fixed)                          & 0 (fixed)                         \\
	$x_{\mathrm{o}}$          & pixels      & $0.4 \times N_{\mathrm{data,xy}}$  & $0.6 \times N_{\mathrm{data,xy}}$ \\
	$y_{\mathrm{o}}$          & pixels      & $0.4 \times N_{\mathrm{data,xy}}$  & $0.6 \times N_{\mathrm{data,xy}}$ \\
	$PA$                      & degrees     & 0                                  & 45                                \\
	$i$                       & degrees     & 5                                  & 85                                \\
	$r_{\mathrm{t}}$          & pixels      & $0.5 \times PSF_{\textsc{FWHM}}$   & $2.0 \times PSF_{\textsc{FWHM}}$  \\
	$r_{\mathrm{0}}$          & pixels      & $1.0 \times PSF_{\textsc{FWHM}}$   & $3.0 \times PSF_{\textsc{FWHM}}$  \\
	$V_{\mathrm{t}}$          & km~s$^{-1}$ & 50                                 & 300                               \\
	$\sigma_{\mathrm{model}}$ & km~s$^{-1}$ & 20                                 & 100                               \\
	\hline
	\end{tabular}
	\label{tab:data-info-moc}
	\end{minipage}
\end{table}

%
%

\subsection{The GHASP data}
\label{sec:data-ghasp}

The Gassendi H$\alpha$ survey of SPirals (GHASP, \citealt{2008MNRAS.388..500E}) consists of 3D H$\alpha$ data of 203 spiral and irregular local ($z\sim0$) galaxies covering a wide range of luminosities and morphological types.

The observations were made with a Fabry-P\'erot interferometer attached at the Cassegrain focus of the 1.93 m telescope at the Observatoire de Haute-Provence in France. Over the course of the observations two different detectors were used: (1) a Thomson IPCS with $256 \times 256$~pixels of 0.96~arcsec coverage each resulting in a field of view of $4 \times 4$~arcmin; (2) a GaAs IPCS with $512 \times 512$~pixels of 0.68~arcsec coverage each resulting in a field of view of $5.8 \times 5.8$~arcmin. The spectral resolution of the instrument is 30~km~s$^{-1}$ while the spatial resolution is limited by the seeing which is  $\sim3$~arcsec and corresponds to 0.02 -- 1.4~kpc depending on the redshift of the observed object.

Adaptive spatial binning techniques based on 2D Voronoi tessellations were applied to the datacubes in order to optimize the spatial resolution to the signal-to-noise ratio. From the resulting cubes a series of two-dimensional images was extracted including radial velocity maps, velocity dispersion maps and continuum maps. 

We have selected a subsample of the GHASP survey consisting of 173 galaxies. The other 30 galaxies were not considered in this work because their kinematic analysis results were not available.

%
%

\subsection{The DYNAMO data}
\label{sec:data-dynamo}

The DYnamics of Newly Assembled Massive Objects (DYNAMO, \citealt{2014MNRAS.437.1070G}) survey consists of spatially resolved local ($z\sim0.1$) galaxies with high star formation rates (SFR), selected primarily by H$\alpha$ luminosity. The main aim of this survey is to study galaxy evolution through comparison with high-redshift galaxies. 

The IFS observations of the DYNAMO survey were obtained using different instruments on different telescopes, with the largest sample (54 galaxies) being the one coming from the SPIRAL Integral Field Unit \citep{2001PASP..113..215K} attached on the 3.9 m Anglo-Australian telescope, located at the Siding Spring Observatory in Australia. The SPIRAL IFU is a lenslet array of $32 \times 16$~pixels with each pixel having 0.7~arcsec coverage resulting in a field of view of $22.4 \times 11.2$~arcsec. SPIRAL was connected to the AAOmega spectrograph \citep{2006SPIE.6269E..0GS} which was configured to provide a coverage of 500{\AA} and a resolving power of R~$\sim12000$. The spatial resolution of the observations was limited by the seeing with a mean of 1.4~arcsec which corresponds to 1.5 -- 3.6~kpc depending on the redshift of the observed object and it is comparable to high-redshift samples observed with adaptive optics. The spectral resolution of the instrument was $\sim30$~km~s$^{-1}$.

Available DYNAMO data include velocity, velocity dispersion and mask maps. For the first two quantities, measurement error maps are also provided. 

For this work we have selected a subset of 44 galaxies observed with the SPIRAL instrument and studied in \citet{2014MNRAS.437.1070G}. We discarded galaxies with complex kinematics (i.e., galaxies whose both velocity and velocity dispersion vary significantly from regular discs) for which kinematic analysis results were not available.

%
%

\section{Fitting Results}
\label{sec:fitting-results}

To examine the consistency and robustness of our method we fit the three different kinds of data which are described in Section~\ref{sec:data}. When fitting data from the GHASP and DYNAMO surveys our method and metrics are adjusted to match the ones found in the literature (\citealt{2008MNRAS.390..466E} and \citealt{2014MNRAS.437.1070G} respectively) as closely as possible. However, some details are not available in the literature, such as the initial model parameter values and the configuration of the optimization algorithms, therefore some difference between the results is expected.

To verify the results, for each observation we run the fitting procedure multiple times using different optimization methods. For all three datasets we use similar optimizer configurations. The LMA optimizer is adjusted to use a finite-differences Jacobian with a step size of $\sim0.1$ for all model parameters. The NS optimizer is configured for $N=50$ active points, a maximum efficiency $e=1.0$ and a tolerance $tol=0.5$. For the NG optimizer we choose an accuracy threshold $a=0.01$ for $r_{\mathrm{t}}$, and $a=0.1$ for the rest of the model parameters. The number of iterations needed to fit using the NG optimizer with all the model parameters free can be quite high and result in prohibitive runtime requirements. To avoid this, we fix the parameters $V_{\mathrm{sys}}$, $x_{\mathrm{o}}$, $y_{\mathrm{o}}$ and $i$ to their actual values (for the mock data) or to a previous best fit (derived by using another optimizer). The grid resolution of the NG optimizer is 7, 11, 11, 5, 5, 5 for the parameters $PA$, $r_{\mathrm{t}}$, $V_{\mathrm{t}}$, $\sigma_{\mathrm{model}}$, $a$, and $g$ respectively. 

%
%

\subsection{Fitting the mock data}
\label{sec:fitting-results-mock}

First, the kinematic model is fitted to the mock data with zero noise. This allows us to evaluate the accuracy of our method but also investigate the robustness of the three optimization techniques. Due to the absence of noise in the data, the method for calculating the initial values for the parameters (Section~\ref{sec:method-models-initialvalues}) may find the best fit for $V_{\mathrm{sys}}$, $x_{\mathrm{o}}$, $y_{\mathrm{o}}$, and $PA$ immediately. When fitting real data, this is rarely the case since the observations include noise and non-circular motions. To account for this, after estimating the initial values for the above parameters, we modified them by a random amount. Also, we assume no prior knowledge about the galaxy's inclination. Therefore, the initial guess for the inclination parameter is set to 45\degr and its prior constraints between 1\degr and 89\degr. During the goodness-of-fit calculation step of the fitting procedure (Section~\ref{sec:method-fitting-procedure-residual}), the dispersion map is taken into account using equation~\ref{eq:goodness-of-fit} with an importance of $W = 1$.

\begin{figure*}
\centering
	\begin{tabular}{cc}
	\includegraphics[width=0.46\textwidth]{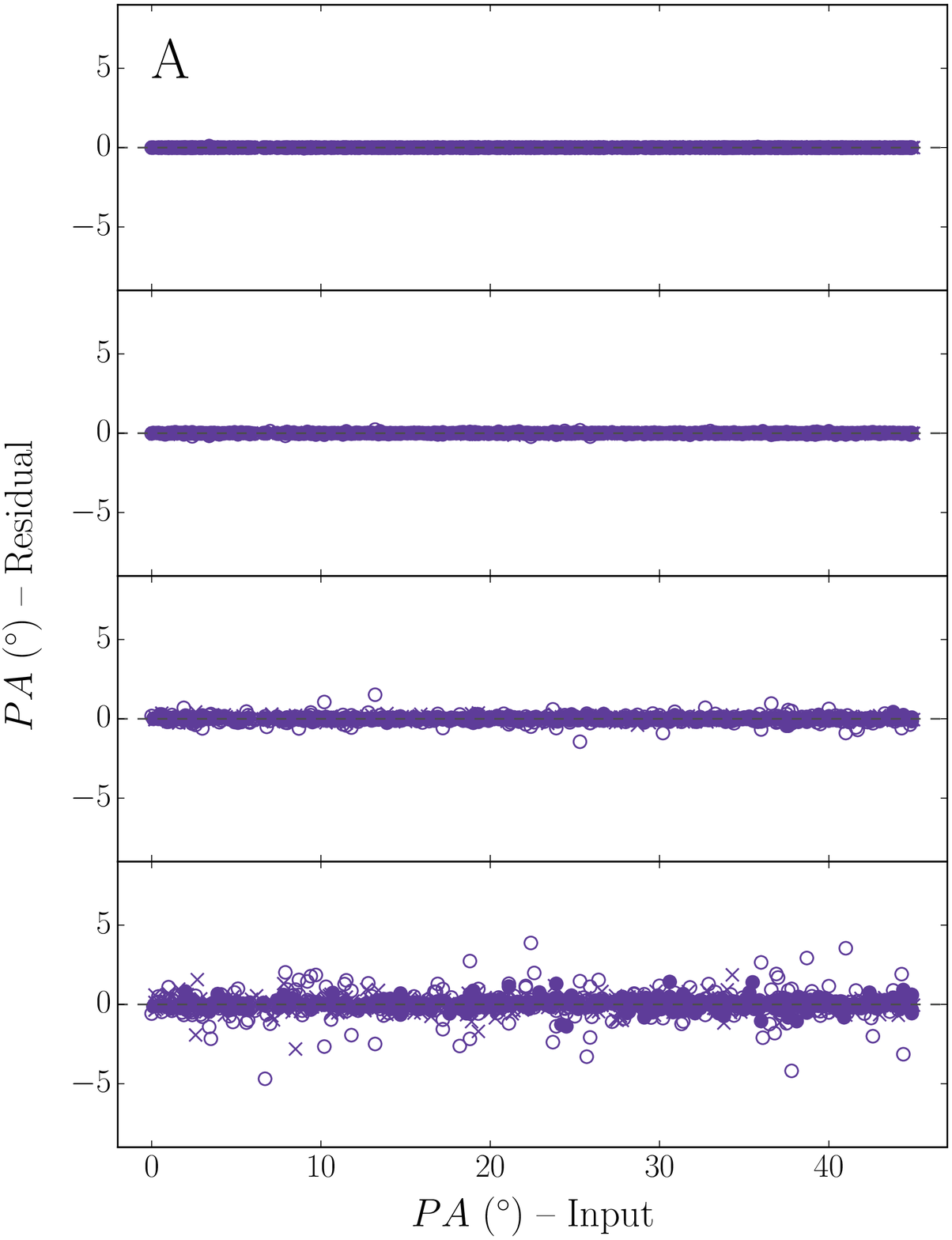} &
	\includegraphics[width=0.46\textwidth]{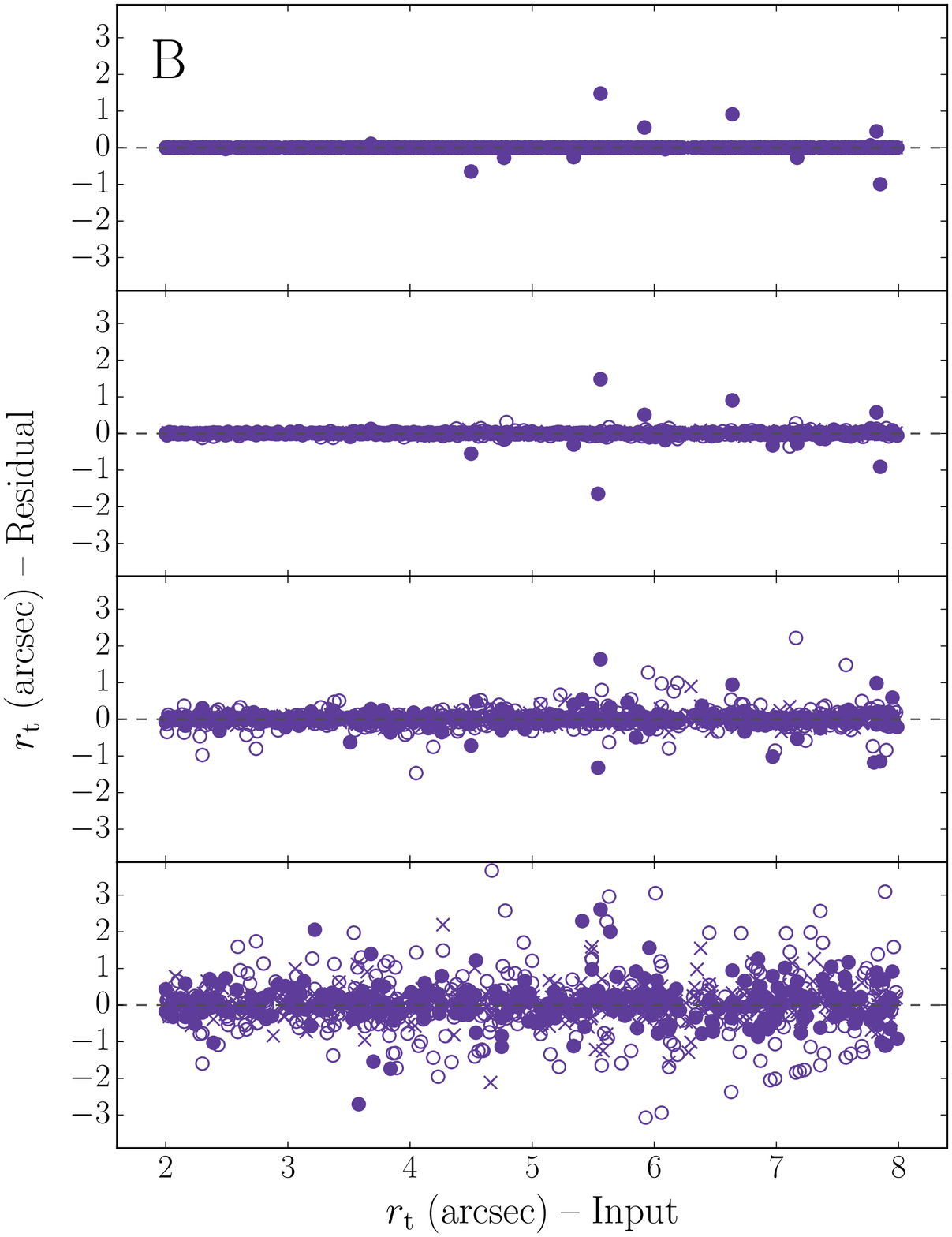} \\
	\includegraphics[width=0.46\textwidth]{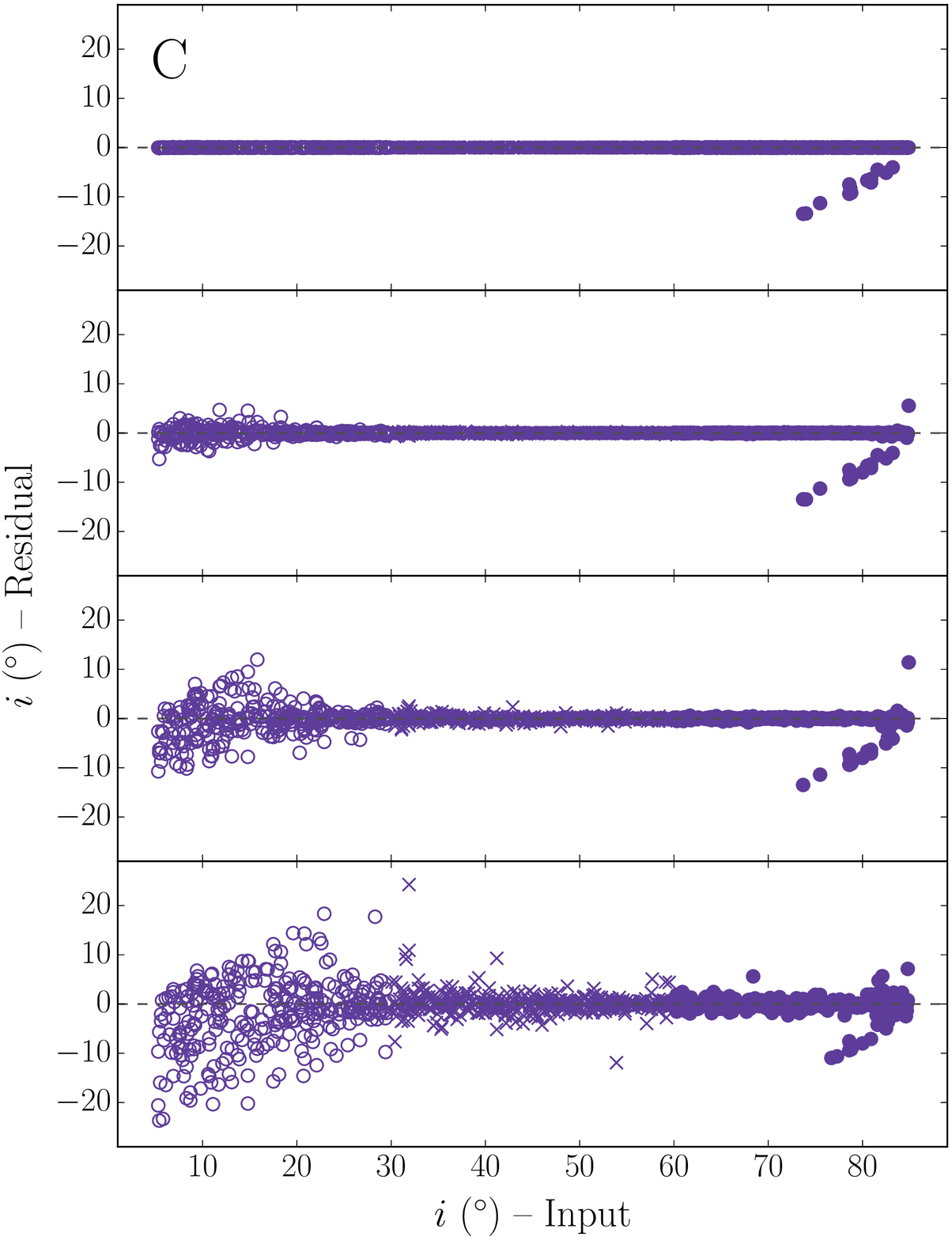} &
	\includegraphics[width=0.46\textwidth]{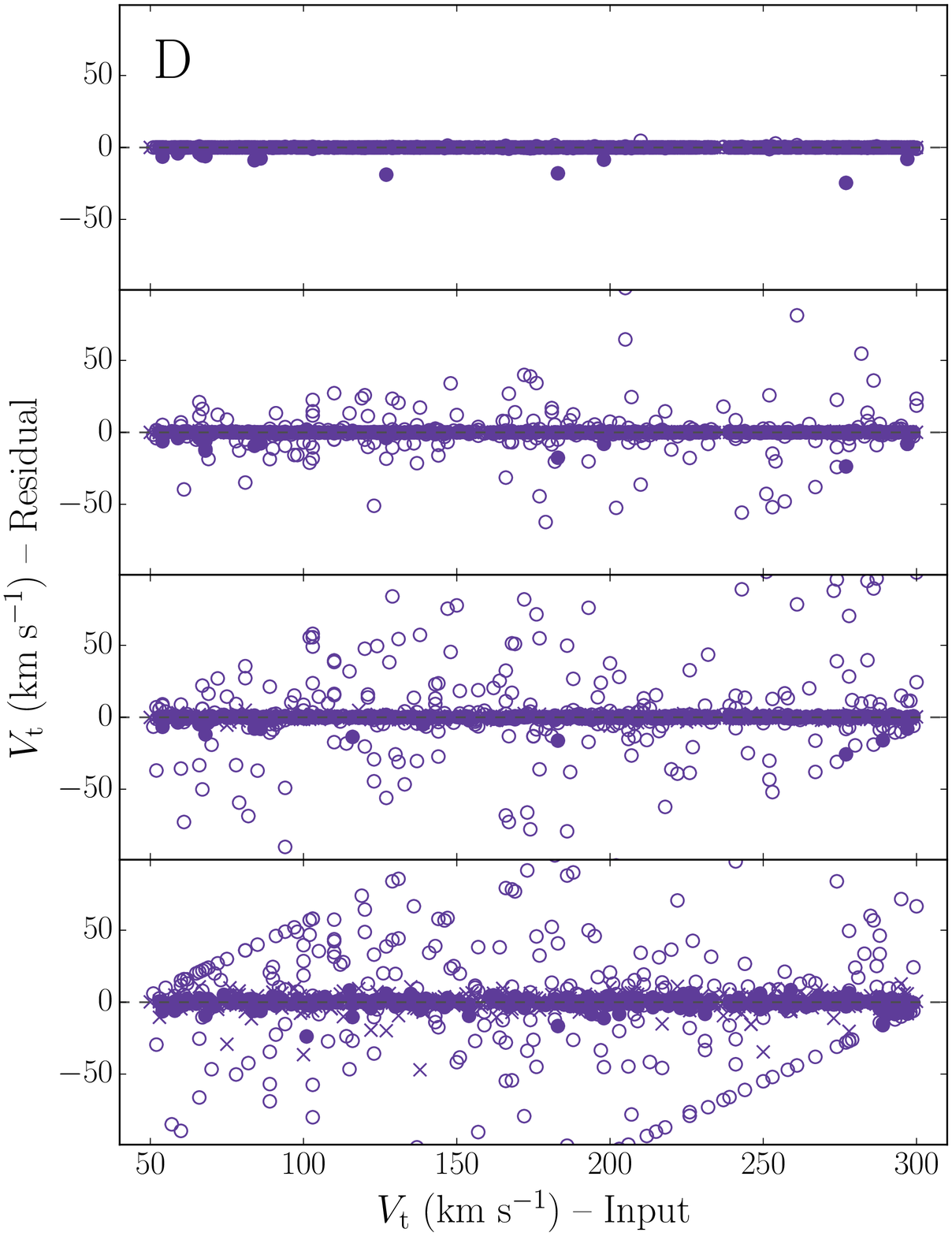} \\
	\end{tabular}
	\caption{Fitting results of the mock data with different levels of noise using the Levenberg--Marquardt Algorithm. We plot the residuals of the fitted parameters as a function of the input parameters. (A)~Position angle. (B)~Turn-over radius. (C)~Disc inclination. (D)~Asymptotic circular velocity. In each of the plots, the rows from top to bottom correspond to datasets with 0, 1, 5, and 20~km~s$^{-1}$ noise on the velocity map, and 0, 1.4, 7, and 28~km~s$^{-1}$ noise on the velocity dispersion map. Different markers are used according to the input inclination: open circles for $ 5^{\circ} \leq i_{\mathrm{Input}} < 30^{\circ}$, crosses for $30^{\circ} \leq i_{\mathrm{Input}} < 60^{\circ}$, and filled circles for $60^{\circ} \leq i_{\mathrm{Input}} < 85^{\circ}$. The error bars are omitted for the sake of clarity.}
	\label{fig:fitting-results-mock}
\end{figure*}

The fitting results for the clean dataset using the LMA optimizer are shown in the first row of Fig.~\ref{fig:fitting-results-mock}A, B, C, and D. Overall, the accuracy of our method with all three optimizers is excellent: The residuals between the actual and fitted parameters for $\gtrsim98$ per cent of the galaxies are less than a reasonable threshold (Table~\ref{tab:fitting-results-mock-1}). In addition, the median residual of each parameter is well below that threshold by 1~--~3 orders of magnitude (Table~\ref{tab:fitting-results-mock-2}). The few inconsistencies seen are mostly for galaxies with high inclinations. 

Low-resolution mock observations of nearly edge-on galaxies, created using a thin and flat disc model, can result in a very small number of pixels in the velocity field. This, in combination with the PSF convolution, can introduce new local minima in the $\chi^2$ landscape and eventually lead the LMA to a wrong solution. In any case, when modelling close to edge-on galaxies, the thickness of the galaxy disc plays an important role, therefore a thin disc model is not a good representative of the observation. Furthermore, the NG optimizer can also converge to a wrong solution if the grid resolution is not sufficient to capture the gradient of the posterior modes. However, we found that we can achieve a $\sim100$ per cent success rate if we re-run the fitting procedure using different optimizer settings or initial model parameter values/constraints.

To assess how our method behaves with noisy observations our kinematic model is fitted to all three noisy versions of the mock data. This allows us to investigate how different levels of noise can affect the final fitting results and if the kinematic properties of the galaxies are actually recoverable. However, as it has already been mentioned in Section~\ref{sec:data-mock}, the Gaussian noise in the noisy datasets does not give any indication on the level of non-circular motions that may also prevent the model to describe correctly the rotating component.

The fitting results for the noisy datasets using the LMA optimizer are shown in the second, third, and fourth rows of Fig.~\ref{fig:fitting-results-mock}A, B, C, and D. Estimating the inclination and/or the circular velocity of nearly face-on galaxies can be quite problematic due to the degeneracy between the two \citep{1987PhDT.......199B}. Indeed, a scatter is observed for galaxies with $i\lesssim20\degr$ (Fig.~\ref{fig:fitting-results-mock}C, empty circles), which subsequently causes a scatter in the circular velocity (Fig.~\ref{fig:fitting-results-mock}D). Overall, the parameters of the majority of the galaxies are recovered with a reasonable accuracy: $\sim68$ and $\sim95$ per cent of the fitted parameters have a standardised residual within $1\sigma$ and $2\sigma$ confidence intervals respectively (Table~\ref{tab:fitting-results-mock-3}).

Further discussion on the fitting results of the mock dataset as well as the degeneracy between the galaxy inclination and circular velocity can be found in Section~\ref{sec:discussion}.

\begin{table*}
	\begin{minipage}{148mm}
		\caption{Number of fits with a parameter residual smaller than the specified threshold for the mock dataset for a total of 1000 zero-noise mock observations.}
		\begin{tabular}{cccccccc}
		\hline
	    Optimization
	    & $x_{\mathrm{o}}$  
	    & $y_{\mathrm{o}}$  
	    & $PA$    
	    & $i$     
	    & $r_{\mathrm{t}}$  
	    & $V_{\mathrm{t}}$                    
	    & $\sigma_{\mathrm{model}}$
	    \\
	    method
	    & ($<1$~pixel) 
	    & ($<1$~pixel) 
	    & ($<1$~\degr) 
	    & ($<1$~\degr) 
	    & ($<1$~pixel) 
	    & ($<5$~km~s$^{-1}$) 
	    & ($<1$~km~s$^{-1}$) 
	    \\
		\hline
		LMA    & 1000  & 1000  & 1000  & 987   & 999   & 990   & 992   \\
		NS     & 1000  & 1000  & 1000  & 1000  & 1000  & 1000  & 1000  \\
		NG     & 1000  & 1000  & 998   & 1000  & 953   & 990   & 971   \\
		\hline
		\end{tabular}
		\label{tab:fitting-results-mock-1}
	\end{minipage}
\end{table*}

\begin{table*}
	\begin{minipage}{148mm}
		\caption{Median of the residuals between actual and fitted parameters for the zero-noise mock dataset.}
		\begin{tabular}{cccccccc}
		\hline
	    Optimization 
	    & $x_{\mathrm{o}}$ 
	    & $y_{\mathrm{o}}$ 
	    & $PA$ 
	    & $i$ 
	    & $r_{\mathrm{t}}$ 
	    & $V_{\mathrm{t}}$ 
	    & $\sigma_{\mathrm{model}}$ 
	    \\
	    method
	    & (pixel) 
	    & (pixel) 
	    & (\degr) 
	    & (\degr) 
	    & (pixel) 
	    & (km~s$^{-1}$) 
	    & (km~s$^{-1}$) 
	    \\
		\hline
		LMA&  $0.7\times10^{-3}$ & $0.6\times10^{-3}$ & $0.1\times10^{-2}$ & $0.5\times10^{-2}$ & $1.3\times10^{-3}$ & $2.2\times10^{-2}$ & $1.6\times10^{-3}$ \\
		NS & $0.7\times10^{-3}$ & $0.7\times10^{-3}$ & $0.1\times10^{-2}$ & $0.5\times10^{-2}$ & $1.4\times10^{-3}$ & $2.5\times10^{-2}$ & $1.8\times10^{-3}$ \\
		NG & --                 & --                 & $0.9\times10^{-1}$ & --                 & $1.1\times10^{-2}$ & $9.8\times10^{-2}$ & $9.7\times10^{-2}$ \\
		\hline
		\end{tabular}
		\label{tab:fitting-results-mock-2}
	\end{minipage}
\end{table*}

\begin{table*}
	\begin{minipage}{152mm}
		\caption{Number of fits with a parameter standardised residual within a specified threshold for a total of 1000 mock observations.}
		\begin{tabular}{ccccccccc}
		\hline
	    Optimization
	    & Noise ($v$/$\sigma$)
	    & $x_{\mathrm{o}}$ 
	    & $y_{\mathrm{o}}$ 
	    & $PA$ 
	    & $i$ 
	    & $r_{\mathrm{t}}$ 
	    & $V_{\mathrm{t}}$ 
	    & $\sigma_{\mathrm{model}}$ 
	    \\
	    method
	    & (km~s$^{-1}$) 
	    & ($1\sigma/2\sigma$) 
	    & ($1\sigma/2\sigma$) 
	    & ($1\sigma/2\sigma$) 
	    & ($1\sigma/2\sigma$) 
	    & ($1\sigma/2\sigma$) 
	    & ($1\sigma/2\sigma$) 
	    & ($1\sigma/2\sigma$) 
	    \\
		\hline
		LMA     & 1/1.4   & 657/939       & 663/943       & 709/957       & 671/934       & 676/937       & 660/927       & 684/944       \\
		NS      & 1/1.4   & 638/935       & 646/942       & 688/938       & 663/941       & 683/939       & 649/933       & 683/947       \\
		NG      & 1/1.4   & 1000/1000     & 1000/1000     & 991/998       & 1000/1000     & 794/896       & 766/878       & 988/993       \\
		
		LMA     & 5/7     & 694/948       & 661/949       & 676/948       & 684/925       & 707/941       & 662/910       & 698/943       \\
		NS      & 5/7     & 660/935       & 641/942       & 657/950       & 655/929       & 679/944       & 647/926       & 682/944       \\
		NG      & 5/7     & 1000/1000     & 1000/1000     & 935/967       & 1000/1000     & 697/877       & 636/812       & 912/955       \\
		
		LMA     & 20/28   & 657/951       & 689/959       & 698/961       & 654/879       & 651/930       & 632/849       & 702/945       \\
		NS      & 20/28   & 642/935       & 678/950       & 684/946       & 641/923       & 635/911       & 657/921       & 678/941       \\
		NG      & 20/28   & 1000/1000     & 1000/1000     & 867/950       & 1000/1000     & 708/933       & 724/912       & 803/951       \\
		\hline
		\end{tabular}
		\label{tab:fitting-results-mock-3}
	\end{minipage}
\end{table*}

%
%

\subsection{Fitting the GHASP data}
\label{sec:fitting-results-ghasp}

To fit the GHASP data we employed the velocity field modelling strategy (Section~\ref{sec:method-fitting-procedure-vfieldval}) and the Epinat rotation curve profile (equation~\ref{eq:model-rcurve-epinat}), resulting in a nine-parameter problem. However, during the fitting procedure we adopted the kinematic centre used in \citet{2008MNRAS.390..466E}, which reduces the number of free parameters in our model to seven. Furthermore, in order to achieve a good fit for some observations with low signal-to-noise ratio or spatial coverage, the position angle and/or the inclination were fixed to their morphological values.

The model parameter errors returned by the optimizer are unrealistically small. To obtain more realistic errors we use a Monte Carlo method which takes into account the residual velocity field resulting from the fitting procedure. The residual velocity field has a non-uniform structure which is mainly caused by non-circular motions (e.g., bars, inflows, outflows, disc warps etc.). We simulate 200 residual velocity fields from the real velocity field by randomly shuffling the positions of its pixels. Then, we generate 200 simulated velocity fields by adding the simulated residuals to the velocity field of the best fit. Finally, we compute the standard deviation of the best-fitting parameters over the 200 simulated velocity fields.

\begin{figure*}
	\centering
	\includegraphics[width=1.0\textwidth]{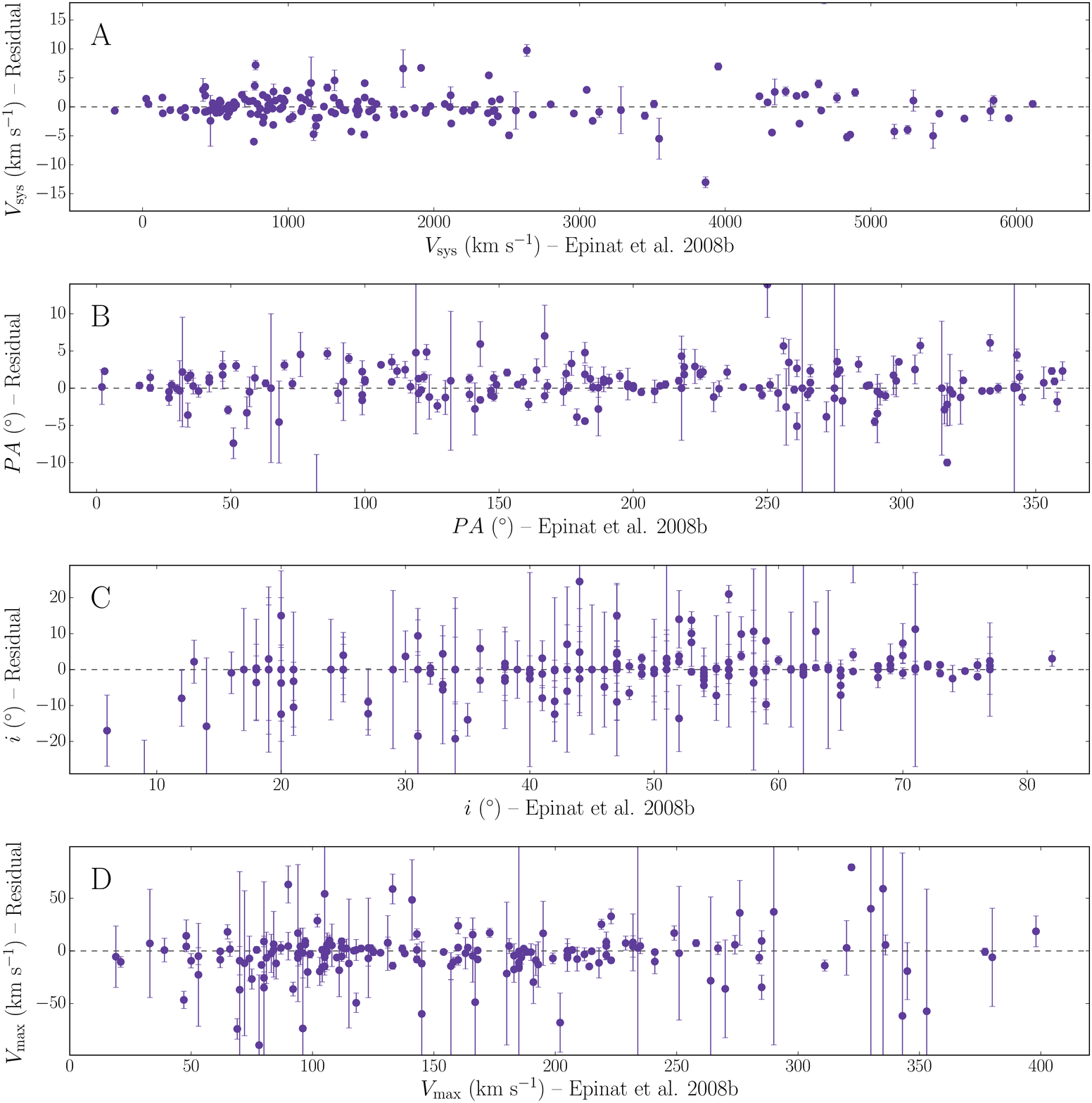}
	\caption{Fitting results of the GHASP sample. We plot the residuals of the fitted parameters as a function of the kinematic analysis results found in \citet{2008MNRAS.390..466E}. (A)~Systemic velocity. (B)~Position angle. (C)~Disc inclination. (D)~Maximum circular velocity.}
	\label{fig:fit-results-ghasp}
\end{figure*}

In Fig.~\ref{fig:fit-results-ghasp} we plot the residuals between our best fits and the best fits of \citet{2008MNRAS.390..466E}. For most of the galaxies there is a good agreement between the results. The median of the residuals for the fitted parameters are: $\sim2$~km~s$^{-1}$ for the systemic velocity, $\sim1\degr$ for the position angle, $\sim2\degr$ for the inclination, and $\sim8$~km~s$^{-1}$ for the maximum circular velocity. Nonetheless, a few inconsistencies between the results are observed. This is not surprising since the optimizer configuration and priors used in this work are most likely different from the ones used in \citet{2008MNRAS.390..466E}. Furthermore, our fitting procedure was completely automated and we did not treat any galaxy differently. For a complete list of the fitting results see Table~\ref{tab:fitting-results-ghasp}.

%
%

\subsection{Fitting the DYNAMO data}
\label{sec:fitting-results-dynamo}

To fit the DYNAMO data we employed the intensity cube modelling strategy (Section~\ref{sec:method-fitting-procedure-cubeeval}) and the arctan rotation curve profile (equation~\ref{eq:model-rcurve-arctan}), resulting in an eight-parameter problem. Due to the low spatial resolution of the data and the degeneracy between the circular velocity and disc inclination, the latter is kept fixed to the photometric value measured by the SDSS photometric pipeline for the \textit{r}-band exponential disc fit. This reduces the number of free parameters in our model to seven. We down-weight the contribution of the velocity dispersion in our goodness-of-fit metric (equation~\ref{eq:goodness-of-fit}) using $W = 0.2$. This is the same value \citet{2014MNRAS.437.1070G} used in their analysis in order to achieve a high quality fit to the velocity field even if the selected velocity dispersion model can not describe the observed data very well.

\begin{figure*}
	\centering
	\includegraphics[width=1.0\textwidth]{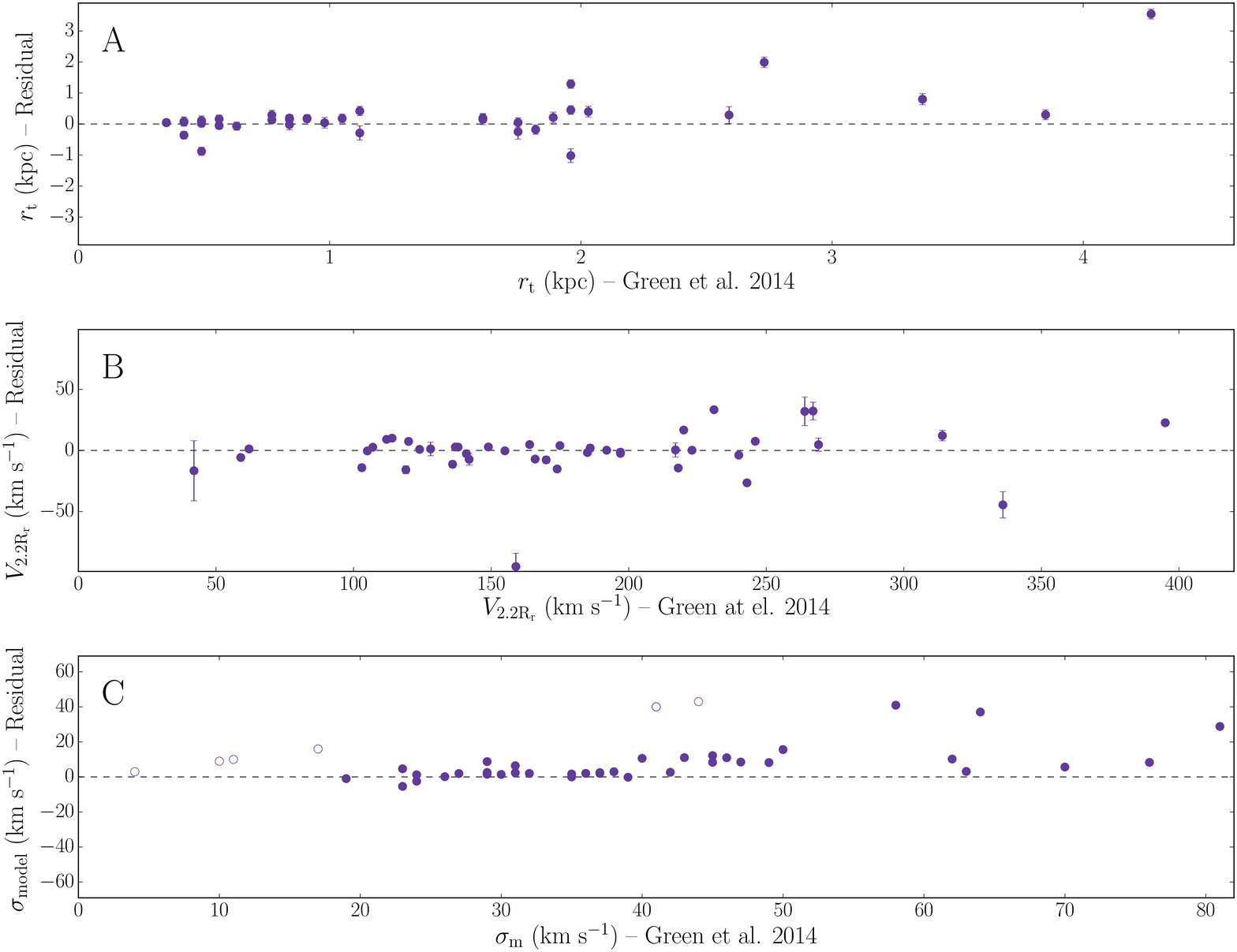}
	\caption{Fitting results of the DYNAMO sample. We plot the residuals of the fitted parameters as a function of the kinematic analysis results found in \citet{2014MNRAS.437.1070G}. (A)~Turn-over radius. (B)~Circular velocity at a radius of 2.2~$\times$~\textit{r}-band exponential disc scale lengths. (C)~Fitted velocity dispersion ($\sigma_{\mathrm{model}}$) against the velocity dispersion ($\sigma_{\mathrm{m}}$) found in \citet{2014MNRAS.437.1070G}. The open circles correspond to galaxies for which our fitting procedure converged to unrealistically small values.}
	\label{fig:fit-results-dynamo}
\end{figure*}

In order to get a better estimate of the uncertainties on the model parameters we use a Monte Carlo method. For each galaxy, we compute the standard deviation of the best-fitting parameters over 100 simulated velocity and velocity dispersion fields. The simulated fields are created by perturbing the pixel values of the actual data based on the measurement error maps. In Fig.~\ref{fig:fit-results-dynamo} we plot the residuals between our best fits and the best fits of \citet{2014MNRAS.437.1070G}.

Due to an error, the values of $r_{\mathrm{t}}$ reported in \citet{2014MNRAS.437.1070G} do not account for the spatial sampling of the instrument which is 0.7~arcsec~pixel$^{-1}$ (A. Green, private communication, 2014), therefore a correction factor of 0.7 is applied prior the residual calculation. A small scatter is observed in the $r_{\mathrm{t}}$ residuals (Fig.~\ref{fig:fit-results-dynamo}A). This is because \citet{2014MNRAS.437.1070G} used a fixed exponential surface brightness profile (i.e., constant scale radius of 7.5~pixels) for all the galaxies in the sample, while we used a varying one based on the scale radii found in the SDSS database ($r$-band photometry). However, the change to a varying scale radius does not affect the results and science conclusions of \citet{2014MNRAS.437.1070G}.

A very good agreement is observed in the plot of the circular velocity (Fig.~\ref{fig:fit-results-dynamo}B). For the circular velocity we used the velocity evaluated at a radius of 2.2~$\times$~scale lengths on the fitted rotation curve profile \citep{1970ApJ...160..811F}.

Our velocity dispersion results appear to be in a good agreement with the previous work (Fig.~\ref{fig:fit-results-dynamo}C). However, a few inconsistencies are observed. This is expected since the velocity dispersion in our case is derived from the fitting procedure, while the velocity dispersion in \citet{2014MNRAS.437.1070G} is calculated using intensity weighted mean of the observed velocity dispersion map ($\sigma_{\mathrm{m}}$, \citealt{2009ApJ...697.2057L}). Furthermore, the intrinsic velocity dispersion of some galaxies is too small to be measured given the instrument's spectral resolution and signal-to-noise ratio, thus our method converges to unrealistically small values (Fig.~\ref{fig:fit-results-dynamo}C, empty circles). For a complete list of the fitting results see Table~\ref{tab:fitting-results-dynamo}.

%
%

\section{Performance Results}
\label{sec:performance-results}

In this section we investigate the performance of our fitting method and assess the benefits of using the graphics processing unit. In order to obtain a better picture of the acceleration achieved by the GPU, a series of tests are performed on different hardware configurations using CPU-based and GPU-based versions of our code. We highlight good and bad uses of the GPU, we investigate how the performance scales with the size of our data and we comment on the memory requirements.

%
%

\subsection{Hardware and software configuration}
\label{sec:performance-results-hardware}

To understand how the performance of our method compares to CPU-based approaches, in addition to the main GPU implementation we have developed a single-threaded and a multi-threaded version of our code which runs on the CPU.

For the multi-threaded CPU implementation we used the \textsc{openmp}\footnote{\url{http://www.openmp.org}} library \citep{660313} to parallelize the code, and we chose a thread count higher than the number of available CPU cores in order to hide the memory access latency \citep{Kurihara91latencytolerance}. The \textsc{fftw3}\footnote{\url{http://www.fftw.org}} library \citep{1386650} was used to perform the FFTs on the CPU. All the allocations on the main system memory were done using a 16-byte alignment in order to give the compiler and the \textsc{fftw3} library the chance to use the SIMD capabilities of the CPU \citep{Eichenberger:2004:VSA:996893.996853}. For the GPU implementation of our method we used the Nvidia Compute Unified Device Architecture (\textsc{cuda}\footnote{\url{http://developer.nvidia.com/about-cuda}}) library to parallelize the code and the \textsc{cufft}\footnote{\url{http://developer.nvidia.com/cufft}} library to perform the FFTs.

The performance tests was run on the following four hardware configurations:
\begin{enumerate}
\renewcommand{\theenumi}{(\arabic{enumi})}
\item 1~$\times$~Intel Xeon E5-2660 in single-threaded mode (ST)
\item 2~$\times$~Intel Xeon E5-2660 in multi-threaded mode (MT)
\item 1~$\times$~Nvidia Tesla C2070
\item 1~$\times$~Nvidia Tesla K40c
\end{enumerate}

Setups 1 and 2 are CPU-based with the first setup using only a single core while the second uses all the available cores from both CPUs. The Intel Xeon E5-2660 processor consists of 8 cores operating at 2.2GHz resulting in a theoretical peak processing power of $\sim140$ GFLOPS. The memory of the CPU-based setups was 64 gigabytes with a maximum theoretical bandwidth of 51.2 GB/s.

The last two setups, 3 and 4, are GPU-based and they represent the low- and high-performance ends of the GPU hardware spectrum respectively. The Tesla C2070 unit consists of 448 cores at 1.15GHz resulting in a theoretical peak processing power of $\sim1$TFLOPS, coupled with 6 gigabytes of memory of 144 GB/s bandwidth. The Tesla K40c unit consists of 2880 cores at 0.75GHz resulting in a theoretical peak processing power of $\sim4.5$TFLOPS coupled with 12 gigabytes of memory of 288 GB/s bandwidth. The two Tesla units communicate with the CPU via the PCIe bus. Tesla C2070 uses a x16 PCIe version 2 slot with a theoretical bandwidth of 8~GB/s, while Tesla K40c uses a x16 PCIe version 3 slot with a theoretical bandwidth of 15.75~GB/s.


%
%

\subsection{Performance analysis}
\label{sec:performance-results-analysis}

To evaluate the performance of our method we measure the runtime of each of the steps involved in a fitting iteration. Since the number of operations in each step is heavily dependent on the size of the data, we create mock observations using a fixed set of morphological and kinematic parameter values but vary the pixel count. The analysis is performed for both the intensity cube and velocity field modelling strategies.

For the intensity cube modelling strategy we perform the analysis three times: one with a fixed pixel count (100~pixels) across the velocity dimension but varying the spatial size; one with a fixed spatial size ($100\times100$~pixels) but varying the number of slices in the cube; and one with a fixed cube size ($50\times50\times$50~pixels), but varying the upsampling factor across all three dimensions. The size of the PSF kernel is set to $16\times16\times16$~pixels (for a seeing of $\sim2$~kpc at a redshift of $\sim0.1$). The above resolution is similar to the resolution of the DYNAMO observations.

\begin{figure*}
	\centering
	\includegraphics[width=1.0\textwidth]{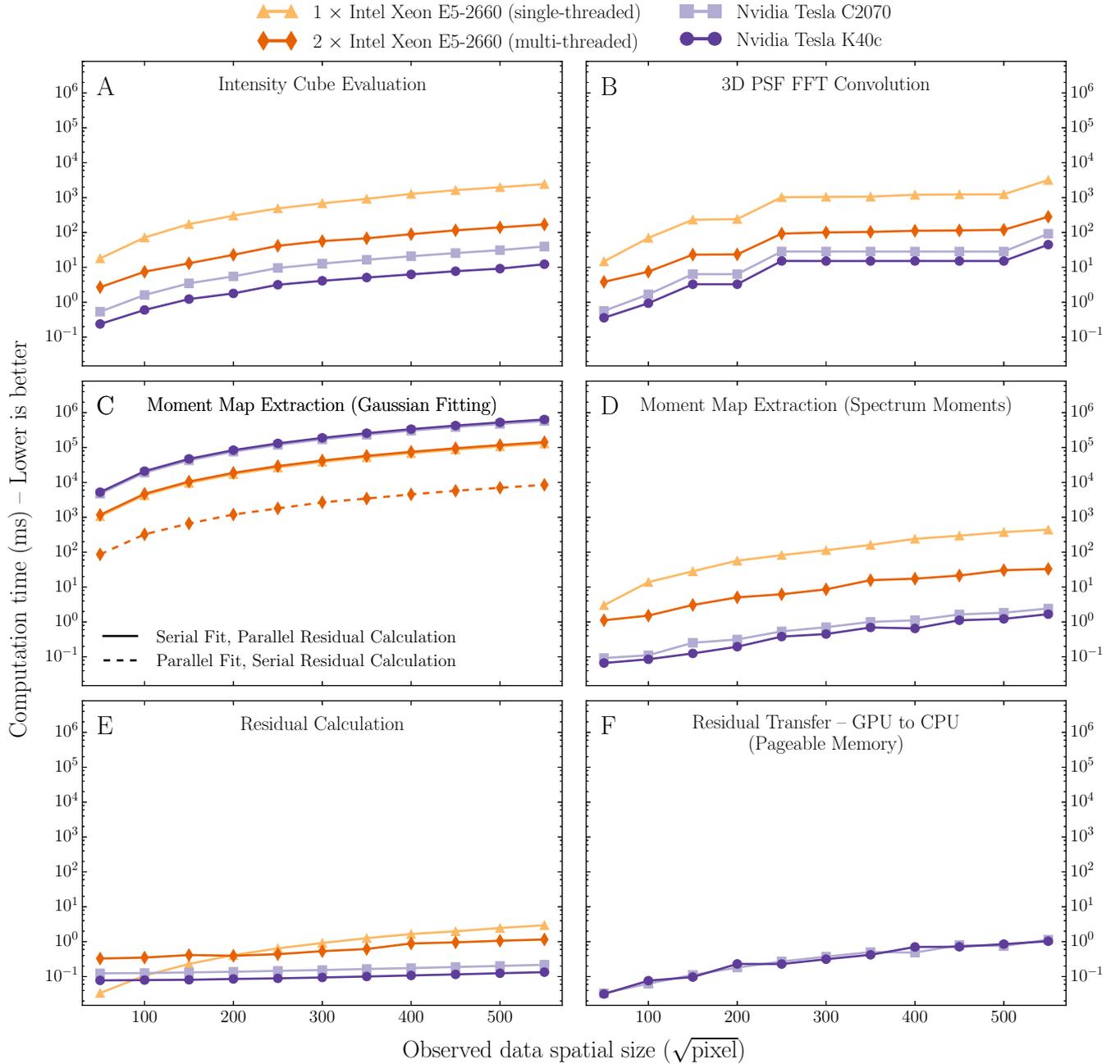}
	\caption{Performance comparison of four different hardware configurations for the steps involved in a single iteration of the fitting procedure using the intensity cube modelling strategy and a varying spatial pixel count. For this comparison we used no upsampling ($u=1$) for all three dimensions of the intensity cube model, and a fixed number of 100~pixels (128 after padding) for the velocity dimension. (A)~Intensity cube model evaluation. (B)~FFT-based convolution with the 3D PSF kernel. (C)~Velocity and velocity dispersion map extraction by fitting Gaussian profiles on the cube's spectral lines. (D)~Velocity and velocity dispersion map extraction by calculating the first- and second-order moments of the cube's spectra. (E)~Calculation of the residual velocity and velocity dispersion maps. (F)~Residual map transfer from the GPU to the CPU memory.}
	\label{fig:performance-results-3d-spat}
\end{figure*}

\begin{figure*}
	\centering
	\includegraphics[width=1.0\textwidth]{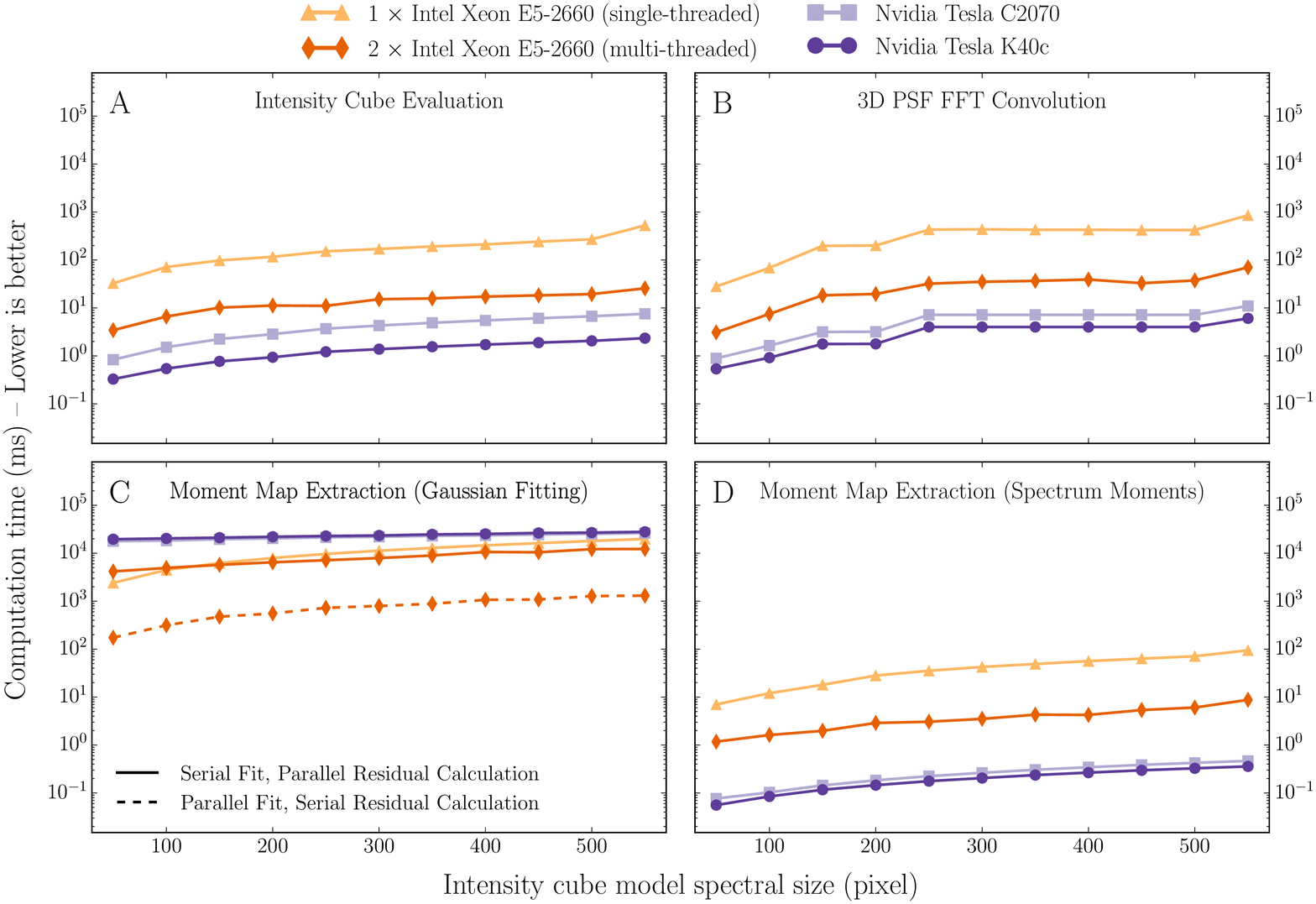}
	\caption{Performance comparison of four different hardware configurations for all the steps involved in a single iteration of the fitting procedure using the intensity cube modelling strategy and a varying spectral pixel count. For this comparison we used no upsampling ($u=1$) for all three dimensions of the intensity cube model, and a fixed number of $100\times100$~pixels ($128\times128$ after padding) for the spatial dimensions. (A)~Intensity cube model evaluation. (B)~FFT-based convolution with the 3D PSF kernel. (C)~Velocity and velocity dispersion map extraction by fitting Gaussian profiles on cube's spectral lines.}
	\label{fig:performance-results-3d-spec}
\end{figure*}

\begin{figure}
	\centering
	\includegraphics[width=0.47\textwidth]{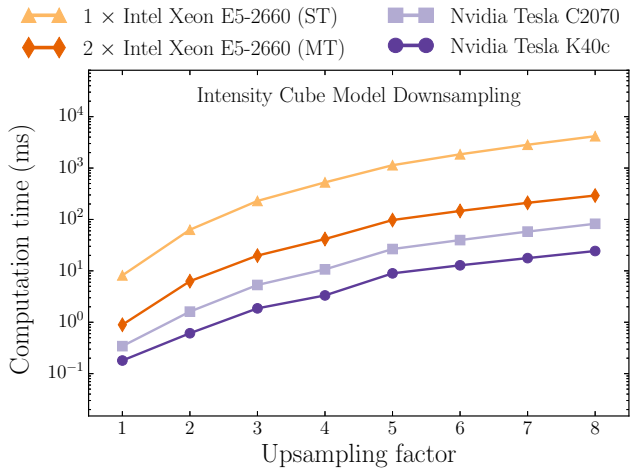}
	\caption{Performance comparison of four different hardware configurations for the downsampling step involved in a single iteration of the fitting procedure using the intensity cube modelling strategy and a varying upsampling factor ($u$). For this comparison we vary the upsampling factor simultaneously across all three timensions, while the initial size of the intensity cube model is fixed to $50\times50\times50$~pixels.}
	\label{fig:performance-results-3d-downsampling}
\end{figure}

\begin{figure*}
	\centering
	\includegraphics[width=1.0\textwidth]{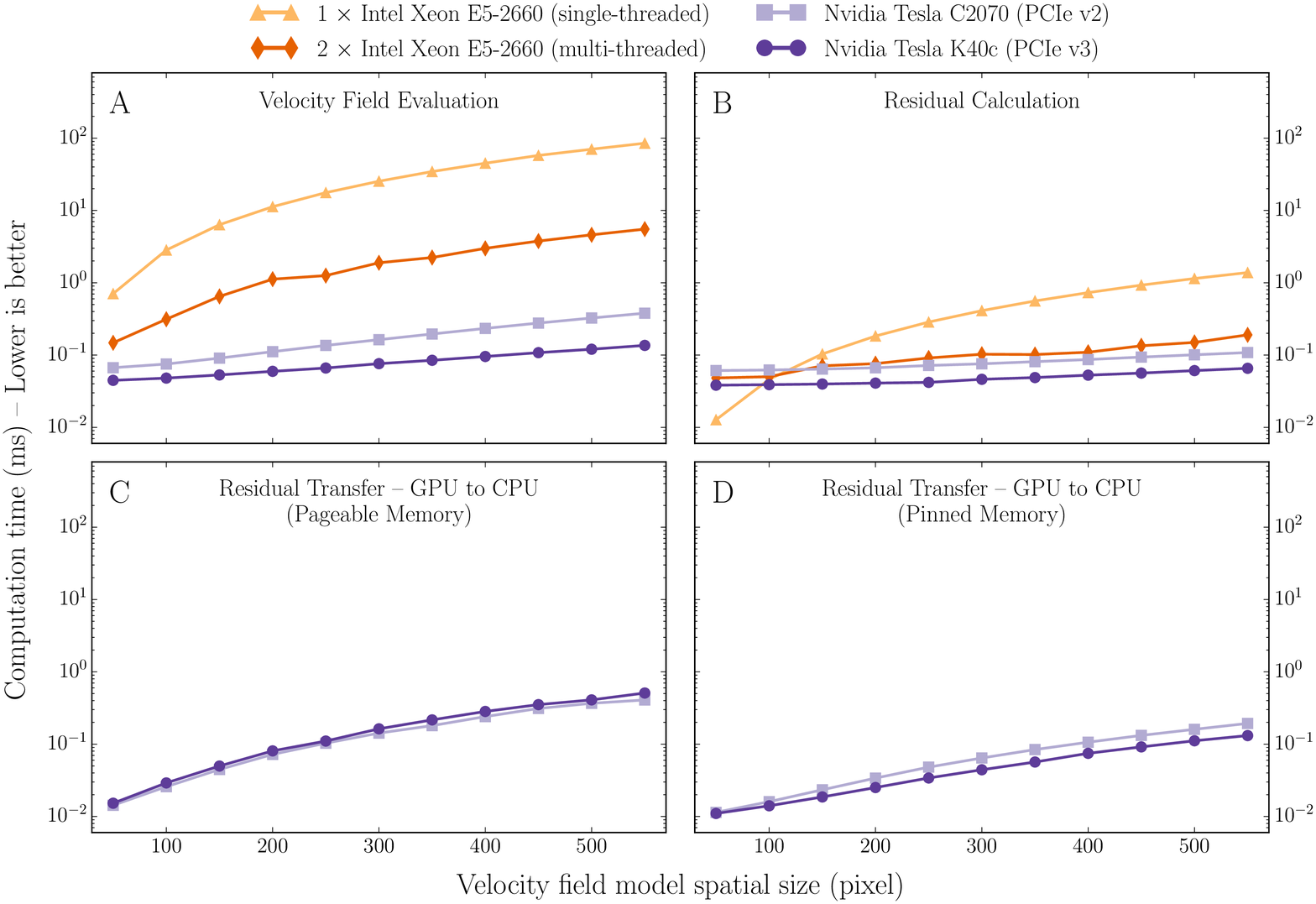}
	\caption{Performance comparison of four different hardware configurations for all the steps involved in a single iteration of the fitting procedure using the velocity field modelling strategy. (A)~Velocity field model evaluation. (B)~Calculation of the residual velocity map. (C)~Residual map transfer from the GPU to the CPU memory. (D)~Residual map transfer from the GPU to the CPU memory using \textsc{cuda} pinned memory.}
	\label{fig:performance-results-2d}
\end{figure*}

The performance analysis results for the both modelling strategies are shown in Fig.~\ref{fig:performance-results-3d-spat}, \ref{fig:performance-results-3d-spec}, \ref{fig:performance-results-3d-downsampling}, and \ref{fig:performance-results-2d}. In most of the figures we observe the same trend: The total execution time of each step increases with the number of pixels in our data but also decreases as we move towards theoretically faster hardware. The GPU can be up to $\sim100$ times faster than a single CPU core and up to $\sim10$ times faster than a multi-threaded double CPU implementation. The intensity cube and velocity field model evaluation steps (Fig.~\ref{fig:performance-results-3d-spat}A, \ref{fig:performance-results-3d-spec}A, and \ref{fig:performance-results-2d}A) clearly demonstrate this substantial performance increase. Nevertheless, in some of the plots we observe performance behaviours that are worth comment.

For the FFT-based 3D convolution (Fig.~\ref{fig:performance-results-3d-spat}B and \ref{fig:performance-results-3d-spec}B) we notice a step-like shape in the performance curves, which is expected since we always round-up the dimensions of the intensity cube model to the closest power of two. The FFT algorithms used by the \textsc{cufft} and \textsc{fftw3} libraries are highly optimized for input sizes that can be factored as $2^a \times 3^b \times 5^c \times 7^d$ (where $a$, $b$, $c$, and $d$ are non-negative integers). In general the smaller the prime factor, the better the performance, i.e., powers of two are the fastest. Our method favours sizes of power of two, but for sizes more than 512, multiples of 256 are used. For example, an input size of 700 will be padded to 768 ($2^8\times3^1$). By doing so we conserve a significant amount of memory while also achieving good performance.

The intensity cube model downsampling step (Fig~\ref{fig:performance-results-3d-downsampling}), although it has a very low arithmetic intensity and is dominated by memory transfers, it is still much faster on the GPU. This is expected since the GPU has a superior memory bandwidth compared to the CPU.

As a first attempt to extract the velocity and velocity dispersion maps from the modelled intensity cube, we fitted Gaussian profiles to the cube's spectra using the \textsc{mpfit} library. Because \textsc{mpfit} is a CPU-based library, the only opportunity for GPU parallelization is during the Gaussian profile evaluation and the calculation of the residuals. This approach has a few major performance issues. Since each spectral line of the model cube consists of only a few tens of pixels the majority of the GPU cores remain inactive and the GPU is underutilised. In addition, although the memory access is coalesced, the arithmetic intensity is so low that it makes the procedure memory bound. As a result of these issues, the GPU implementation of this step is slower than the CPU version (Fig.~\ref{fig:performance-results-3d-spat}C and \ref{fig:performance-results-3d-spec}C). Furthermore, an extra overhead is created because the LMA logic of \textsc{mpfit} runs on the CPU serially. This overhead overshadows the time requirements of the parallelized residual calculation and as a result we observe a negligible performance difference between the high- and low-end GPUs and between the single- and multi-threaded CPU configurations. A much better performance for this step was achieved by performing the Gaussian profile evaluation and residual calculation serially but fitting multiple spectral lines in parallel on the CPU. To do the same on the GPU would require a GPU-specific implementation of the Levenberg--Marquardt algorithm. \citet{2013PLoSO...876665Z} showed that implementing the LMA to run solely on the GPU can achieve a performance significantly superior to the CPU but such code is not yet publicly available. 

Finally, we decided to approach the problem of the velocity and velocity map extraction differently by calculating the first- and second-order moments of the cube's spectral lines. This method is much simpler compared to an iterative one (e.g., the Levenberg--Marquardt algorithm) and also highly parallelizable. Consequently, a parallelized implementation of this approach on the GPU was trivial and we were able to achieve a substantial performance boost compared to the CPU implementation (Fig.~\ref{fig:performance-results-3d-spat}D and \ref{fig:performance-results-3d-spec}D).

The residual calculation step features an atypical performance behavior. The computational time remains almost constant regardless of the number of pixels in our dataset (Fig.~\ref{fig:performance-results-3d-spat}E and \ref{fig:performance-results-2d}B). During the development of our GPU code we wrote a series of reusable \textsc{cuda} programs (\emph{kernels}). Instead of writing a kernel which calculates the residual of the velocity and dispersion maps specifically, we wrote a kernel that calculates the residual of any image and we call it twice, once for each map. Each time we call a \textsc{cuda} kernel there is an associated overhead. If the computation time requirements of our kernel are very low then they are overshadowed by that overhead. In our case, not only are the computational requirements of the kernel low, but also the kernel call overhead occurs twice. However, the computational time of this step is negligible compared to the computational time requirements of the rest of the steps in the iteration.

The runtime of the residual map transfer from the GPU to the CPU memory is much higher than the expected (Fig.~\ref{fig:performance-results-3d-spat}F). It also appears to be the same for both GPU setups, even though the Tesla K40c system has twice the theoretical bandwidth of the Tesla C2070 system. While this poor performance is not an important issue for the intensity cube modelling strategy, it becomes noticeable when the velocity field modelling strategy is used (Fig.~\ref{fig:performance-results-2d}C). We were able improve the runtime of the residual map transfer by making use of the \textsc{cuda} pinned memory (Fig.~\ref{fig:performance-results-2d}D). By using \textsc{cuda} pinned memory we avoid an extra copy between the CPU's virtual and physical memory that the \textsc{cuda} driver has to make in order to ensure that all CPU$\leftrightarrow$GPU transfers are done through CPU's non-pageable memory \citep{Fatica:2009:ALC:1513895.1513901}.

\begin{figure}
	\centering
	\includegraphics[width=0.47\textwidth]{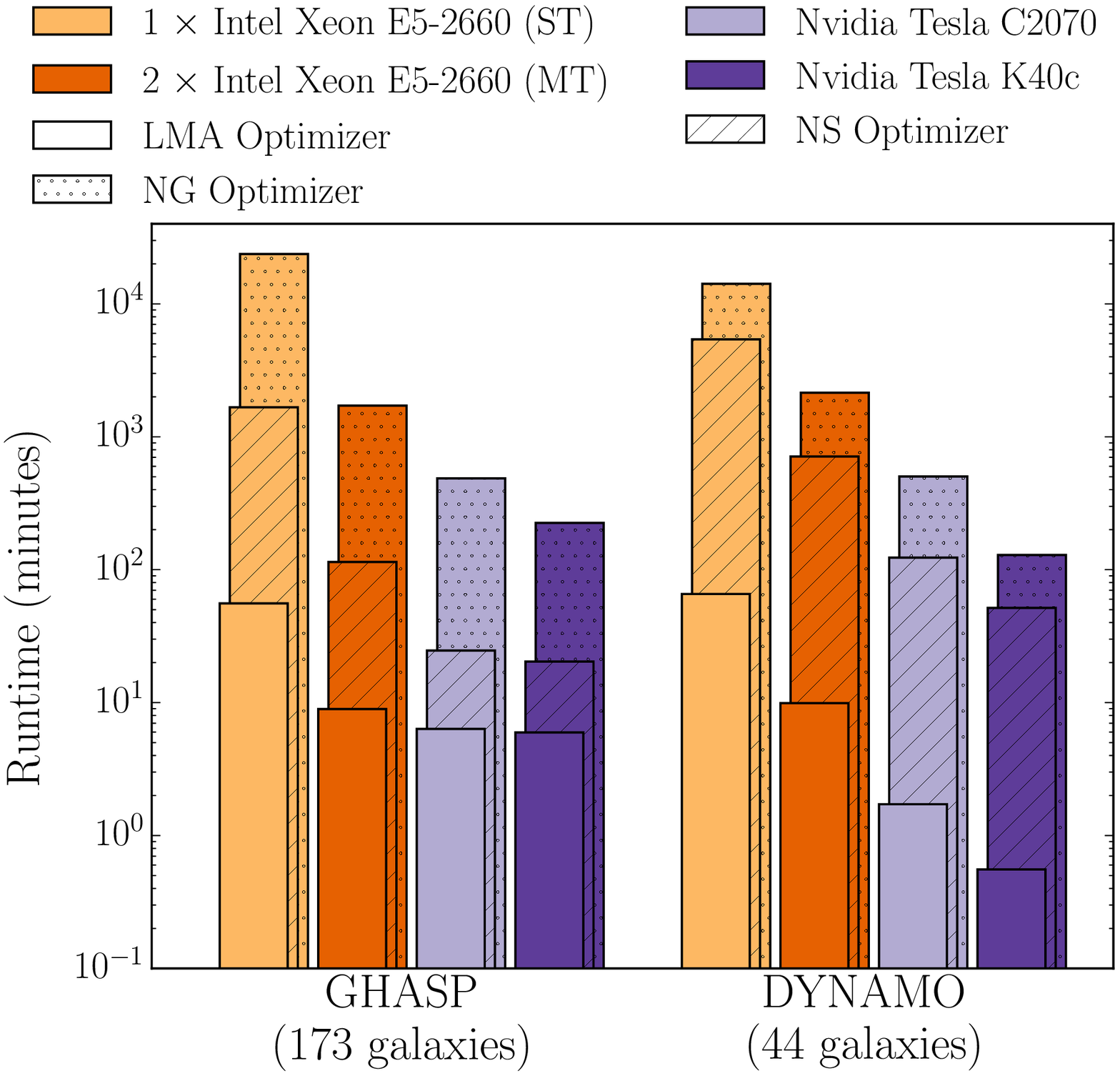}
	\caption{Total time spent to fit the entire GHASP and DYNAMO surveys using different hardware configurations and optimization techniques.}
	\label{fig:performance-results-total}
\end{figure}

To get a better idea about the time required to fit real observations, we fit the GHASP and DYNAMO data using different optimization techniques and hardware configurations. An upsampling factor of $u=2$ across all three dimensions was used while modelling the DYNAMO data. In Fig.~\ref{fig:performance-results-total} we plot the total time spent to fit all the galaxies from both samples. For the DYNAMO sample, the speedup achieved by the GPU is between $\sim10$ and $\sim100$ depending on the hardware configuration. When fitting the GHASP sample, however, the speedup difference between the two GPUs is not as big as expected. This happens because the runtime requirements of the velocity field modelling strategy are comparable to the runtime requirements of the optimizer. In Table~\ref{tab:performance-fitting-surveys} we quote the average number of iterations and the time required to fit a galaxy, but we also report how much of that time is spent on the optimization logic.  

In general, fitting real observations tends to be slower than fitting artificial observations because most of the times the optimizer requires more iterations before converging to a solution. This can be explained due to the fact that, unlike real data, the simulated observations used in this work do not include any random motions or correlated noise, and thus the resulting posterior distribution of the parameters is usually well behaved.

\newcommand{\cpuone}{1 $\times$ Intel Xeon E5-2660 (ST)}
\newcommand{\cputwo}{2 $\times$ Intel Xeon E5-2660 (MT)}
\newcommand{\gpuone}{Nvidia Tesla C2070}
\newcommand{\gputwo}{Nvidia Tesla K40c}

\begin{table*}
	\begin{minipage}{155mm}
	\caption{Average number of iterations and time spent to fit a GHASP or DYNAMO galaxy using different optimization techniques and hardware configurations.}
	\label{tab:performance-fitting-surveys}
	\begin{tabular}{cccrrrrr}

	\hline
      \multirow{2}{*}{Survey}
    & \multirow{2}{*}{Optimizer}
    & \multirow{2}{*}{Hardware}
    & \multirow{2}{*}{$N_{\mathrm{median}}$}
    & \multirow{2}{*}{$N_{\mathrm{mad}}$}
    & \multicolumn{1}{c}{$t_{\mathrm{opt}}$}
    & \multicolumn{1}{c}{$t_{\mathrm{total}}$}
    & \multirow{2}{*}{$t_{\mathrm{opt}}/t_{\mathrm{total}}$}
    \\

    &
    &
    &
    &
    & \multicolumn{1}{c}{(seconds)}
    & \multicolumn{1}{c}{(seconds)}
    &
    \\

      \multicolumn{1}{c}{(1)}
    & \multicolumn{1}{c}{(2)}
    & \multicolumn{1}{c}{(3)}
    & \multicolumn{1}{c}{(4)}
    & \multicolumn{1}{c}{(5)}
    & \multicolumn{1}{c}{(6)}
    & \multicolumn{1}{c}{(7)}
    & \multicolumn{1}{c}{(8)}
    \\

	\hline
	\multirow{12}{*}{GHASP}  & \multirow{4}{*}{LMA} & \multicolumn{1}{l}{\cpuone} & 234        & 63         & 0.77       & 7.48       & 10.35\%    \\
						     &					    & \multicolumn{1}{l}{\cputwo} & 234        & 63         & 0.74       & 1.18       & 62.84\%    \\
						     &					    & \multicolumn{1}{l}{\gpuone} & 242        & 70         & 0.83       & 0.89       & 93.56\%    \\
						     &					    & \multicolumn{1}{l}{\gputwo} & 242        & 70         & 0.79       & 0.82       & 95.74\%    \\

						     & \multirow{4}{*}{NS}  & \multicolumn{1}{l}{\cpuone} & 12030      & 4739       & 2.89       & 351.21     & 0.82\%     \\
						     &					    & \multicolumn{1}{l}{\cputwo} & 11852      & 4832       & 3.01       & 24.96      & 12.07\%    \\
						     &					    & \multicolumn{1}{l}{\gpuone} & 12207      & 4777       & 2.61       & 7.09       & 36.84\%    \\
						     &					    & \multicolumn{1}{l}{\gputwo} & 13496      & 5818       & 3.20       & 5.16       & 62.09\%    \\
						    
						     & \multirow{4}{*}{NG}  & \multicolumn{1}{l}{\cpuone} & 232925     & 21175      & 31.12      & 6063.23    & 0.51\%     \\
						     &					    & \multicolumn{1}{l}{\cputwo} & 232925     & 21175      & 32.70      & 412.87     & 7.92\%     \\
  						     &					    & \multicolumn{1}{l}{\gpuone} & 232925     & 21175      & 35.52      & 144.93     & 24.51\%    \\
  						     &					    & \multicolumn{1}{l}{\gputwo} & 232925     & 21175      & 30.48      & 58.80      & 51.83\%    \\
	\hline						   
    \multirow{12}{*}{DYNAMO} & \multirow{4}{*}{LMA} & \multicolumn{1}{l}{\cpuone} & 148        & 25         & 0.01       & 75.61      & 0.01\%     \\
    					     &					    & \multicolumn{1}{l}{\cputwo} & 148        & 25         & 0.01       & 11.73      & 0.08\%     \\
    					     &					    & \multicolumn{1}{l}{\gpuone} & 153        & 30         & 0.01       & 2.22       & 0.35\%     \\
    					     &					    & \multicolumn{1}{l}{\gputwo} & 153        & 30         & 0.01       & 0.69       & 1.11\%     \\
	    
    					     & \multirow{4}{*}{NS}  & \multicolumn{1}{l}{\cpuone} & 10706      & 2230       & 1.18       & 5505.42    & 0.02\%     \\
    					     &					    & \multicolumn{1}{l}{\cputwo} & 9623       & 2869       & 1.30       & 797.38     & 0.16\%     \\
    					     &					    & \multicolumn{1}{l}{\gpuone} & 12308      & 3600       & 0.78       & 110.27     & 0.71\%     \\
    					     &					    & \multicolumn{1}{l}{\gputwo} & 10921      & 2423       & 0.90       & 51.02      & 1.76\%     \\
    					    
    					     & \multirow{4}{*}{NG}  & \multicolumn{1}{l}{\cpuone} & 42350      & 0          & 0.30       & 20373.03   & 0.00\%     \\
 						     &					    & \multicolumn{1}{l}{\cputwo} & 42350      & 0          & 0.45       & 2948.69    & 0.02\%     \\
 						     &					    & \multicolumn{1}{l}{\gpuone} & 38115      & 4235       & 0.18       & 699.01     & 0.03\%     \\
 						     &					    & \multicolumn{1}{l}{\gputwo} & 38115      & 4235       & 0.20       & 189.17     & 0.11\%     \\
	\hline

	\end{tabular}
	\end{minipage}
	\begin{minipage}{175mm}
	(1,~2,~3)~The dataset, optimization technique, and hardware configuration used in the fitting procedure.
	(4,~5)~Median and median absolute deviation of the number of iterations performed. 
	(6)~Median of the time spent on optimization logic.
	(7)~Median of the total runtime.
	\end{minipage}	
\end{table*}

%
%

\subsection{Memory considerations}
\label{sec:performance-results-memory}

Apart from our main goal to accelerate the fitting procedure, it is also desirable to keep the memory requirements of our method at a low level. When running CPU-based software this is hardly an issue because of the large main system memory available on today's computers. In this work we did not encounter any GPU memory problems because we used enterprise-grade Tesla GPU units which feature a relatively large amount of memory (up to 12GB). However, it is desirable to run our method on the commodity versions of these cards, which have a much more limited amount of memory (e.g., 2GB).

During the runtime of our code most of the utilised memory is occupied by: 
\begin{enumerate}
\renewcommand{\theenumi}{(\arabic{enumi})}
\item The data of the intensity cube model.
\item The data of the intensity cube model in Fourier space.
\item The data of the PSF kernel in Fourier space.
\item The data buffers allocated by the \textsc{cufft} library.
\end{enumerate}
The expected size (in bytes) of a 3D dataset such as the intensity cube model is:
\begin{equation}
\label{eq:mem-data-size}
M_{\mathrm{3D}} = N_{\mathrm{z}} \times N_{\mathrm{y}} \times N_{\mathrm{x}} \times 4,
\end{equation}
where $N_{\mathrm{z}}$, $N_{\mathrm{y}}$, $N_{\mathrm{x}}$ the length of each dimension in pixels and the size of a single precision real number is 4 bytes. The expected size (in bytes) for the same 3D dataset but in frequency space is calculated using:
\begin{equation}
\label{eq:mem-data-size-fft}
M_{\mathrm{FFT,3D}} = N_{\mathrm{z}} \times N_{\mathrm{y}} \times (N_{\mathrm{x}}/2 + 1) \times 8,
\end{equation}
for single precision complex numbers.

\textsc{cufft} allows us to perform the FFT in-place without allocating additional space for the frequency space data. In order to do so we have to make sure that the memory space we allocate for the data is padded to the size required to accommodate the data in frequency space.

To perform an FFT, the \textsc{cufft} library has to create a so-called \emph{plan}. This object contains all the data \textsc{cufft} needs to compute the FFT. Internally the plan allocates memory buffers which are used for interim results. The size of these buffers depends on the size of the FFT and it can be quite large. \citet{5577905} showed that \textsc{cufft} needs about four times the space of the input for its intermediate results for real-to-complex transforms when transforming two-dimensional data. The same plan can be applied to different datasets as long as they have the same length.

\begin{figure}
	\centering
	\includegraphics[width=0.47\textwidth]{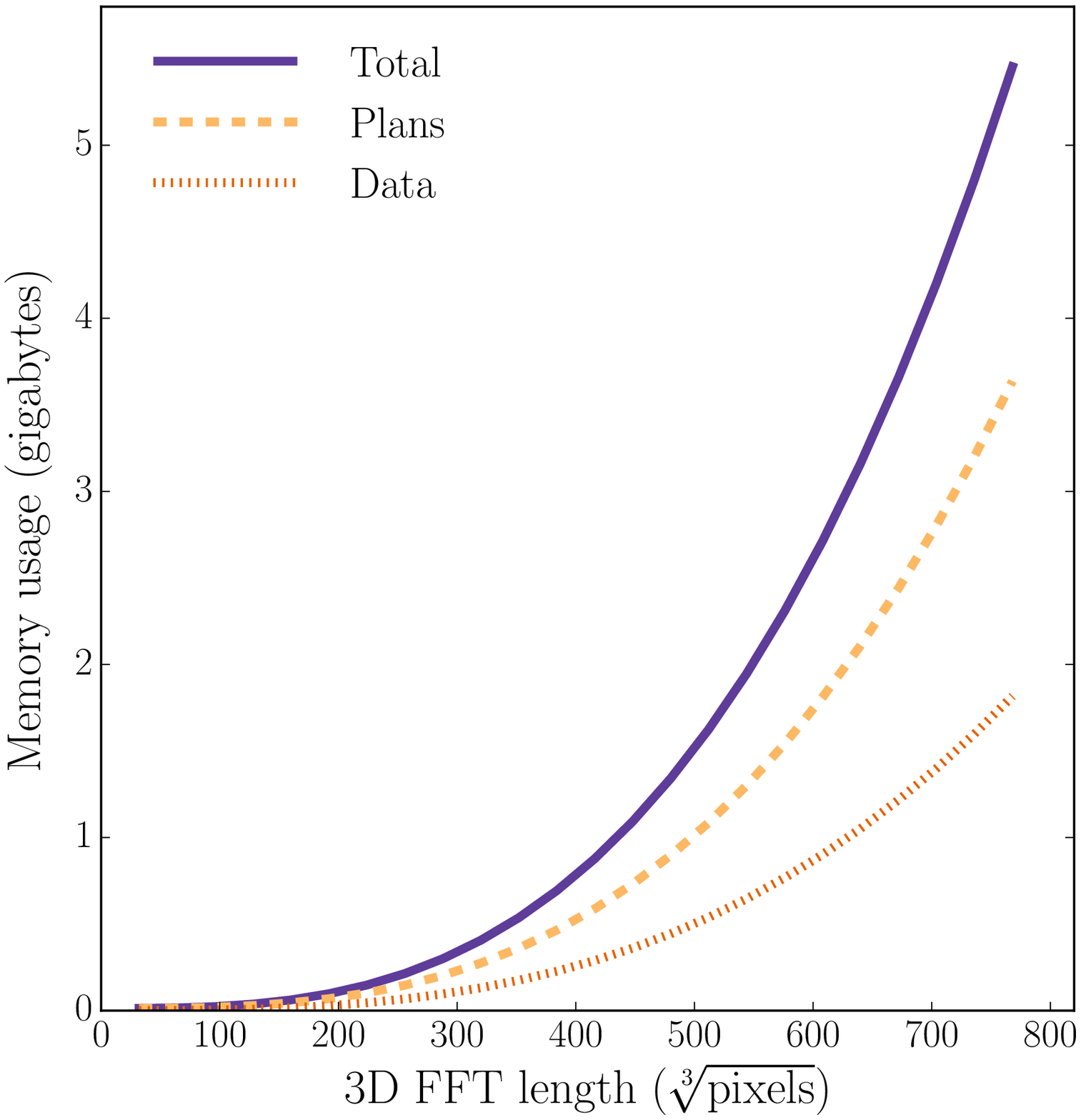}
	\caption{Total GPU memory consumption for varying-size three-dimensional real-to-complex FFT transforms followed by inverse FFT transforms using the \textsc{cufft} library. All transforms use single precision data and are done in-place.}
	\label{fig:mem-fft-3d}
\end{figure}

To get a better picture of the memory requirements of the \textsc{cufft} library we perform a series of three-dimensional FFTs followed by inverse FFTs. More specifically, we generate 3D datacubes of different sizes and for each one of them we perform an in-place real-to-complex 3D FFT followed by a complex-to-real inverse 3D FFT. Since two different types of transforms are performed for each dataset, two plans are needed. Both plans appear to occupy the same amount of memory. The results are plotted in Fig.~\ref{fig:mem-fft-3d}.

For datasets with a number of pixels below $\sim200^3$ the memory requirements of the two FFTs are much smaller than the memory available on todays low-range commodity GPUs. At a resolution of $\sim550^3$ the required size exceeds the typical memory size available on a mid-range commodity GPU. In addition, we notice that the plans require more memory than the actual data itself. The maximum FFT size the \textsc{cufft} library can compute is $\sim512$ million single precision elements, which corresponds to a 3D FFT of $\sim800^3$ or a plan of $\sim2$ gigabytes. 

Here we present an easy way to reduce the size of the FFT plans and enable us to perform large 3D FFTs on commodity GPUs. The 3D FFT is decomposed into a series of 2D FFTs across the two dimensions of the data followed by a series of 1D FFTs across the third dimension. Although this results in much smaller memory requirements for the plans, the total computation time increases. This is expected since we now have to call the appropriate \textsc{cufft} functions for plan execution multiple times instead of one, which creates an additional overhead. We overcome this issue by using \textsc{cufft}'s batched plans which can perform multiple FFTs of the same length with one function call. However, the size of the batched plans increases with the number of transforms in the batch. Hence, in order to achieve balance between performance and memory requirements, one has to carefully select the batch size of the plans.

\begin{figure}
	\centering
	\includegraphics[width=0.47\textwidth]{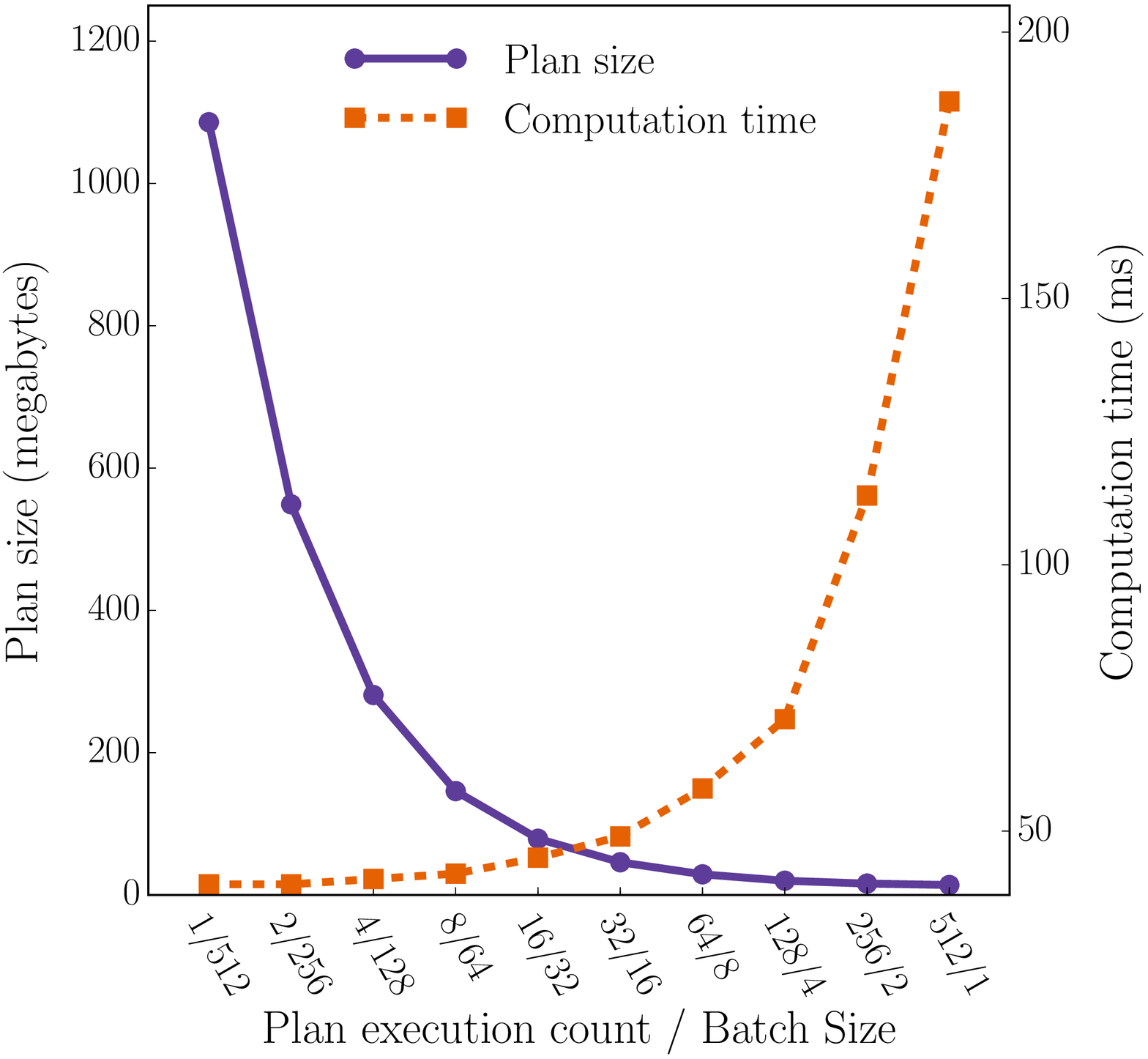}
	\caption{FFT plan memory consumption and execution time for two-dimensional $512^2$ real-to-complex FFT transforms followed by inverse FFT transforms using the \textsc{cufft} library. All transforms use single precision data and are done in-place. A total of 512 2D FFTs are computed, each time with a different plan configuration.}
	\label{fig:mem-fft-2d-batch}
\end{figure}

Fig.~\ref{fig:mem-fft-2d-batch} shows how the performance and the plan size vary when calculating 512 two-dimensional FFTs with different batch configurations. A compromise of $10-20$ per cent in performance can decrease the memory requirements by one order of magnitude. Finally, the 1D FFT that has to be applied across the third dimension of the data in order to calculate the 3D FFT, behaves in a similar manner hence it is not discussed here.

To lower the memory requirements even further, we perform the convolution in the frequency domain using the point (2D) and line (1D) spread functions instead of the 3D PSF kernel. As a result, the memory requirements for the PSF data are reduced from $M_{\mathrm{FFT,3D}}$ (equation~\ref{eq:mem-data-size-fft}) to:
\begin{equation}
M_{\mathrm{FFT,2D+1D}} = [N_{\mathrm{y}} \times (N_{\mathrm{x}}/2 + 1) + (N_{\mathrm{z}}/2 + 1)] \times 8.
\end{equation}

To get a better understanding of the memory savings after applying the changes discussed in this section, we run the fitting procedure for low- and high-resolution mock data on a commodity laptop equipped with a 2GB Nvidia GeForce 750M GPU. The low-resolution data have the same resolution as the DYNAMO observations, while the high-resolution data have $700\times700$ spatial pixels and a velocity resolution of 15~km~s$^{-1}$. For both datasets the upsampling factor was set to $u=1$ for all three dimensions. The GPU memory requirements are now $\sim2$~megabytes down from $\sim11$~megabytes for the low-resolution dataset, and $\sim0.7$~gigabytes down from $\sim3$~gigabytes for the high-resolution dateset. 

%
%

\section{Discussion}
\label{sec:discussion}

In the previous sections we demonstrated how the graphics processing unit can be used as an accelerator for the kinematic fitting procedure. Our method adopts a similar pattern to the one followed by most existing kinematic fitting software. Two different modelling strategies were explored: One uses a velocity field and it is more appropriate for high resolution observations, while the other one uses an intensity cube model and the knowledge of the point and line spread functions in order to account for beam smearing effects.

We focused on parallelizing the steps involved in the fitting procedure but did not attempt to parallelize the optimizer's logic on the GPU. This allowed us to use pre-existing optimization libraries and accelerated the development of the software implementation. Due to the image nature of the data involved, we were able to use data-parallel algorithms and exploit the power of the GPU's massively parallel architecture.

To gain a better idea of the acceleration the GPU has to offer we also implemented and ran our method on the CPU. In most cases, our tests showed a remarkable performance increase when the GPU was used. We achieved a maximum of $\sim100$ times increase over a single-threaded CPU implementation and a peak of $\sim10$ times increase over a multi-threaded dual CPU implementation.

Nevertheless, we encountered a case were the GPU performed worse than the CPU. This was the step where Gaussian profiles were fitted to each spectrum of the intensity cube model in order to extract the velocity and velocity dispersion maps. To achieve a good performance for this task we would need to implement an LMA optimizer that runs entirely on the GPU \citep{2013PLoSO...876665Z}, the development of which is still ongoing. Instead, to improve the performance of this step we followed a different approach using the cube's spectral line first- and second-order moments.

A peculiar performance behaviour was also noticed in the velocity field modelling strategy: Although the speedup achieved by the GPU for the velocity field evaluation step is remarkable, the same amount of speedup was not observed on the total time required to fit a GHASP galaxy. The reason for this behaviour is that the parallelized steps involved in the velocity field modelling strategy have very low runtime requirements and comparable to the time spent on the optimization logic.

By looking at the theoretical computational performance of our hardware (Section~\ref{sec:performance-results-hardware}) we notice that in several cases we are below the expected speedup. This is not a surprise since the theoretical performance of the graphics processing unit is rarely achieved in real-world applications. In addition, achieving the peak performance of the GPU can require a high level of expertise and result in complicated or hard-to-read code \citep{Ryoo:2008:OPA:1345206.1345220}. Nonetheless, the speedup achieved by our GPU-based method is remarkable.

Memory requirements are also an important factor when developing GPU software. Although the amount of memory on the GPU increases every year, so does the resolution of the data produced by the new generation astronomical instruments. We found that the FFT-based 3D PSF convolution step in the intensity cude modelling strategy can require a large amount of memory when working with high resolution data, and we proposed a simple solution to substantially decrease the memory requirements using 2D+1D FFT and convolution decomposition (Section~\ref{sec:performance-results-memory}).

An issue one might face when developing software for the GPU is that existing CPU-codes has to be rewritten \citep{2011PASA...28...15F}. Although this was a bigger issue in the past, the situation has improved. Libraries and tools that ease the GPU code development or the porting of CPU code to the GPU have been developed. \textsc{cuda} unified memory relieves the developer from managing the GPU memory \citep{7040988}. The \textsc{thrust}\footnote{\url{http://developer.nvidia.com/thrust}} library \citep{thrustandcuda} can be used to perform data parallel operations which run on the GPU using an interface that resembles parts of the C++ standard library. With \textsc{openacc}\footnote{\url{http://www.openacc.org}} one can port existing CPU code to the GPU by just adding a series of compiler directives \citep{openacc2012} to the code. We expect that technologies and tools like these will continue to improve and further decrease the effort required to write GPU software.

Another interesting point one has to consider is that the total cost of a GPU solution is much lower than the CPU cost for the same amount of computational power. Although in this work we used enterprise-level Tesla GPUs, the commodity versions of these cards, which have very similar performance capabilities and energy requirements, cost one order of magnitude less.

Our method was tested by fitting artificial observations with zero noise. We were able to recover the actual kinematic parameter values for $\gtrsim98$ per cent of the galaxies with high accuracy (Table~\ref{tab:fitting-results-mock-1}). In order to successfully recover the rest $\sim2$ per cent the fitting procedure was re-run using a different optimizer configuration. This translates to better initial parameter values or prior constraints for the LMA optimizer, lower tolerance threshold ($tol$) and more active points ($N$) for the NS optimizer, and higher grid resolution for the NG optimizer.

To fit the mock data using the NG optimizer we had to fix parameters $V_{\mathrm{sys}}$, $x_{\mathrm{o}}$, $y_{\mathrm{o}}$ and $i$ to their true values in order to reduce the number of iterations required by the fitting procedure. This is still relevant of a real-world scenario since it is sometimes preferred to derive the kinematic centre and inclination of the galaxy from high-resolution photometry data rather than the velocity field \citep{2010MNRAS.401.2113E}. By using the GPU we were able to fit the data in a reasonable amount of time regardless of the inefficiency of the algorithm.

Our method was also able to fit mock observations with low and high levels of noise with great success: Regardless of the noise level, $\sim68$ and $\sim95$ per cent of the the galaxy parameters were enclosed within the 1 and 2$\sigma$ confidence intervals of our fits respectively. 

Furthermore, our method verified successfully the kinematic modelling results of the low-redshift surveys GHASP and DYNAMO as found in the literature. Although in this work we only fitted data from optical surveys, our method can potentially be used at other wavelengths as long as an appropriate model is employed. For example, when fitting models to 21-cm data it is preferable to use a model that accounts for the warping observed at the outer parts of the neutral hydrogen layer in the majority of galaxies (\citealt{1981AJ.....86.1791B}; \citealt{1990ApJ...352...15B}; \citealt{2002A&A...394..769G}).

%
%

\subsection{The degeneracy between $V_{\mathrm{t}}$ and $i$}
\label{sec:discussion-degeneracy}

Caution is advised when fitting near face-on galaxies, especially in the case of low-resolution data and/or observations with low signal-to-noise ratio. In such case, recovering the disc inclination from the kinematics can be quite problematic due to the near-degeneracy between the circular velocity and inclination. 

\begin{figure*}
	\centering
	\includegraphics[width=1.0\textwidth]{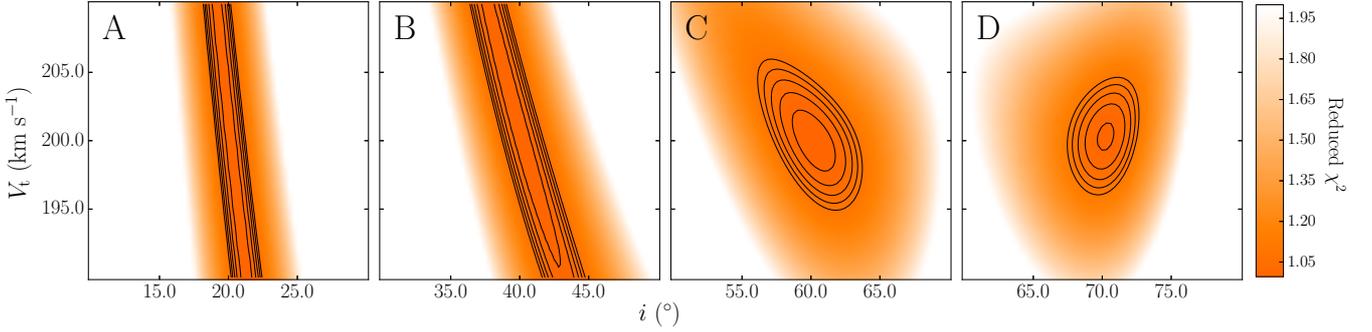}
	\caption{Reduced $\chi^2$ space as a function of disc inclination ($i$) and asymptotic circular velocity ($V_{\mathrm{t}}$). The contours correspond to 1, 2, 3, 4, and 5$\sigma$ confidence intervals. Panels A, B, C, D correspond to galaxies with $V_{\mathrm{t}} = 200$~km~s$^{-1}$ and $i = $~20\degr, 40\degr, 60\degr, and 70\degr~respectively.}
	\label{fig:degeneracy}
\end{figure*}

To better demonstrate this degeneracy we generated four mock galaxies with varying inclinations (20\degr, 40\degr, 60\degr, and 70\degr) and constant circular velocity $V_{\mathrm{t}}=200$~km~s$^{-1}$. Gaussian noise of 5~km~s$^{-1}$ was added to the mock velocity and velocity dispersion maps. In Fig.~\ref{fig:degeneracy} we plot the reduced $\chi^2$ space as a function of the disc inclination ($i$) and circular velocity ($V_{\mathrm{t}}$). This figure illustrates that the degeneracy between the two parameters is quite strong for inclinations less then 40\degr. 

\begin{figure}
	\centering
	\includegraphics[width=0.47\textwidth]{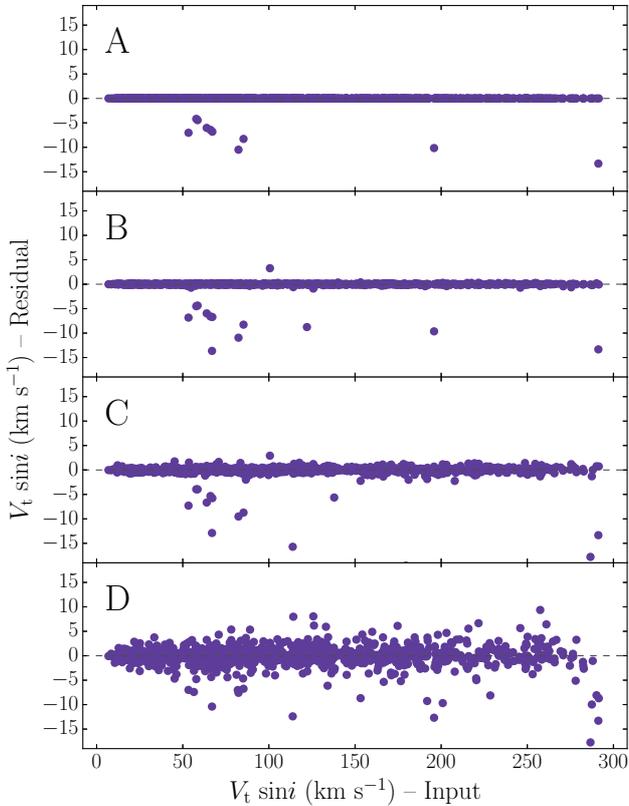}
	\caption{Fitting results of the mock data with different levels of noise using the Levenberg--Marquardt Algorithm. We plot the residuals of the fitted $V_{\mathrm{t}}\sin{i}$ products as a function of the input parameters. The rows from top to bottom correspond to datasets with 0, 1, 5, and 20~km~s$^{-1}$ noise on the velocity map, and 0, 1.4, 7, and 28~km~s$^{-1}$ noise on the velocity dispersion map. The error bars are omitted for the sake of clarity.}
	\label{fig:fitting-results-mock-vtsini}
\end{figure}

To investigate whether the scatter observed in Fig.~\ref{fig:fitting-results-mock}B and \ref{fig:fitting-results-mock}C is due to this degeneracy we plot the residual between the actual and the fitted product $V_{\mathrm{t}}\sin{i}$ (Fig~\ref{fig:fitting-results-mock-vtsini}). Indeed the combination of the two parameters reduces the scatter and their product is recovered with a good accuracy.

\subsection{Moments vs Gaussian fitting}

\begin{figure*}
	\centering
	\includegraphics[width=1.0\textwidth]{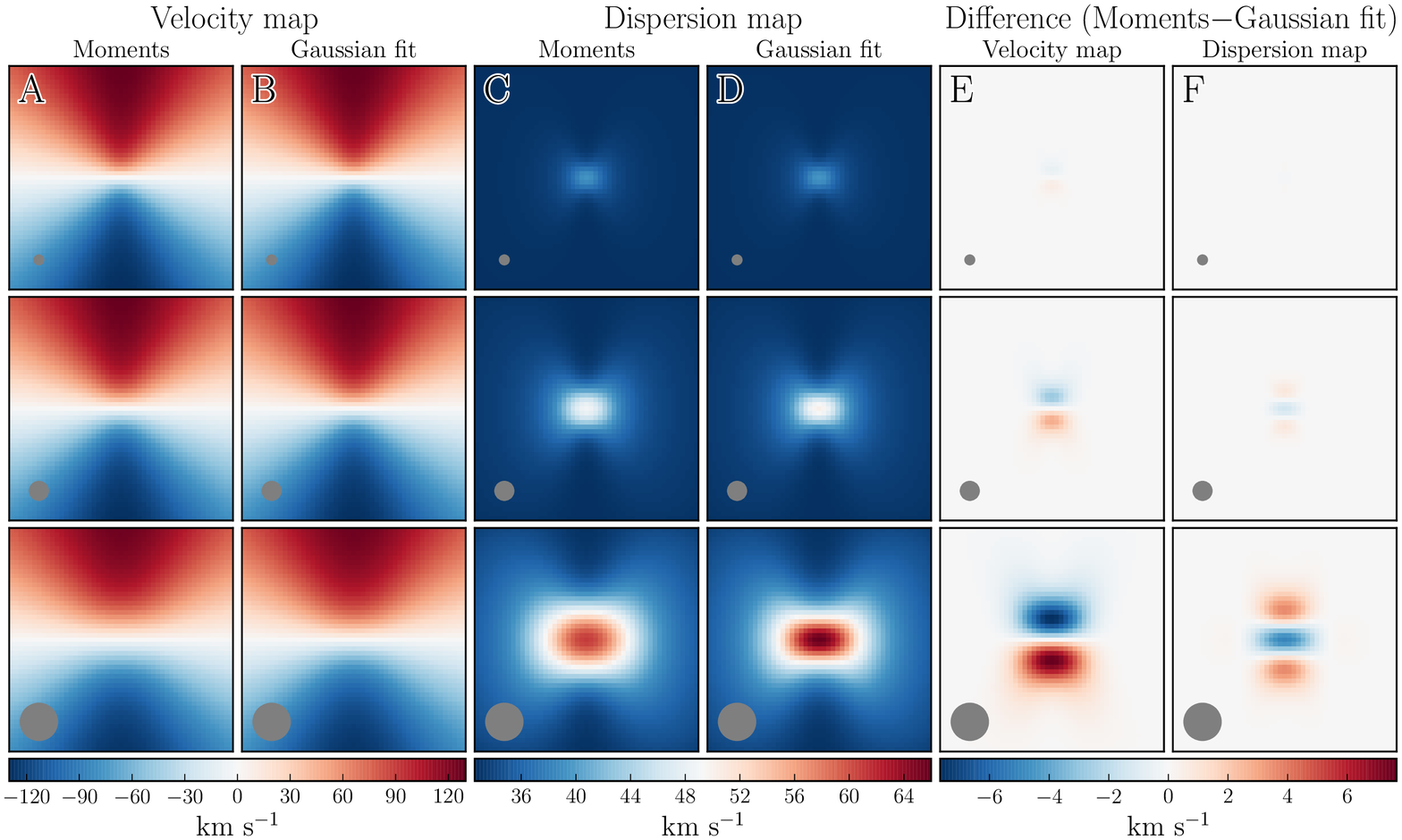}
	\caption{Comparison of two different methods for extracting velocity and velocity dispersion maps from intensity datacubes. (A,~C)~Velocity and velocity dispersion maps extracted using the first- and second-order moments of the cube's spectral lines. (B,~D)~Velocity and velocity dispersion maps extracted by fitting Gaussian profiles to the cube's spectral lines. (E,~F)~The differences in the velocity and velocity dispersion maps between the two methods. The seeing FWHM from top to bottom corresponds to 2, 4, and 8~pixels respectively. The spatial size of the data is 49$\times$49~pixels and the velocity resolution is 30~km~s$^{-1}$. The scale radius is 9~pixels, the turn-over radius is 3~pixels, the disc inclination is 45\degr, the maximum rotation velocity is 200~km~s$^{-1}$, and the intrinsic velocity dispersion is 30~km~s$^{-1}$.}
	\label{fig:moments-1}
\end{figure*}

There are cases where the input velocity and velocity dispersion maps are extracted using the Gaussian fitting method but the modelling of the kinematics is done using the spectral moments approach due to the much lower runtime requirements and the simplicity of the implementation (e.g., \citealt{2014MNRAS.437.1070G}; Section~\ref{sec:fitting-results-dynamo}, this paper). To get a better picture on how the two methods compare to each other we create several mock observations with the same properties but varying the seeing size. Then we extract the velocity and velocity dispersion maps using both methods and we calculate the difference between them (Fig~\ref{fig:moments-1}).

Under good seeing conditions we observe no difference between the two methods. However, as the size of the PSF increases, low-amplitude features appear on both maps. More specifically, for the maps extracted with the Gaussian fitting approach, the line-of-sight velocity gradient is steeper in the inner regions of the galaxy, while the velocity dispersion at the centre is slightly higher. Instead of increasing the PSF size, the same behaviour would be observed by reducing the turn-over radius of the galaxy.

\begin{figure*}
	\centering
	\includegraphics[width=1.0\textwidth]{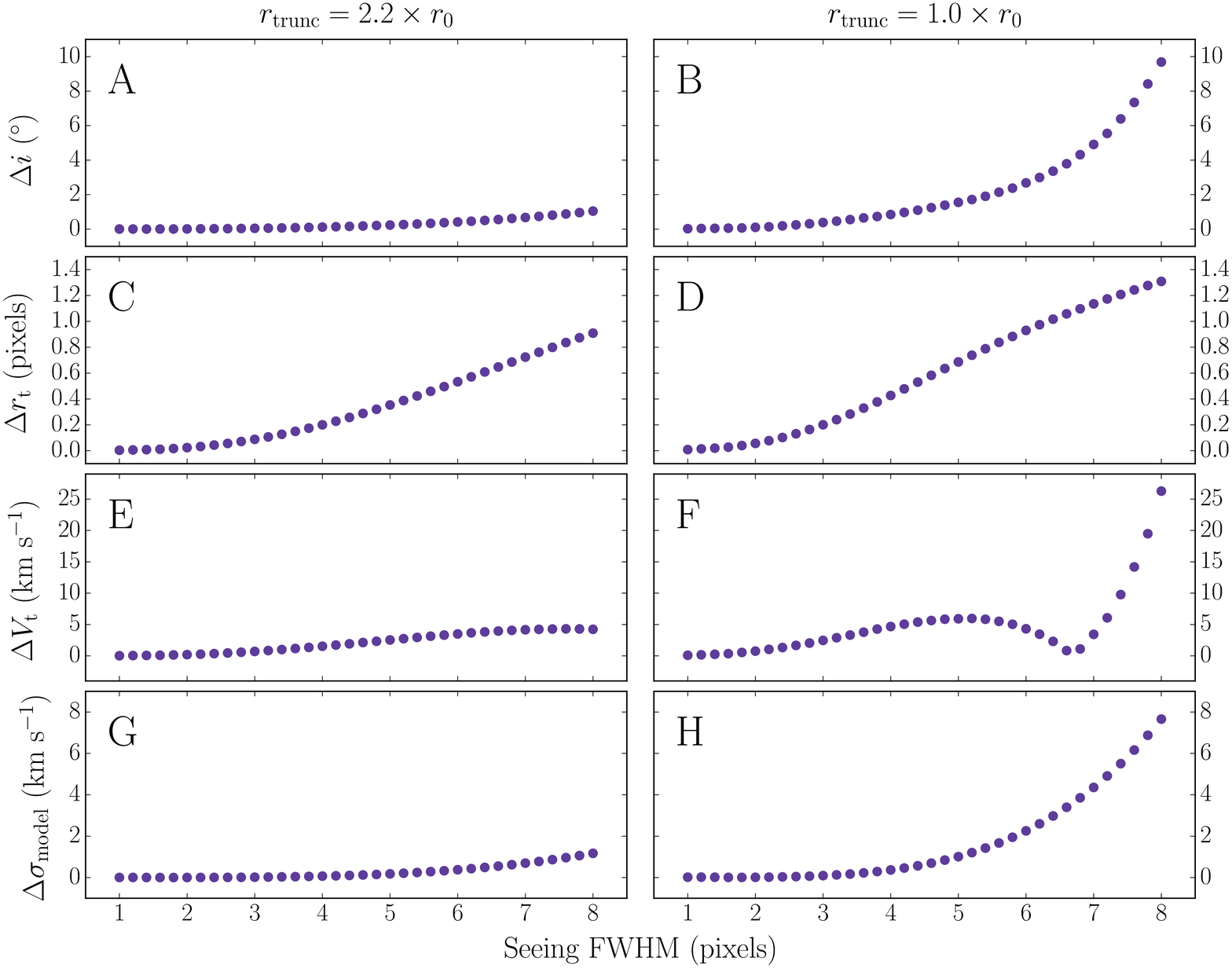}
	\caption{The error between the actual and fitted parameters as a function of seeing FWHM. We fit a model that uses the spectral moments approach to data that have been extracted using Gaussian profile fitting. The rows from top to bottom correspond to the error in the inclination, turn-over radius, circular velocity and intrinsic velocity dispersion respectively. The scale radius is 9~pixels, the turn-over radius is 3~pixels, the disc inclination is 45\degr, the maximum rotation velocity is 200~km~s$^{-1}$, and the intrinsic velocity dispersion is 30~km~s$^{-1}$. The left column corresponds to data truncated at 2.2~$\times$~exponential disc scale lengths, while the right column corresponds to data truncated at 1.0~$\times$~exponential disc scale lengths. }
	\label{fig:moments-2}
\end{figure*}

To understand how the above differences can affect the modelling results we fit a model that uses the spectral moments approach to data that have been extracted using Gaussian fitting. The procedure is performed twice, with the data truncated at $r_{\mathrm{trunc}} = 1.0 \times r_{\mathrm{0}}$ and $r_{\mathrm{trunc}} = 2.2 \times r_{\mathrm{0}}$. The scale radius of the modelled galaxy has chosen to be $r_{\mathrm{0}} = 3.0 \times r_{\mathrm{t}}$ In Fig.~{\ref{fig:moments-2}} we plot the absolute error between the actual and the fitted parameters as a function of PSF FWHM.

We observe that when the flat part of the rotation curve is detected in the data (Fig.~\ref{fig:moments-2}, left column), the error in the inclination, circular velocity, and intrinsic velocity dispersion is very small, even under bad seeing conditions. However, if the data are truncated at a much smaller radius (Fig.~\ref{fig:moments-2}, right column), the error on the fitted parameters is noticeable, especially for observations with severe beam smearing. The errors on the rest of the model parameters are not presented here because they are negligible.

While the above suggest caution in selecting the right modelling approach for a given dataset, they are not necessarily an issue if the affected parameters are not essential for any resulting analysis. For example, if we are interested in the systemic velocity, kinematic centre or position angle of a galaxy, choosing one approach over the other will not affect our results. However, if we want to recover the disc inclination from a heavily beam-smeared observation truncated at small radius, we will have to choose our modelling approach carefully. In the case of the DYNAMO sample, since the galaxies are measured up to a large radius, both approaches yielded similar results.

Finally, we remind the reader that, as shown in Section~\ref{sec:fitting-results-mock}, if the moments method is used consistently (in both data extraction and fitting), none of the above issues arise and the fitted parameters are recovered correctly.

%
%

\subsection{Choosing an optimizer}
\label{sec:discussion-choosing-optimizer}

In our tests, the NS optimizer appears to be more robust than the LMA and NG optimizers, but this does not necessarily make it the best choice. 

The LMA fails to converge to the right solution for 2 per cent of the mock observations from cases where it gets trapped in a local minimum. Due to the fast convergence properties of the algorithm, we can achieve a $\sim100$ per cent success rate by repeating the fitting procedure (using better initial estimates for the model parameters) for this 2 per cent in only a few seconds. However, as mentioned in Section~\ref{sec:fitting-results-mock}, the majority of the failed fits occur for highly inclined galaxies, for which a thin disc model is not a good representation anyway. 

The LMA's fast convergence properties can also help to tackle one of its weaknesses: parameter error estimation. The derived error of the fitted parameters may not be robust since they are based on an elliptical Gaussian approximation for the likelihood at the minimum. To overcome this issue, one can perform further $\chi^2$ analysis around the best fit, or perform a Monte Carlo simulation by repeating the fitting procedure multiple times using different noise realizations.

On the contrary, the NS optimizer does not suffer from any of the above issues because, instead of blindly following the $\chi^2$ gradient or curvature of the parameter space, it explores the likelihood of the posterior probability distribution thoroughly but efficiently. This make NS less likely to converge to a wrong solution and also allows for robust error estimation. Indeed, in our tests with zero-noise mock data, the NS optimizer was never trapped into a local optimum and achieved a $\sim100$ per cent success rate. However, the above advantages come with the price of higher computational cost. The NS algorithm requires two orders of magnitude more iterations compared to the LMA.

Lastly, the NG optimizer was also able to fit the majority of the galaxies in the zero-noise mock dataset with great success. However, a direct comparison with the other two optimizers would not be fair because only four of the model parameters were left free when fitting with the NG optimizer. Reducing the number of free parameters in our model was necessary in order to perform the fitting procedure in a reasonable amount of time. Despite its inefficiency, as already mentioned in Section~\ref{sec:method-optimization-algorithms-ng}, the NG optimizer is trivial to implement as a first solution when other optimization algorithms are not available. The time required to develop the NG code can be a few orders of magnitude less than the time required to develop sophisticated optimization software such as the \textsc{mpfit} and \textsc{multinest} libraries.

When it comes to choosing an optimization technique, there is no such thing as `one-size-fits-all'. When fitting simple models that are unlikely to cause multimodalities on the parameter space, the LMA optimizer should suffice. In case we fit complex models with a large number of parameters that may result multimodalities or strong degeneracies, the NS optimizer will most likely give better results. 

%
%

\subsection{Accelerating science}
\label{sec:discussion-accelerating-science}

By accelerating the computations involved in galaxy kinematic modelling using the GPU we can increase the pace of scientific discovery. Besides the obvious benefit of analysing results faster, the GPU may benefit kinematic modelling in several other important ways.

By providing much shorter computation times it is now feasible to fit multiple models to the data in the same compute time. This can be beneficial when no a priori information about the kinematics of the galaxy is available. Alternatively, one could fit a very complicated model using the GPU in the same amount of time that one could fit a simpler model using the CPU. Using more complicated models can eventually lead to a better understanding of the kinematics of the galaxy and potentially provide better constraints on galaxy's physical quantities.

It is known that the majority of the optimization methods used for model fitting in astronomy are not global optimizers but local ones. As a result there is a chance that the optimizer will not converge to the global optimal solution leading to the wrong results and eventually to the wrong science conclusions. This can happen for different reasons including parameter space multimodalities, low resolution data and bad initial model parameter values. With the acceleration the GPU provides, one can now fit the same model to the same data multiple times using either different optimizers or the same optimizer but starting from a different position in the parameter space. This allows us to cross-check the fitting results before proceeding to further analysis.

The GPU can also benefit the process of calculating uncertainties in the model parameters. This task usually requires the additional model evaluations, or even the repetition of the fitting procedure, for the same object hundreds or thousands of times.

Lastly, the use of the GPU could potentially enable different approaches to galaxy kinematic modelling, such as interactive visualization-directed model fitting which allows the user to build and fit complex kinematic models interactively while exploring the data \citep{2010ASPC..434..167F}, in near real-time. In addition, it could potentially allow the user to provide better model parameter priors to the optimization algorithm. This approach to modelling can be useful when the observed galaxies feature exotic or unique kinematic structures that cannot be described by any of the models available in the automated modelling procedure. Although it may be demanding in terms of human interaction, it will allow scientist to model selected galaxies in greater detail. \citet{2011ITVCG..17..454S} developed a similar software which allows the user to model planetary nebulae interactively, and mentioned the need for GPU acceleration in such tools.

%
%

\subsection{Fitting large-scale surveys}
\label{sec:discussion-future-surveys}


To obtain a first glimpse of the degree of speedup the graphics processing unit has to offer when fitting large-scale surveys, we estimate the computational time that is required to fit the entire SAMI \citep{2015MNRAS.447.2857B} and MaNGA \citep{2015ApJ...798....7B} surveys. We do so by fitting simulated observations that match the properties of the real survey data.

The simulated SAMI observations have a spatial sampling of 0.5~arcsec~pixel$^{-1}$ across a field of view of $\sim15$~arcsec. The typical size of the point spread function is $\sim2.1$~arcsec, while the velocity resolution is $\sim67$~km~s$^{-1}$. This configuration, along with an upsampling factor of $u = 2$ for all three dimensions, result in an intensity cube model of $128\times128\times256$ pixels.

The simulated MaNGA observations have a spatial sampling of 0.5~arcsec~pixel$^{-1}$, but the field of view can vary depending on the IFU size. MaNGA uses six different IFU configurations of 7, 19, 37, 61, 91 and 127 fibres, with the first one used for flux calibration. Each fibre has a diameter of 2~arcsec resulting in IFUs with diameters between 12 and 32~arcsec on-sky. We assume that the number of galaxies observed is uniformly distributed among the five different IFU configurations (2000 galaxies with each IFU setup). The typical size of the point spread function is $\sim2.5$~arcsec, while the velocity resolution is $\sim120$~km~s$^{-1}$. This configuration, along with an upsampling factor of $u = 2$ for all three dimensions, result in intensity cube models with spatial dimensions of $128\times128$ and $256\times256$~pixels, and a spectral dimension of 128~pixels.

We investigate the execution time for all four hardware configurations using the Levenberg--Marquardt and Nested Sampling optimizers. For the LMA scenario we assume that a mean of $\sim200$ iterations is required before converging to a solution, while for the NS optimizer we assume a mean of $\sim20000$ iterations.

\begin{figure}
	\centering
	\includegraphics[width=0.47\textwidth]{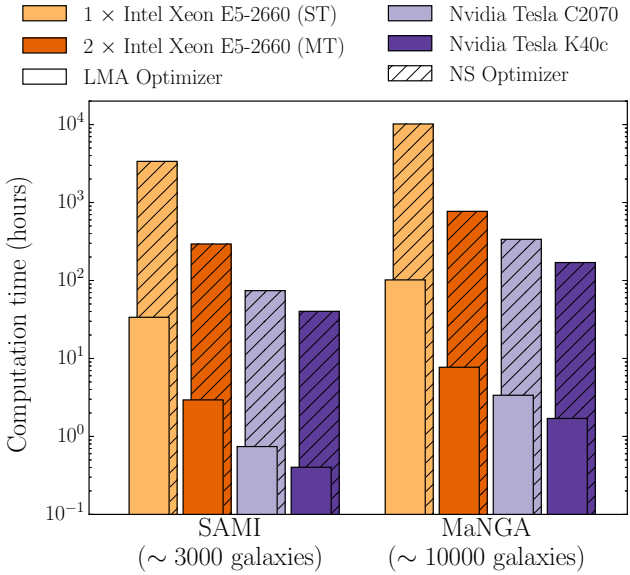}
	\caption{Computational time across different hardware configurations and optimization methods for fitting SAMI and MaNGA surveys. The computation time does not include the time spent on the optimizer's logic.}
	\label{fig:dis-fut-tim}
\end{figure}

The execution time for the two surveys is shown in Fig.~\ref{fig:dis-fut-tim}. It clearly shows the improvement achieved by the GPU implementation. While a single-threaded CPU implementation would only allow for analysis using the LMA optimizer in a reasonable amount of time ($\lesssim200$~hours), such performance can be reached for any optimizer using GPU programming. However, the speed achieved using the LMA optimizer and the GPU ($\lesssim 2$~hours) shows that more sophisticated models could be developed. Moreover, if further decrease of the computation time is desired, astronomers may have to shift from the traditional approach of analysing data on the desktop, to the use of remote services and resources in the form of web applications and GPU-enabled supercomputers. 

%
%

\subsection{Next steps}
\label{sec:discussion-next-steps}

The substantial speedup gained by using the graphics processing unit opens up new directions in galaxy kinematic modelling. It allows us to perform analysis that was hard to carry out before due to the enormous computational requirements. In future work we will investigate:

\begin{enumerate}
\renewcommand{\theenumi}{(\arabic{enumi})}
\item The implementation of more complex methods and kinematic models that would previously take a prohibitive amount of time to fit on the CPU. These could include methods to fit directly to datacubes and models that account for non-circular or asymmetric motions, disc warps and multiple galactic components.
\item Interactive and visualization-directed approaches to galaxy kinematic modelling. This will allow scientists to build complex galaxy models, featuring unusual and peculiar kinematic structures, while exploring their data.
\item Distributed kinematic analysis of large-scale 3D spectroscopic surveys using remote HPC facilities and web services.
\end{enumerate}

An implementation of a gradient- and curvature-based optimizer, such as the Levenberg--Marquardt Algorithm, running solely on the GPU would also be beneficial. This would allow the extraction of the velocity and velocity dispersion maps from intensity cubes with complex spectra, where the first- and second-order moments approach may fail (e.g., multiple emission and absorption lines, asymmetric continuum etc.).

Moreover, we believe that the majority of kinematic modelling software and tools, such as \textsc{tirific}, \textsc{$^{\mathrm{3D}}$barolo}, \textsc{galpak$^{\mathrm{3D}}$}, \textsc{diskfit}, \textsc{kinemetry} and others, can benefit from a GPU implementation because they all perform a high number of independent pixel operations on two- and three-dimensional data.

As part of this work we have developed a high-performance software tool for galaxy kinematic modelling: \textsc{gbkfit}. The software along with its source code, documentation and other supporting material is available at: \mbox{\url{http://supercomputing.swin.edu.au/gbkfit}}. For more information about the development and implementation of \textsc{gbkfit}, see Appendix~\ref{app:gbkfit}.

%
%

\section{Summary and conclusion}
\label{sec:summary-and-conclusion}

We have investigated the use of the graphics processing unit (GPU) as an accelerator for the computationally demanding fitting procedure in the context of galaxy kinematic modelling. 

For our analysis we used three different optimization techniques including the Levenberg--Marquardt algorithm, Nested Sampling, and a naive brute-force technique based on Nested Grids.

To assess the consistency of our method we fitted kinematic models to data from three different sources including artificially generated observations and observations from the low redshift surveys GHASP and DYNAMO. The last two were chosen as representatives of high- and low-spatial-resolution surveys respectively. For all three datasets only a minimal fine-tuning of our fitting code was required.

We found that our method can successfully recover the parameters of zero-noise mock galaxies with a success rate up to $\sim100$ per cent. In addition, our method was able to fit mock observations with low and high levels of noise with great success: $\sim68$ and $\sim95$ per cent of the the galaxy parameters are enclosed within the 1 and 2$\sigma$ confidence intervals of our fits respectively. Furthermore, we were able to verify the kinematic modelling results of the GHASP and DYNAMO surveys by comparing our results to the literature. 

For our performance tests we used four different hardware configurations: two high-end CPUs in single-threaded and multi-threaded mode, a low-end GPU and a high-end GPU configuration. 

Our tests and analysis led to the following findings:
 
\begin{enumerate}
\renewcommand{\theenumi}{(\arabic{enumi})}
\item The GPU can substantially accelerate most of the steps involved in the model fitting procedure. We found that our GPU code can perform up to $\sim100$ times faster than a single-threaded CPU implementation, and up to $\sim10$ times faster than a multi-threaded dual-CPU solution.
\item Taking advantage of the massive processing power of the GPU allows us to use simpler, naive and less efficient approaches which are conceptually easier to implement.
\item The substantial acceleration the GPU provides allows us to perform additional analysis on our data. This can include the fitting of multiple models to the same data or the use of multiple optimization methods to verify results.
\item Although to achieve the highest possible performance on a GPU might require a high amount of expertise, we can still achieve substantial performance improvement (compared to the CPU) by following a few basic GPU programming guidelines.
\item Caution is advised when fitting high-resolution data since the memory available on GPU is much less than the main system memory.
\end{enumerate}

We are poised on the brink of a revolution of our understanding of galaxies in the nearby universe. The currently underway or upcoming large-scale 3D spectroscopic surveys (e.g., SAMI, MaNGA, WALLABY) will reveal the internal structure and kinematics of several thousands of galaxies. This vast amount of data will push the computational requirements for analysis up by several orders of magnitude. We expect that massively parallel architectures, such as the graphics processing unit, will play a crucial role in analysing and understanding this fire hose of new data, which will eventually lead to new scientific discoveries.

%
%

\section*{Acknowledgements}
We thank Andrew Green for providing us with the data of the DYNAMO survey. G.B. thanks Sylvia Mackie for providing linguistic advice. We are grateful for the constructive comments of the referee, which have helped to improve this paper. This work was performed on the gSTAR national facility at Swinburne University of Technology. gSTAR is funded by Swinburne and the Australian Government's Education Investment Fund. This research has made use of the Fabry Perot database, operated at CeSAM/LAM, Marseille, France. We acknowledge the usage of the HyperLeda database (\url{http://leda.univ-lyon1.fr}). This research was supported under Australian Research Council's Discovery Projects funding scheme (project number DP130101460).

%
%

\newcommand{\jpdc}{JPDC} 

\bibliographystyle{mnras}
\bibliography{gbekiaris_paper_01}

%
%

\appendix

%
%

\section{Using an MCMC optimizer}
\label{app:mcmc}

Bayesian model parameter estimation through Markov Chain Monte Carlo (MCMC) techniques has been used extensively in the field of astronomy (\citealt{2002PhRvD..66j3511L}; \citealt{2011MNRAS.417.2601W}; \citealt{2012MNRAS.425..405B}; \citealt{2015AJ....150...92B}). The main strength of the MCMC techniques is their ability to explore the parameter space efficiently, and provide robust estimates of the posterior probability distribution of the model parameters.

Standard MCMC methods, such as the Metropolis--Hastings algorithm (M--H, \citealt{HASTINGS01041970}), can be problematic or experience slow convergence when the likelihood landscape of the parameter space features multiple modes or strong degeneracies. In addition, simple and easy-to-implement MCMC techniques, such as the Random-Walk Metropolis algorithm (RWM, \citealt{metropolis1953equation}), can be difficult to use because they require manual fine tuning of the proposal distribution in order to achieve high sampling efficiency. 

Over the years several MCMC algorithms have been developed to address the above issues. Adaptive Metropolis (AM, \citealt{haario2001adaptive}) is an extension to the RWM and adapts its proposals based on the history of the chain, thus no manual tuning of the proposal distribution is required. Metropolis-Coupled MCMC (MC$^3$, \citealt{geyer1991markov}) is better able to approximate multimodal distributions by running multiple chains of different temperatures.

Unfortunately only a few MCMC codes are publicly available and in a form of reusable libraries compatible with our software (i.e., C/C++ libraries). The majority of them are simple implementations of the standard M--H and RWM algorithms, and suffer from the issues mentioned above. In our case, while multimodalities are rarely formed on the likelihood landscape of the parameter space, the inclination and circular velocity parameters of our models will always form a strong degeneracy for galaxies close to face-on.

We used the publicly available \textsc{gbpmcmc}\footnote{\url{http://github.com/gbpoole/gbpCode}} library which implements a RWM sampler and provides an auto-tune facility to ease the tuning of the proposal distribution prior to the MCMC burn-in and integration phases. This library has been used successfully in \citet{2011MNRAS.417.2601W} to model the kinematics of high redshift galaxies. In that work, to remove the degeneracy between the inclination ($i$) and the circular velocity ($V_{\mathrm{t}}$), they simplified the kinematic model by using the product of $V_{\mathrm{t}}\times\sin{i}$ as a single parameter.

After fitting several mock observations with both the MCMC and NS optimizers we concluded that the latter is superior for the following reasons:

\begin{enumerate}
\renewcommand{\theenumi}{(\arabic{enumi})}
\item Although the \textsc{gbpmcmc} library provides a facility to ease the tuning of the proposal distribution, careful adjustment of this facility is still required in order to build a proposal distribution that will result in good sampling efficiency. 
\item With the \textsc{multinest} library we were able to achieve high quality fits without any fine tuning of the algorithm and only by using the default configuration of the library.
\item The auto-tune facility of \textsc{gbpmcmc} requires a few hundreds or thousands likelihood evaluations which can add to the total runtime of the fitting procedure.
\item An initial set of model parameter values is required when using the MCMC optimizer, while this is not the case with the NS sampler.
\item The MCMC optimizer requires a higher number of iterations in order to achieve a precision similar to NS.
\item The MCMC optimizer does not check for the convergence of the chain during the runtime. The user has to perform convergence tests on the resulting chain after the integration phase is complete. On the other hand, NS convergence is checked during the runtime and it is controlled through a single criterion: $tol$.
\item Finally, the \textsc{multinest} library is modified compared to the original Nested Sampling algorithm specifically for sampling multimodal distributions and posteriors with pronounced degeneracies efficiently.
\end{enumerate}

Since a `vanilla' implementation of the RWM sampler has no advantage over the Nested Sampling implementation of the \textsc{multinest} library, we decided not to use the MCMC optimizer in this work.

%
%

\section{GBKFIT development and implementation notes}
\label{app:gbkfit}

\textsc{gbkfit} is an open source high-performance software for galaxy kinematic modelling which was developed as part of this work. It is written in C++ and supports all major operating systems such as Linux, Windows and \mbox{Mac OS}. To achieve this cross-platform compatibility we used free and open source  tools, libraries and technologies while also conforming with the C++ coding standards and avoiding compiler-specific language extensions. To minimize the dependencies to external libraries we take advantage of C++~11/14 features when possible. Furthermore, in order to provide a common unified compilation and installation procedure we make use of the CMake\footnote{\url{http://www.cmake.org/}} build system.

\textsc{gbkfit} is not a stand-alone executable but a library which can be linked with an executable and provide the necessary facilities for galaxy kinematic modelling. For this work, in order to communicate with the \textsc{gbkfit} library and fit models to data we have developed a simple command-line application in C++. 

The \textsc{gbkfit} library features an extensible object-oriented architecture which allows the user to add support for arbitrary models and optimization techniques in the form of modules without modifying its source code. In addition, the modular architecture of \textsc{gbkfit} allows the user to load new model and optimization modules during the runtime in the form of plugins without the need for recompilation. 

So far we have developed the two kinematic models described in Section~\ref{sec:method-models}. Both models use \textsc{cuda} in order to run on the GPU, but they also provide a multi-threaded CPU code path in case there is no Nvidia GPU available on the system. Furthermore, adapting \textsc{gbkfit} to different Nvidia GPU models is done automatically by \textsc{cuda}, which compiles the GPU code to binary code specifically for the underlying GPU architecture. 

Finally, we have developed three optimizer modules which implement the three optimization techniques discussed in Section~\ref{sec:method-optimization-algorithms}.

\section{Fitting Results}

%
%

\begin{table*}
	\begin{minipage}{151mm}
	\caption{Kinematic properties of the GHASP sample (see Section~\ref{sec:fitting-results-ghasp}).}
	\label{tab:fitting-results-ghasp}
	\begin{tabular}{ccccccccc}
	\hline
	Galaxy	 & $V_{\mathrm{sys}}$ & $V_{\mathrm{sys,Epinat}}$ & $PA$   & $PA_{\mathrm{Epinat}}$ & $i$	  & $i_{\mathrm{Epinat}}$ & $V_{\mathrm{max}}$ & $V_{\mathrm{max,Epinat}}$ \\
	(UGC ID) & (km~s$^{-1}$) 	  & (km~s$^{-1}$)	          &(\degr) & (\degr)				& (\degr) & (\degr)			      & (km~s$^{-1}$)	   & (km~s$^{-1}$)             \\
	(1)		 & (2)				  & (3)						  & (4)	   & (5)					& (6)	  & (7)					  & (8)				   & (9)					   \\
	\hline
	IC476 & $4765\pm{1}$ & $4767\pm{3}$ & $73\pm{6}^{}$ & $68\pm{6}^{}$ & $55\pm{15}^{}$ & $55\pm{24}^{}$ & $107\pm{113}$ & $70\pm{22}$\\
	IC2542 & $6111\pm{1}$ & $6111\pm{2}$ & $174\pm{2}^{}$ & $174\pm{3}^{}$ & $24\pm{11}^{}$ & $20\pm{15}^{}$ & $253\pm{127}$ & $290\pm{192}$\\
	NGC5296 & $2255\pm{1}$ & $2254\pm{2}$ & $2\pm{3}^{}$ & $2\pm{3}^{}$ & $72\pm{10}^{}$ & $65\pm{4}^{}$ & $115\pm{101}$ & $80\pm{9}$\\
	89 & $4513\pm{1}$ & $4510\pm{5}$ & $174\pm{2}^{}$ & $177\pm{4}^{}$ & $29\pm{8}^{}$ & $33\pm{13}^{}$ & $405\pm{155}$ & $343\pm{117}$\\
	94 & $4546\pm{1}$ & $4548\pm{2}$ & $90\pm{1}^{}$ & $94\pm{2}^{}$ & $42\pm{3}^{}$ & $42\pm{5}^{}$ & $217\pm{12}$ & $209\pm{21}$\\
	508 & $4637\pm{1}$ & $4641\pm{2}$ & $118\pm{2}^{}$ & $123\pm{2}^{}$ & $25\pm{7}^{*}$ & $25\pm{7}^{*}$ & $537\pm{27}$ & $553\pm{127}$\\
	528 & $627\pm{1}$ & $628\pm{1}$ & $49\pm{1}^{}$ & $52\pm{3}^{}$ & $24\pm{6}^{}$ & $21\pm{14}^{}$ & $78\pm{32}$ & $84\pm{52}$\\
	763 & $1147\pm{1}$ & $1148\pm{2}$ & $117\pm{1}^{}$ & $117\pm{3}^{}$ & $56\pm{3}^{}$ & $54\pm{6}^{}$ & $103\pm{3}$ & $104\pm{11}$\\
	1117 & $-190\pm{1}$ & $-191\pm{1}$ & $198\pm{1}^{}$ & $198\pm{3}^{}$ & $56\pm{16}^{*}$ & $56\pm{16}^{*}$ & $92\pm{2}$ & $79\pm{17}$\\
	1256 & $429\pm{1}$ & $428\pm{1}$ & $72\pm{1}^{}$ & $73\pm{2}^{}$ & $78\pm{1}^{}$ & $76\pm{2}^{}$ & $120\pm{9}$ & $105\pm{9}$\\
	12632 & $412\pm{2}$ & $415\pm{2}$ & $44\pm{3}^{}$ & $47\pm{4}^{}$ & $46\pm{16}^{*}$ & $46\pm{16}^{*}$ & $143\pm{10}$ & $69\pm{19}$\\
	12754 & $743\pm{1}$ & $742\pm{2}$ & $342\pm{1}^{}$ & $342\pm{2}^{}$ & $43\pm{4}^{}$ & $53\pm{5}^{}$ & $130\pm{8}$ & $123\pm{11}$\\
	12893 & $1097\pm{1}$ & $1097\pm{2}$ & $260\pm{6}^{}$ & $257\pm{5}^{}$ & $16\pm{18}^{}$ & $19\pm{19}^{}$ & $84\pm{69}$ & $72\pm{67}$\\
	1317 & $3092\pm{1}$ & $3090\pm{2}$ & $103\pm{1}^{}$ & $106\pm{1}^{}$ & $72\pm{1}^{}$ & $73\pm{1}^{}$ & $211\pm{4}$ & $205\pm{9}$\\
	1437 & $4863\pm{1}$ & $4858\pm{2}$ & $301\pm{2}^{}$ & $307\pm{2}^{}$ & $43\pm{4}^{}$ & $47\pm{4}^{}$ & $229\pm{15}$ & $218\pm{15}$\\
	1655 & $5432\pm{3}$ & $5427\pm{7}$ & $318\pm{5}^{}$ & $318\pm{6}^{}$ & $45\pm{18}^{*}$ & $45\pm{18}^{*}$ & $209\pm{16}$ & $205\pm{64}$\\
	1736 & $1527\pm{1}$ & $1522\pm{2}$ & $27\pm{1}^{}$ & $27\pm{3}^{}$ & $49\pm{5}^{}$ & $35\pm{14}^{}$ & $206\pm{22}$ & $193\pm{68}$\\
	1886 & $4841\pm{1}$ & $4836\pm{3}$ & $33\pm{1}^{}$ & $35\pm{2}^{}$ & $63\pm{2}^{}$ & $62\pm{2}^{}$ & $264\pm{6}$ & $267\pm{8}$\\
	1913 & $537\pm{1}$ & $536\pm{2}$ & $288\pm{1}^{}$ & $288\pm{3}^{}$ & $47\pm{2}^{}$ & $48\pm{9}^{}$ & $111\pm{3}$ & $105\pm{16}$\\
	2023 & $594\pm{1}$ & $593\pm{2}$ & $315\pm{9}^{*}$ & $315\pm{9}^{*}$ & $19\pm{18}^{*}$ & $19\pm{18}^{*}$ & $58\pm{9}$ & $53\pm{49}$\\
	2034 & $567\pm{1}$ & $567\pm{2}$ & $342\pm{17}^{*}$ & $342\pm{17}^{*}$ & $19\pm{23}^{*}$ & $19\pm{23}^{*}$ & $38\pm{12}$ & $39\pm{46}$\\
	2045 & $1523\pm{1}$ & $1525\pm{3}$ & $320\pm{1}^{}$ & $319\pm{4}^{}$ & $61\pm{1}^{}$ & $61\pm{8}^{}$ & $196\pm{4}$ & $185\pm{16}$\\
	2080 & $893\pm{1}$ & $893\pm{1}$ & $336\pm{1}^{}$ & $336\pm{2}^{}$ & $25\pm{9}^{*}$ & $25\pm{9}^{*}$ & $123\pm{6}$ & $131\pm{42}$\\
	2141 & $964\pm{1}$ & $965\pm{3}$ & $190\pm{1}^{}$ & $191\pm{5}^{}$ & $77\pm{4}^{}$ & $74\pm{23}^{}$ & $264\pm{80}$ & $157\pm{20}$\\
	2183 & $1475\pm{1}$ & $1475\pm{4}$ & $158\pm{1}^{}$ & $159\pm{3}^{}$ & $49\pm{4}^{}$ & $41\pm{10}^{}$ & $136\pm{8}$ & $160\pm{32}$\\
	2193 & $519\pm{1}$ & $519\pm{1}$ & $302\pm{2}^{}$ & $305\pm{6}^{}$ & $23\pm{10}^{}$ & $6\pm{15}^{}$ & $34\pm{7}$ & $174\pm{423}$\\
	2455 & $368\pm{1}$ & $367\pm{2}$ & $263\pm{21}^{*}$ & $263\pm{21}^{*}$ & $51\pm{30}^{*}$ & $51\pm{30}^{*}$ & $31\pm{6}$ & $21\pm{12}$\\
	2503 & $2372\pm{1}$ & $2377\pm{2}$ & $33\pm{1}^{}$ & $34\pm{1}^{}$ & $48\pm{4}^{}$ & $49\pm{2}^{}$ & $320\pm{12}$ & $285\pm{12}$\\
	2800 & $1179\pm{2}$ & $1174\pm{2}$ & $294\pm{4}^{}$ & $291\pm{4}^{}$ & $38\pm{8}^{}$ & $52\pm{13}^{}$ & $122\pm{14}$ & $103\pm{20}$\\
	2855 & $1190\pm{1}$ & $1188\pm{2}$ & $99\pm{1}^{}$ & $100\pm{2}^{}$ & $67\pm{1}^{}$ & $68\pm{2}^{}$ & $222\pm{7}$ & $229\pm{9}$\\
	3013 & $2452\pm{1}$ & $2453\pm{6}$ & $193\pm{1}^{}$ & $195\pm{4}^{}$ & $58\pm{8}^{*}$ & $58\pm{8}^{*}$ & $215\pm{7}$ & $212\pm{21}$\\
	3273 & $615\pm{1}$ & $615\pm{2}$ & $41\pm{2}^{}$ & $42\pm{4}^{}$ & $79\pm{3}^{}$ & $82\pm{7}^{}$ & $106\pm{18}$ & $106\pm{7}$\\
	3334 & $3945\pm{1}$ & $3952\pm{13}$ & $275\pm{1}^{}$ & $277\pm{6}^{}$ & $47\pm{14}^{*}$ & $47\pm{14}^{*}$ & $378\pm{4}$ & $377\pm{85}$\\
	3382 & $4488\pm{1}$ & $4490\pm{2}$ & $183\pm{2}^{}$ & $184\pm{2}^{}$ & $22\pm{11}^{}$ & $18\pm{6}^{}$ & $276\pm{180}$ & $335\pm{111}$\\
	3429 & $899\pm{1}$ & $868\pm{4}$ & $327\pm{1}^{}$ & $317\pm{3}^{}$ & $55\pm{2}^{}$ & $54\pm{8}^{}$ & $243\pm{4}$ & $322\pm{30}$\\
	3463 & $2680\pm{1}$ & $2679\pm{3}$ & $109\pm{1}^{}$ & $110\pm{2}^{}$ & $62\pm{1}^{}$ & $63\pm{3}^{}$ & $168\pm{2}$ & $168\pm{9}$\\
	3521 & $4412\pm{1}$ & $4415\pm{2}$ & $255\pm{4}^{}$ & $258\pm{3}^{}$ & $62\pm{9}^{}$ & $58\pm{5}^{}$ & $171\pm{26}$ & $166\pm{12}$\\
	3528 & $4337\pm{3}$ & $4340\pm{5}$ & $220\pm{3}^{}$ & $223\pm{4}^{}$ & $54\pm{8}^{}$ & $42\pm{12}^{}$ & $240\pm{31}$ & $276\pm{66}$\\
	3574 & $1437\pm{1}$ & $1433\pm{1}$ & $101\pm{2}^{}$ & $99\pm{3}^{}$ & $54\pm{6}^{}$ & $19\pm{10}^{}$ & $74\pm{7}$ & $202\pm{96}$\\
	3685 & $1796\pm{1}$ & $1795\pm{1}$ & $297\pm{3}^{}$ & $298\pm{4}^{}$ & $20\pm{8}^{}$ & $12\pm{17}^{}$ & $74\pm{15}$ & $133\pm{177}$\\
	3691 & $2204\pm{1}$ & $2203\pm{2}$ & $249\pm{1}^{}$ & $248\pm{3}^{}$ & $63\pm{2}^{}$ & $64\pm{4}^{}$ & $151\pm{4}$ & $143\pm{10}$\\
	3708 & $5165\pm{2}$ & $5161\pm{4}$ & $231\pm{2}^{}$ & $230\pm{4}^{}$ & $47\pm{11}^{}$ & $44\pm{16}^{}$ & $231\pm{198}$ & $234\pm{69}$\\
	3709 & $5291\pm{2}$ & $5292\pm{4}$ & $232\pm{2}^{}$ & $232\pm{2}^{}$ & $62\pm{3}^{}$ & $55\pm{4}^{}$ & $251\pm{12}$ & $241\pm{14}$\\
	3734 & $965\pm{1}$ & $966\pm{1}$ & $140\pm{1}^{}$ & $139\pm{2}^{}$ & $36\pm{6}^{}$ & $43\pm{7}^{}$ & $103\pm{21}$ & $108\pm{16}$\\
	3740 & $2417\pm{1}$ & $2416\pm{2}$ & $247\pm{1}^{}$ & $247\pm{4}^{}$ & $55\pm{2}^{}$ & $48\pm{14}^{}$ & $84\pm{6}$ & $87\pm{20}$\\
	3809 & $2201\pm{1}$ & $2200\pm{1}$ & $356\pm{1}^{}$ & $357\pm{1}^{}$ & $59\pm{2}^{}$ & $58\pm{2}^{}$ & $251\pm{4}$ & $258\pm{9}$\\
	3826 & $1724\pm{1}$ & $1724\pm{2}$ & $255\pm{3}^{}$ & $254\pm{5}^{}$ & $5\pm{13}^{}$ & $20\pm{19}^{}$ & $266\pm{91}$ & $74\pm{66}$\\
	3876 & $855\pm{1}$ & $854\pm{2}$ & $360\pm{2}^{}$ & $358\pm{3}^{}$ & $59\pm{3}^{}$ & $59\pm{5}^{}$ & $103\pm{4}$ & $112\pm{10}$\\
	3915 & $4660\pm{1}$ & $4659\pm{3}$ & $30\pm{1}^{}$ & $30\pm{2}^{}$ & $45\pm{3}^{}$ & $47\pm{4}^{}$ & $205\pm{7}$ & $205\pm{16}$\\
	4026 & $4890\pm{1}$ & $4892\pm{3}$ & $138\pm{1}^{}$ & $139\pm{2}^{}$ & $54\pm{2}^{}$ & $56\pm{4}^{}$ & $290\pm{6}$ & $284\pm{14}$\\
	4165 & $505\pm{1}$ & $504\pm{1}$ & $266\pm{1}^{}$ & $265\pm{2}^{}$ & $42\pm{4}^{}$ & $41\pm{10}^{}$ & $71\pm{6}$ & $80\pm{18}$\\
	4256 & $5256\pm{1}$ & $5252\pm{3}$ & $291\pm{3}^{}$ & $291\pm{6}^{}$ & $40\pm{15}^{}$ & $38\pm{21}^{}$ & $120\pm{48}$ & $123\pm{59}$\\
	4273 & $2397\pm{1}$ & $2398\pm{2}$ & $211\pm{1}^{}$ & $212\pm{2}^{}$ & $57\pm{2}^{}$ & $60\pm{4}^{}$ & $194\pm{5}$ & $219\pm{11}$\\
	4274 & $428\pm{1}$ & $430\pm{1}$ & $173\pm{2}^{}$ & $175\pm{3}^{}$ & $39\pm{5}^{}$ & $27\pm{17}^{}$ & $73\pm{7}$ & $102\pm{57}$\\
	4284 & $535\pm{1}$ & $536\pm{2}$ & $176\pm{1}^{}$ & $176\pm{3}^{}$ & $69\pm{2}^{}$ & $59\pm{9}^{}$ & $167\pm{10}$ & $118\pm{14}$\\
	4305 & $140\pm{1}$ & $139\pm{1}$ & $188\pm{1}^{}$ & $189\pm{4}^{}$ & $40\pm{27}^{*}$ & $40\pm{27}^{*}$ & $44\pm{3}$ & $48\pm{28}$\\
	4325 & $508\pm{1}$ & $508\pm{2}$ & $57\pm{2}^{}$ & $57\pm{3}^{}$ & $52\pm{9}^{}$ & $63\pm{14}^{}$ & $97\pm{15}$ & $85\pm{13}$\\
	4393 & $2122\pm{1}$ & $2119\pm{4}$ & $236\pm{5}^{}$ & $250\pm{7}^{}$ & $50\pm{9}^{*}$ & $50\pm{9}^{*}$ & $94\pm{9}$ & $47\pm{10}$\\
	4422 & $4325\pm{1}$ & $4321\pm{2}$ & $36\pm{2}^{}$ & $36\pm{2}^{}$ & $21\pm{7}^{}$ & $25\pm{8}^{}$ & $410\pm{116}$ & $353\pm{94}$\\
	\hline
	\end{tabular}
	\end{minipage}
\end{table*}

\begin{table*}
	\begin{minipage}{151mm}
	\contcaption{}
	\begin{tabular}{ccccccccc}
	\hline
	Galaxy	 & $V_{\mathrm{sys}}$ & $V_{\mathrm{sys,Epinat}}$ & $PA$   & $PA_{\mathrm{Epinat}}$ & $i$	  & $i_{\mathrm{Epinat}}$ & $V_{\mathrm{max}}$ & $V_{\mathrm{max,Epinat}}$ \\
	(UGC ID) & (km~s$^{-1}$) 	  & (km~s$^{-1}$)	          &(\degr) & (\degr)				& (\degr) & (\degr)			      & (km~s$^{-1}$)	   & (km~s$^{-1}$)             \\
	(1)		 & (2)				  & (3)						  & (4)	   & (5)					& (6)	  & (7)					  & (8)				   & (9)					   \\
	\hline
	4456 & $5471\pm{1}$ & $5470\pm{1}$ & $125\pm{3}^{}$ & $124\pm{3}^{}$ & $43\pm{14}^{}$ & $9\pm{14}^{}$ & $51\pm{40}$ & $211\pm{321}$\\
	4499 & $680\pm{1}$ & $682\pm{1}$ & $144\pm{4}^{}$ & $141\pm{4}^{}$ & $50\pm{14}^{*}$ & $50\pm{14}^{*}$ & $70\pm{8}$ & $62\pm{13}$\\
	4543 & $1949\pm{1}$ & $1948\pm{2}$ & $319\pm{2}^{}$ & $316\pm{4}^{}$ & $66\pm{10}^{}$ & $52\pm{15}^{}$ & $79\pm{14}$ & $70\pm{15}$\\
	4555 & $4233\pm{1}$ & $4235\pm{2}$ & $91\pm{2}^{}$ & $90\pm{2}^{}$ & $36\pm{9}^{}$ & $38\pm{7}^{}$ & $200\pm{137}$ & $185\pm{30}$\\
	4770 & $7042\pm{3}$ & $7026\pm{3}$ & $280\pm{4}^{}$ & $278\pm{3}^{}$ & $32\pm{19}^{}$ & $20\pm{13}^{}$ & $290\pm{165}$ & $330\pm{194}$\\
	4820 & $1349\pm{1}$ & $1350\pm{2}$ & $157\pm{1}^{}$ & $157\pm{2}^{}$ & $37\pm{2}^{}$ & $38\pm{3}^{}$ & $330\pm{10}$ & $336\pm{20}$\\
	4936 & $1729\pm{1}$ & $1728\pm{1}$ & $295\pm{1}^{}$ & $294\pm{2}^{}$ & $11\pm{6}^{}$ & $13\pm{12}^{}$ & $292\pm{80}$ & $264\pm{227}$\\
	5045 & $7664\pm{1}$ & $7667\pm{2}$ & $147\pm{1}^{}$ & $148\pm{2}^{}$ & $17\pm{6}^{}$ & $16\pm{9}^{}$ & $410\pm{118}$ & $429\pm{228}$\\
	5175 & $3046\pm{1}$ & $3049\pm{2}$ & $145\pm{1}^{}$ & $143\pm{2}^{}$ & $56\pm{2}^{}$ & $56\pm{3}^{}$ & $189\pm{4}$ & $188\pm{10}$\\
	5228 & $1869\pm{1}$ & $1869\pm{2}$ & $119\pm{1}^{}$ & $120\pm{2}^{}$ & $71\pm{1}^{}$ & $72\pm{2}^{}$ & $125\pm{5}$ & $125\pm{9}$\\
	5251 & $1466\pm{1}$ & $1465\pm{3}$ & $260\pm{1}^{}$ & $260\pm{3}^{}$ & $74\pm{1}^{}$ & $73\pm{6}^{}$ & $126\pm{3}$ & $125\pm{9}$\\
	5253 & $1323\pm{1}$ & $1322\pm{2}$ & $354\pm{1}^{}$ & $356\pm{2}^{}$ & $41\pm{2}^{}$ & $40\pm{4}^{}$ & $230\pm{8}$ & $235\pm{17}$\\
	5316 & $1033\pm{1}$ & $1031\pm{2}$ & $131\pm{3}^{}$ & $130\pm{3}^{}$ & $76\pm{8}^{}$ & $77\pm{4}^{}$ & $205\pm{69}$ & $145\pm{9}$\\
	5319 & $2441\pm{1}$ & $2439\pm{1}$ & $346\pm{2}^{}$ & $345\pm{2}^{}$ & $26\pm{8}^{}$ & $30\pm{9}^{}$ & $202\pm{68}$ & $180\pm{47}$\\
	5373 & $291\pm{1}$ & $291\pm{2}$ & $58\pm{3}^{}$ & $51\pm{8}^{}$ & $85\pm{3}^{}$ & $10\pm{18}^{}$ & $27\pm{18}$ & $90\pm{162}$\\
	5414 & $592\pm{1}$ & $592\pm{2}$ & $216\pm{3}^{}$ & $219\pm{4}^{}$ & $60\pm{13}^{}$ & $71\pm{13}^{}$ & $81\pm{22}$ & $74\pm{10}$\\
	5510 & $1299\pm{1}$ & $1298\pm{2}$ & $200\pm{1}^{}$ & $200\pm{3}^{}$ & $22\pm{5}^{}$ & $31\pm{10}^{}$ & $216\pm{52}$ & $167\pm{44}$\\
	5532 & $2802\pm{1}$ & $2802\pm{1}$ & $148\pm{1}^{}$ & $147\pm{1}^{}$ & $32\pm{2}^{}$ & $32\pm{3}^{}$ & $379\pm{15}$ & $398\pm{24}$\\
	5721 & $528\pm{1}$ & $527\pm{6}$ & $275\pm{21}^{*}$ & $275\pm{21}^{*}$ & $62\pm{30}^{*}$ & $62\pm{30}^{*}$ & $102\pm{4}$ & $99\pm{29}$\\
	5786 & $989\pm{1}$ & $992\pm{4}$ & $151\pm{1}^{}$ & $153\pm{5}^{}$ & $39\pm{3}^{}$ & $53\pm{11}^{}$ & $106\pm{6}$ & $80\pm{15}$\\
	5789 & $729\pm{1}$ & $730\pm{2}$ & $28\pm{1}^{}$ & $27\pm{3}^{}$ & $70\pm{3}^{}$ & $68\pm{10}^{}$ & $124\pm{27}$ & $131\pm{10}$\\
	5829 & $626\pm{1}$ & $626\pm{2}$ & $197\pm{3}^{}$ & $198\pm{6}^{}$ & $53\pm{10}^{}$ & $34\pm{20}^{}$ & $34\pm{16}$ & $48\pm{26}$\\
	5840 & $582\pm{1}$ & $580\pm{1}$ & $333\pm{1}^{}$ & $333\pm{3}^{}$ & $18\pm{4}^{}$ & $18\pm{11}^{}$ & $253\pm{64}$ & $251\pm{138}$\\
	5842 & $1245\pm{1}$ & $1245\pm{1}$ & $293\pm{2}^{}$ & $292\pm{2}^{}$ & $32\pm{9}^{}$ & $47\pm{9}^{}$ & $127\pm{62}$ & $115\pm{18}$\\
	5931 & $1606\pm{1}$ & $1604\pm{3}$ & $358\pm{2}^{}$ & $360\pm{5}^{}$ & $56\pm{6}^{}$ & $54\pm{16}^{}$ & $171\pm{13}$ & $157\pm{32}$\\
	5982 & $1572\pm{1}$ & $1573\pm{2}$ & $28\pm{1}^{}$ & $28\pm{2}^{}$ & $55\pm{2}^{}$ & $55\pm{4}^{}$ & $206\pm{6}$ & $199\pm{13}$\\
	6118 & $1521\pm{1}$ & $1525\pm{2}$ & $339\pm{1}^{}$ & $343\pm{3}^{}$ & $39\pm{8}^{*}$ & $39\pm{8}^{*}$ & $135\pm{4}$ & $137\pm{24}$\\
	6277 & $1194\pm{2}$ & $1191\pm{3}$ & $71\pm{3}^{}$ & $76\pm{3}^{}$ & $17\pm{17}^{*}$ & $17\pm{17}^{*}$ & $306\pm{47}$ & $270\pm{258}$\\
	6419 & $1382\pm{1}$ & $1381\pm{2}$ & $38\pm{2}^{}$ & $34\pm{7}^{}$ & $22\pm{20}^{}$ & $66\pm{19}^{}$ & $76\pm{49}$ & $53\pm{11}$\\
	6521 & $5841\pm{1}$ & $5842\pm{2}$ & $19\pm{1}^{}$ & $20\pm{2}^{}$ & $51\pm{3}^{}$ & $46\pm{4}^{}$ & $232\pm{7}$ & $249\pm{18}$\\
	6523 & $5949\pm{1}$ & $5947\pm{2}$ & $352\pm{3}^{}$ & $353\pm{3}^{}$ & $24\pm{14}^{*}$ & $24\pm{14}^{*}$ & $117\pm{10}$ & $118\pm{63}$\\
	6537 & $854\pm{1}$ & $856\pm{2}$ & $200\pm{1}^{}$ & $200\pm{2}^{}$ & $47\pm{1}^{}$ & $47\pm{5}^{}$ & $185\pm{2}$ & $187\pm{17}$\\
	6628 & $863\pm{1}$ & $863\pm{1}$ & $183\pm{2}^{}$ & $179\pm{2}^{}$ & $20\pm{20}^{*}$ & $20\pm{20}^{*}$ & $201\pm{13}$ & $183\pm{168}$\\
	6702 & $7334\pm{1}$ & $7332\pm{2}$ & $250\pm{2}^{}$ & $256\pm{2}^{}$ & $41\pm{6}^{}$ & $38\pm{6}^{}$ & $178\pm{31}$ & $195\pm{23}$\\
	6778 & $951\pm{1}$ & $951\pm{1}$ & $343\pm{1}^{}$ & $343\pm{2}^{}$ & $50\pm{1}^{}$ & $49\pm{4}^{}$ & $232\pm{3}$ & $223\pm{14}$\\
	6787 & $1153\pm{5}$ & $1157\pm{3}$ & $110\pm{1}^{}$ & $112\pm{2}^{}$ & $71\pm{2}^{}$ & $70\pm{2}^{}$ & $228\pm{11}$ & $232\pm{10}$\\
	7021 & $1976\pm{1}$ & $1976\pm{3}$ & $264\pm{1}^{}$ & $266\pm{2}^{}$ & $56\pm{7}^{*}$ & $56\pm{7}^{*}$ & $190\pm{8}$ & $223\pm{18}$\\
	7045 & $757\pm{1}$ & $758\pm{1}$ & $100\pm{1}^{}$ & $99\pm{2}^{}$ & $68\pm{1}^{}$ & $68\pm{2}^{}$ & $157\pm{2}$ & $160\pm{9}$\\
	7154 & $1011\pm{1}$ & $1009\pm{1}$ & $276\pm{1}^{}$ & $275\pm{2}^{}$ & $67\pm{1}^{}$ & $65\pm{3}^{}$ & $157\pm{3}$ & $145\pm{9}$\\
	7323 & $506\pm{1}$ & $505\pm{1}$ & $38\pm{1}^{}$ & $38\pm{3}^{}$ & $51\pm{11}^{*}$ & $51\pm{11}^{*}$ & $83\pm{2}$ & $84\pm{15}$\\
	7766 & $807\pm{1}$ & $807\pm{1}$ & $322\pm{1}^{}$ & $323\pm{2}^{}$ & $68\pm{1}^{}$ & $69\pm{3}^{}$ & $118\pm{4}$ & $120\pm{9}$\\
	7831 & $134\pm{1}$ & $136\pm{3}$ & $295\pm{1}^{}$ & $290\pm{5}^{}$ & $35\pm{3}^{}$ & $56\pm{12}^{}$ & $128\pm{7}$ & $92\pm{15}$\\
	7853 & $529\pm{1}$ & $530\pm{2}$ & $216\pm{1}^{}$ & $217\pm{4}^{}$ & $58\pm{28}^{*}$ & $58\pm{28}^{*}$ & $116\pm{2}$ & $110\pm{35}$\\
	7861 & $599\pm{1}$ & $598\pm{1}$ & $295\pm{3}^{}$ & $297\pm{4}^{}$ & $47\pm{24}^{*}$ & $47\pm{24}^{*}$ & $59\pm{7}$ & $50\pm{21}$\\
	7876 & $944\pm{1}$ & $944\pm{1}$ & $343\pm{1}^{}$ & $344\pm{3}^{}$ & $45\pm{7}^{}$ & $53\pm{9}^{}$ & $118\pm{15}$ & $98\pm{14}$\\
	7901 & $788\pm{1}$ & $788\pm{2}$ & $255\pm{1}^{}$ & $254\pm{2}^{}$ & $54\pm{1}^{}$ & $53\pm{2}^{}$ & $215\pm{3}$ & $215\pm{10}$\\
	7971 & $458\pm{1}$ & $457\pm{2}$ & $30\pm{8}^{}$ & $32\pm{11}^{}$ & $50\pm{23}^{}$ & $31\pm{27}^{}$ & $26\pm{52}$ & $33\pm{27}$\\
	7985 & $642\pm{1}$ & $642\pm{2}$ & $274\pm{1}^{}$ & $276\pm{3}^{}$ & $46\pm{2}^{}$ & $49\pm{6}^{}$ & $109\pm{2}$ & $112\pm{13}$\\
	8334 & $484\pm{1}$ & $484\pm{1}$ & $99\pm{1}^{}$ & $100\pm{1}^{}$ & $67\pm{1}^{}$ & $66\pm{1}^{}$ & $229\pm{2}$ & $214\pm{9}$\\
	8403 & $974\pm{1}$ & $975\pm{2}$ & $121\pm{1}^{}$ & $121\pm{2}^{}$ & $53\pm{2}^{}$ & $57\pm{4}^{}$ & $130\pm{2}$ & $128\pm{10}$\\
	8490 & $191\pm{1}$ & $190\pm{1}$ & $168\pm{1}^{}$ & $167\pm{3}^{}$ & $43\pm{7}^{}$ & $40\pm{15}^{}$ & $86\pm{14}$ & $90\pm{29}$\\
	8709 & $2406\pm{1}$ & $2405\pm{3}$ & $330\pm{1}^{}$ & $330\pm{2}^{}$ & $75\pm{1}^{}$ & $76\pm{1}^{}$ & $206\pm{2}$ & $207\pm{9}$\\
	8852 & $2075\pm{1}$ & $2075\pm{1}$ & $62\pm{1}^{}$ & $63\pm{2}^{}$ & $50\pm{2}^{}$ & $52\pm{3}^{}$ & $192\pm{7}$ & $186\pm{10}$\\
	8863 & $1782\pm{4}$ & $1789\pm{4}$ & $218\pm{7}^{*}$ & $218\pm{7}^{*}$ & $77\pm{13}^{*}$ & $77\pm{13}^{*}$ & $221\pm{21}$ & $191\pm{13}$\\
	8898 & $3450\pm{1}$ & $3448\pm{2}$ & $31\pm{5}^{}$ & $31\pm{6}^{}$ & $36\pm{10}^{}$ & $27\pm{20}^{}$ & $47\pm{7}$ & $65\pm{45}$\\
	8900 & $3511\pm{1}$ & $3511\pm{3}$ & $163\pm{1}^{}$ & $161\pm{2}^{}$ & $47\pm{5}^{}$ & $57\pm{10}^{}$ & $364\pm{28}$ & $345\pm{37}$\\
	8937 & $2962\pm{1}$ & $2961\pm{5}$ & $185\pm{1}^{}$ & $185\pm{3}^{}$ & $33\pm{4}^{}$ & $32\pm{12}^{}$ & $317\pm{26}$ & $320\pm{105}$\\
	9013 & $263\pm{1}$ & $262\pm{1}$ & $162\pm{2}^{}$ & $164\pm{4}^{}$ & $21\pm{16}^{*}$ & $21\pm{16}^{*}$ & $62\pm{3}$ & $62\pm{45}$\\
	9179 & $295\pm{1}$ & $293\pm{2}$ & $52\pm{1}^{}$ & $49\pm{4}^{}$ & $30\pm{6}^{}$ & $36\pm{14}^{}$ & $129\pm{26}$ & $111\pm{36}$\\
	9248 & $3878\pm{1}$ & $3865\pm{2}$ & $266\pm{2}^{}$ & $261\pm{2}^{}$ & $47\pm{6}^{}$ & $58\pm{4}^{}$ & $151\pm{16}$ & $166\pm{11}$\\
	9358 & $1905\pm{1}$ & $1912\pm{3}$ & $186\pm{1}^{}$ & $182\pm{2}^{}$ & $57\pm{1}^{}$ & $54\pm{4}^{}$ & $213\pm{4}$ & $221\pm{14}$\\
	\hline
	\end{tabular}
	\end{minipage}
\end{table*}

\begin{table*}
	\begin{minipage}{151mm}
	\contcaption{}
	\begin{tabular}{ccccccccc}
	\hline
	Galaxy	 & $V_{\mathrm{sys}}$ & $V_{\mathrm{sys,Epinat}}$ & $PA$   & $PA_{\mathrm{Epinat}}$ & $i$	  & $i_{\mathrm{Epinat}}$ & $V_{\mathrm{max}}$ & $V_{\mathrm{max,Epinat}}$ \\
	(UGC ID) & (km~s$^{-1}$) 	  & (km~s$^{-1}$)	          &(\degr) & (\degr)				& (\degr) & (\degr)			      & (km~s$^{-1}$)	   & (km~s$^{-1}$)             \\
	(1)		 & (2)				  & (3)						  & (4)	   & (5)					& (6)	  & (7)					  & (8)				   & (9)					   \\
	\hline
	9363 & $1577\pm{1}$ & $1577\pm{1}$ & $147\pm{1}^{}$ & $147\pm{3}^{}$ & $18\pm{14}^{*}$ & $18\pm{14}^{*}$ & $142\pm{6}$ & $143\pm{105}$\\
	9366 & $2109\pm{1}$ & $2109\pm{2}$ & $223\pm{1}^{}$ & $225\pm{2}^{}$ & $64\pm{1}^{}$ & $62\pm{2}^{}$ & $242\pm{2}$ & $241\pm{9}$\\
	9406 & $2281\pm{1}$ & $2281\pm{2}$ & $131\pm{10}^{}$ & $132\pm{12}^{}$ & $51\pm{24}^{}$ & $59\pm{25}^{}$ & $25\pm{30}$ & $19\pm{10}$\\
	9465 & $1486\pm{1}$ & $1485\pm{2}$ & $129\pm{1}^{}$ & $127\pm{3}^{}$ & $65\pm{2}^{}$ & $65\pm{4}^{}$ & $88\pm{3}$ & $97\pm{9}$\\
	9576 & $1556\pm{1}$ & $1555\pm{2}$ & $120\pm{1}^{}$ & $122\pm{3}^{}$ & $38\pm{5}^{}$ & $41\pm{11}^{}$ & $120\pm{14}$ & $104\pm{25}$\\
	9649 & $441\pm{1}$ & $440\pm{1}$ & $233\pm{1}^{}$ & $235\pm{3}^{}$ & $54\pm{6}^{*}$ & $54\pm{6}^{*}$ & $97\pm{3}$ & $94\pm{11}$\\
	9736 & $3136\pm{1}$ & $3135\pm{2}$ & $217\pm{2}^{}$ & $219\pm{2}^{}$ & $48\pm{5}^{}$ & $51\pm{5}^{}$ & $201\pm{15}$ & $192\pm{16}$\\
	9753 & $770\pm{1}$ & $764\pm{2}$ & $1\pm{1}^{}$ & $3\pm{2}^{}$ & $69\pm{1}^{}$ & $69\pm{1}^{}$ & $140\pm{2}$ & $138\pm{9}$\\
	9858 & $2628\pm{2}$ & $2638\pm{3}$ & $67\pm{1}^{}$ & $70\pm{2}^{}$ & $75\pm{1}^{}$ & $75\pm{2}^{}$ & $169\pm{5}$ & $160\pm{9}$\\
	9866 & $427\pm{1}$ & $430\pm{1}$ & $149\pm{1}^{}$ & $148\pm{2}^{}$ & $58\pm{3}^{}$ & $56\pm{6}^{}$ & $112\pm{9}$ & $114\pm{11}$\\
	9943 & $1947\pm{1}$ & $1946\pm{1}$ & $266\pm{1}^{}$ & $266\pm{2}^{}$ & $58\pm{1}^{}$ & $54\pm{2}^{}$ & $201\pm{4}$ & $185\pm{10}$\\
	9969 & $2521\pm{1}$ & $2516\pm{2}$ & $16\pm{1}^{}$ & $16\pm{1}^{}$ & $61\pm{2}^{}$ & $61\pm{1}^{}$ & $325\pm{6}$ & $311\pm{9}$\\
	10075 & $830\pm{1}$ & $827\pm{1}$ & $210\pm{1}^{}$ & $210\pm{1}^{}$ & $61\pm{1}^{}$ & $62\pm{2}^{}$ & $176\pm{2}$ & $168\pm{9}$\\
	10310 & $701\pm{2}$ & $702\pm{2}$ & $190\pm{4}^{}$ & $187\pm{7}^{}$ & $42\pm{20}^{*}$ & $42\pm{20}^{*}$ & $64\pm{7}$ & $66\pm{27}$\\
	10359 & $910\pm{1}$ & $911\pm{2}$ & $281\pm{2}^{}$ & $284\pm{3}^{}$ & $44\pm{12}^{*}$ & $44\pm{12}^{*}$ & $127\pm{5}$ & $143\pm{30}$\\
	10445 & $961\pm{1}$ & $961\pm{2}$ & $106\pm{2}^{}$ & $110\pm{4}^{}$ & $46\pm{3}^{}$ & $47\pm{12}^{}$ & $76\pm{4}$ & $77\pm{17}$\\
	10470 & $1354\pm{1}$ & $1354\pm{2}$ & $287\pm{1}^{}$ & $287\pm{2}^{}$ & $34\pm{9}^{*}$ & $34\pm{9}^{*}$ & $161\pm{3}$ & $164\pm{39}$\\
	10502 & $4290\pm{1}$ & $4291\pm{2}$ & $97\pm{2}^{}$ & $99\pm{2}^{}$ & $51\pm{7}^{}$ & $50\pm{5}^{}$ & $165\pm{15}$ & $163\pm{14}$\\
	10521 & $834\pm{1}$ & $832\pm{2}$ & $20\pm{1}^{}$ & $20\pm{2}^{}$ & $59\pm{2}^{}$ & $59\pm{3}^{}$ & $121\pm{4}$ & $124\pm{9}$\\
	10546 & $1265\pm{1}$ & $1268\pm{2}$ & $177\pm{2}^{}$ & $182\pm{3}^{}$ & $51\pm{6}^{}$ & $42\pm{10}^{}$ & $100\pm{8}$ & $106\pm{22}$\\
	10564 & $1119\pm{1}$ & $1120\pm{2}$ & $149\pm{2}^{}$ & $149\pm{3}^{}$ & $75\pm{3}^{}$ & $77\pm{6}^{}$ & $102\pm{10}$ & $75\pm{8}$\\
	10652 & $1090\pm{1}$ & $1089\pm{1}$ & $223\pm{1}^{}$ & $225\pm{3}^{}$ & $31\pm{8}^{}$ & $21\pm{13}^{}$ & $93\pm{39}$ & $141\pm{82}$\\
	10757 & $1212\pm{1}$ & $1210\pm{2}$ & $59\pm{3}^{}$ & $56\pm{6}^{}$ & $39\pm{13}^{}$ & $44\pm{22}^{}$ & $88\pm{25}$ & $81\pm{33}$\\
	10791 & $1313\pm{2}$ & $1318\pm{3}$ & $91\pm{6}^{}$ & $92\pm{4}^{}$ & $34\pm{20}^{*}$ & $34\pm{20}^{*}$ & $91\pm{18}$ & $95\pm{48}$\\
	10897 & $1311\pm{1}$ & $1313\pm{1}$ & $113\pm{2}^{}$ & $115\pm{3}^{}$ & $31\pm{17}^{*}$ & $31\pm{17}^{*}$ & $111\pm{7}$ & $113\pm{56}$\\
	11012 & $24\pm{1}$ & $25\pm{1}$ & $295\pm{1}^{}$ & $299\pm{2}^{}$ & $71\pm{1}^{}$ & $72\pm{2}^{}$ & $117\pm{2}$ & $117\pm{9}$\\
	11124 & $1606\pm{1}$ & $1606\pm{2}$ & $180\pm{2}^{}$ & $182\pm{3}^{}$ & $49\pm{8}^{}$ & $51\pm{10}^{}$ & $98\pm{24}$ & $96\pm{15}$\\
	11218 & $1476\pm{1}$ & $1477\pm{2}$ & $41\pm{1}^{}$ & $42\pm{2}^{}$ & $59\pm{1}^{}$ & $58\pm{2}^{}$ & $186\pm{2}$ & $185\pm{9}$\\
	11269 & $2564\pm{4}$ & $2563\pm{6}$ & $276\pm{3}^{}$ & $272\pm{3}^{}$ & $66\pm{5}^{}$ & $69\pm{4}^{}$ & $270\pm{29}$ & $202\pm{13}$\\
	11283 & $1945\pm{1}$ & $1944\pm{4}$ & $121\pm{2}^{}$ & $120\pm{5}^{}$ & $34\pm{17}^{*}$ & $34\pm{17}^{*}$ & $156\pm{5}$ & $173\pm{73}$\\
	11300 & $481\pm{1}$ & $482\pm{1}$ & $168\pm{1}^{}$ & $168\pm{2}^{}$ & $66\pm{3}^{}$ & $70\pm{3}^{}$ & $117\pm{5}$ & $112\pm{9}$\\
	11407 & $2405\pm{1}$ & $2402\pm{8}$ & $65\pm{10}^{*}$ & $65\pm{10}^{*}$ & $64\pm{22}^{*}$ & $64\pm{22}^{*}$ & $169\pm{5}$ & $158\pm{30}$\\
	11429 & $4660\pm{2}$ & $4679\pm{7}$ & $208\pm{2}^{}$ & $208\pm{6}^{}$ & $61\pm{16}^{*}$ & $61\pm{16}^{*}$ & $224\pm{28}$ & $232\pm{35}$\\
	11466 & $824\pm{1}$ & $826\pm{3}$ & $224\pm{1}^{}$ & $226\pm{3}^{}$ & $62\pm{2}^{}$ & $66\pm{5}^{}$ & $147\pm{4}$ & $133\pm{10}$\\
	11470 & $3552\pm{4}$ & $3546\pm{5}$ & $45\pm{2}^{}$ & $47\pm{3}^{}$ & $47\pm{7}^{}$ & $47\pm{7}^{}$ & $386\pm{47}$ & $380\pm{40}$\\
	11496 & $2113\pm{2}$ & $2115\pm{2}$ & $160\pm{5}^{}$ & $167\pm{4}^{}$ & $20\pm{14}^{}$ & $44\pm{16}^{}$ & $170\pm{85}$ & $96\pm{29}$\\
	11498 & $3285\pm{5}$ & $3284\pm{4}$ & $251\pm{1}^{}$ & $251\pm{2}^{}$ & $71\pm{2}^{}$ & $71\pm{2}^{}$ & $268\pm{9}$ & $274\pm{9}$\\
	11557 & $1393\pm{1}$ & $1392\pm{1}$ & $272\pm{2}^{}$ & $276\pm{3}^{}$ & $29\pm{22}^{*}$ & $29\pm{22}^{*}$ & $109\pm{6}$ & $105\pm{72}$\\
	11597 & $40\pm{1}$ & $40\pm{2}$ & $241\pm{1}^{}$ & $241\pm{3}^{}$ & $41\pm{1}^{}$ & $40\pm{10}^{}$ & $155\pm{3}$ & $154\pm{32}$\\
	11670 & $769\pm{1}$ & $776\pm{3}$ & $327\pm{2}^{}$ & $333\pm{2}^{}$ & $69\pm{2}^{}$ & $65\pm{2}^{}$ & $191\pm{8}$ & $190\pm{9}$\\
	11707 & $900\pm{1}$ & $897\pm{2}$ & $58\pm{2}^{}$ & $59\pm{3}^{}$ & $63\pm{6}^{}$ & $70\pm{4}^{}$ & $91\pm{7}$ & $97\pm{5}$\\
	11852 & $5822\pm{2}$ & $5821\pm{3}$ & $188\pm{2}^{}$ & $189\pm{3}^{}$ & $56\pm{6}^{}$ & $47\pm{7}^{}$ & $216\pm{14}$ & $221\pm{27}$\\
	11861 & $1477\pm{1}$ & $1476\pm{2}$ & $214\pm{1}^{}$ & $218\pm{3}^{}$ & $49\pm{4}^{}$ & $43\pm{12}^{}$ & $176\pm{8}$ & $181\pm{39}$\\
	11872 & $1138\pm{1}$ & $1140\pm{1}$ & $81\pm{1}^{}$ & $86\pm{2}^{}$ & $42\pm{4}^{}$ & $47\pm{3}^{}$ & $187\pm{12}$ & $183\pm{12}$\\
	11891 & $468\pm{5}$ & $466\pm{6}$ & $114\pm{11}^{}$ & $119\pm{10}^{}$ & $43\pm{23}^{*}$ & $43\pm{23}^{*}$ & $85\pm{19}$ & $83\pm{35}$\\
	11914 & $944\pm{1}$ & $945\pm{1}$ & $265\pm{1}^{}$ & $266\pm{2}^{}$ & $37\pm{2}^{}$ & $33\pm{4}^{}$ & $276\pm{10}$ & $285\pm{26}$\\
	11951 & $1083\pm{1}$ & $1085\pm{2}$ & $258\pm{2}^{}$ & $261\pm{4}^{}$ & $75\pm{2}^{}$ & $76\pm{8}^{}$ & $100\pm{8}$ & $106\pm{7}$\\
	12060 & $878\pm{1}$ & $879\pm{2}$ & $187\pm{2}^{}$ & $187\pm{3}^{}$ & $39\pm{4}^{}$ & $36\pm{11}^{}$ & $97\pm{5}$ & $107\pm{27}$\\
	12082 & $793\pm{1}$ & $792\pm{2}$ & $137\pm{4}^{}$ & $143\pm{5}^{}$ & $30\pm{20}^{}$ & $14\pm{19}^{}$ & $51\pm{74}$ & $105\pm{137}$\\
	12101 & $766\pm{1}$ & $770\pm{2}$ & $319\pm{3}^{}$ & $317\pm{4}^{}$ & $58\pm{9}^{*}$ & $58\pm{9}^{*}$ & $101\pm{8}$ & $94\pm{12}$\\
	12212 & $896\pm{2}$ & $899\pm{2}$ & $101\pm{11}^{}$ & $82\pm{7}^{}$ & $71\pm{27}^{*}$ & $71\pm{27}^{*}$ & $168\pm{67}$ & $78\pm{15}$\\
	12276 & $5644\pm{1}$ & $5642\pm{2}$ & $323\pm{4}^{}$ & $322\pm{5}^{}$ & $39\pm{16}^{}$ & $33\pm{15}^{}$ & $77\pm{66}$ & $94\pm{37}$\\
	12343 & $2372\pm{1}$ & $2371\pm{2}$ & $204\pm{1}^{}$ & $203\pm{2}^{}$ & $48\pm{2}^{}$ & $52\pm{4}^{}$ & $225\pm{4}$ & $221\pm{14}$\\
	\hline
	\end{tabular}
	\end{minipage}
	\begin{minipage}{175mm}
	(1)~Name of the galaxy in the Uppsala General Catalogue, except for IC 476, IC 2542, and NGC 5296 which do not have a UGC name.
	(2)~Systemic velocity derived by our fitting method. 
	(3)~Systemic velocity as found in \citet{2008MNRAS.390..466E}. 
	(4)~Position angle of the receding side derived by our fitting method. 
	(5)~Position angle of the receding side as found in  \citet{2008MNRAS.390..466E}. 
	Those marked with an asterisk (*) have been fixed to the morphological value from HyperLeda. 
	(6)~Disc inclination derived by our fitting method. 
	(7)~Disc inclination as found in \citet{2008MNRAS.390..466E}. 
	Those marked with an asterisk (*) have been fixed to the morphological value from HyperLeda, except for UGC 6118, UGC 9013, UGC 9363, and UGC 10791 for which we used the inclinations derived from HI data (\citealt{2008MNRAS.390..466E}).
	(8)~Peak circular velocity derived by our fitting method. 
	(9)~Peak circular velocity as found in \citet{2008MNRAS.390..466E}.
	\end{minipage}
\end{table*}

%
%

\begin{table*}
	\begin{minipage}{166mm}
	\caption{Kinematic properties of the DYNAMO sample (see Section~\ref{sec:fitting-results-dynamo}).}
	\label{tab:fitting-results-dynamo}
	\begin{tabular}{cccccccccc}
	\hline
	Galaxy		& Seeing   & $r_{\mathrm{exp,r}}$ & $r_{\mathrm{t}}$ & $r_{\mathrm{t,Green}}$	& $i$	  & $V_{\mathrm{2.2R_{r}}}$	& $V_{\mathrm{2.2R_{r},\mathrm{Green}}}$ & $\sigma_{\mathrm{model}}$ & $\sigma_{\mathrm{m,Green}}$ \\
	(DYNAMO ID)	& (arcsec) & (kpc)			      & (kpc)	         & (kpc)					& (\degr) & (km~s$^{-1}$)      	    & (km~s$^{-1}$)	                         & (km~s$^{-1}$)	         & (km~s$^{-1}$)               \\
	(1)			& (2)	   & (3)	              & (4)			     & (5)	                    & (6)	  & (7)			 	        & (8)						             & (9)                       & (10)                        \\
	\hline
	A 04-3 & $2.1$ & $4.9$ & $1.37\pm{0.08}$ & $0.5\pm{0.1}$ & $58$ & $232\pm{3}$ & $218\pm{2}$ & $1$ & $10$ \\
	A 04-4 & $1.4$ & $6.2$ & $2.98\pm{0.17}$ & $2.0\pm{0.1}$ & $54$ & $135\pm{4}$ & $119\pm{4}$ & $20$ & $29$ \\
	A 08-3 & $1.3$ & $5.9$ & $2.00\pm{0.10}$ & $1.8\pm{0.0}$ & $59$ & $146\pm{2}$ & $149\pm{3}$ & $20$ & $19$ \\
	A 08-4 & $4.0$ & $7.5$ & $10.29\pm{0.40}$ & $11.9\pm{0.1}$ & $80$ & $381\pm{11}$ & $336\pm{23}$ & $1$ & $44$ \\
	A 10-1 & $1.0$ & $7.2$ & $2.00\pm{0.19}$ & $1.8\pm{0.1}$ & $76$ & $217\pm{6}$ & $217\pm{3}$ & $39$ & $39$ \\
	A 10-2 & $4.0$ & $7.4$ & $6.04\pm{0.35}$ & $12.4\pm{0.1}$ & $81$ & $232\pm{12}$ & $264\pm{19}$ & $1$ & $41$ \\
	A 13-2 & $1.1$ & $5.6$ & $2.30\pm{0.22}$ & $2.6\pm{0.1}$ & $35$ & $127\pm{6}$ & $128\pm{12}$ & $18$ & $23$ \\
	B 10-1 & $1.5$ & $4.7$ & $1.68\pm{0.12}$ & $1.9\pm{0.1}$ & $66$ & $135\pm{3}$ & $138\pm{4}$ & $25$ & $27$ \\
	B 14-1 & $1.5$ & $5.9$ & $0.78\pm{0.07}$ & $0.4\pm{0.1}$ & $78$ & $199\pm{2}$ & $197\pm{2}$ & $1$ & $17$ \\
	B 15-1 & $1.0$ & $5.6$ & $1.70\pm{0.10}$ & $1.8\pm{0.1}$ & $44$ & $187\pm{3}$ & $185\pm{6}$ & $29$ & $30$ \\
	B 15-2 & $1.0$ & $5.4$ & $0.94\pm{0.12}$ & $1.0\pm{0.1}$ & $55$ & $123\pm{2}$ & $124\pm{3}$ & $26$ & $24$ \\
	B 20-1 & $1.1$ & $5.2$ & $1.46\pm{0.12}$ & $1.6\pm{0.1}$ & $35$ & $105\pm{2}$ & $105\pm{5}$ & $27$ & $29$ \\
	C 00-1 & $1.4$ & $3.0$ & $1.63\pm{0.12}$ & $2.0\pm{0.1}$ & $56$ & $104\pm{3}$ & $107\pm{8}$ & $30$ & $32$ \\
	C 13-1 & $1.3$ & $5.3$ & $1.42\pm{0.10}$ & $1.6\pm{0.1}$ & $31$ & $223\pm{3}$ & $223\pm{8}$ & $26$ & $29$ \\
	C 13-3 & $1.4$ & $5.1$ & $0.38\pm{0.10}$ & $0.5\pm{0.1}$ & $72$ & $203\pm{2}$ & $220\pm{2}$ & $1$ & $11$ \\
	C 14-2 & $1.4$ & $2.7$ & $0.40\pm{0.08}$ & $0.6\pm{0.0}$ & $58$ & $159\pm{2}$ & $164\pm{4}$ & $26$ & $26$ \\
	C 20-2 & $1.9$ & $2.6$ & $0.70\pm{0.08}$ & $0.6\pm{0.1}$ & $45$ & $178\pm{3}$ & $170\pm{6}$ & $1$ & $4$ \\
	C 21-1 & $1.4$ & $1.9$ & $0.47\pm{0.06}$ & $0.5\pm{0.1}$ & $58$ & $239\pm{2}$ & $246\pm{9}$ & $1$ & $41$ \\
	C 22-2 & $1.3$ & $2.8$ & $0.87\pm{0.09}$ & $1.0\pm{0.1}$ & $35$ & $198\pm{3}$ & $197\pm{9}$ & $29$ & $40$ \\
	D 00-2 & $1.3$ & $2.8$ & $0.48\pm{0.10}$ & $0.8\pm{0.1}$ & $35$ & $61\pm{2}$ & $62\pm{2}$ & $35$ & $38$ \\
	D 10-4 & $1.4$ & $2.8$ & $13.37\pm{0.05}$ & $13.4\pm{0.1}$ & $68$ & $65\pm{2}$ & $59\pm{23}$ & $60$ & $63$ \\
	D 13-1 & $1.4$ & $3.6$ & $2.56\pm{0.13}$ & $3.4\pm{0.1}$ & $76$ & $113\pm{3}$ & $120\pm{10}$ & $34$ & $36$ \\
	D 13-5 & $1.4$ & $4.4$ & $0.40\pm{0.05}$ & $0.5\pm{0.1}$ & $49$ & $192\pm{1}$ & $192\pm{2}$ & $35$ & $46$ \\
	D 15-1 & $1.4$ & $2.0$ & $0.34\pm{0.10}$ & $0.4\pm{0.1}$ & $55$ & $149\pm{5}$ & $142\pm{5}$ & $23$ & $24$ \\
	D 15-2 & $1.6$ & $0.6$ & $0.30\pm{0.05}$ & $0.3\pm{0.1}$ & $38$ & $59\pm{25}$ & $42\pm{22}$ & $39$ & $42$ \\
	D 15-3 & $1.2$ & $4.7$ & $0.61\pm{0.06}$ & $0.6\pm{0.0}$ & $47$ & $244\pm{1}$ & $240\pm{3}$ & $25$ & $31$ \\
	D 20-1 & $1.1$ & $2.9$ & $0.74\pm{0.07}$ & $0.9\pm{0.0}$ & $39$ & $134\pm{2}$ & $137\pm{5}$ & $35$ & $37$ \\
	D 21-3 & $1.3$ & $4.2$ & $0.31\pm{0.05}$ & $0.3\pm{0.1}$ & $57$ & $171\pm{1}$ & $175\pm{2}$ & $28$ & $23$ \\
	D 22-1 & $1.3$ & $2.0$ & $0.36\pm{0.05}$ & $0.4\pm{0.1}$ & $20$ & $144\pm{1}$ & $141\pm{10}$ & $29$ & $31$ \\
	D 22-2 & $1.3$ & $5.0$ & $3.55\pm{0.11}$ & $3.8\pm{0.1}$ & $54$ & $155\pm{2}$ & $155\pm{6}$ & $33$ & $35$ \\
	D 23-1 & $1.4$ & $3.0$ & $1.51\pm{0.09}$ & $2.0\pm{0.1}$ & $54$ & $184\pm{3}$ & $186\pm{10}$ & $35$ & $35$ \\
	E 04-1 & $0.9$ & $3.5$ & $0.73\pm{0.07}$ & $0.9\pm{0.2}$ & $35$ & $372\pm{3}$ & $395\pm{22}$ & $17$ & $58$ \\
	E 09-1 & $1.3$ & $4.2$ & $1.41\pm{0.18}$ & $1.1\pm{0.1}$ & $19$ & $235\pm{8}$ & $267\pm{46}$ & $35$ & $37$ \\
	E 10-1 & $1.6$ & $3.9$ & $0.68\pm{0.08}$ & $0.8\pm{0.2}$ & $31$ & $198\pm{2}$ & $231\pm{10}$ & $33$ & $45$ \\
	G 03-2 & $1.7$ & $2.8$ & $0.64\pm{0.06}$ & $0.8\pm{0.2}$ & $24$ & $189\pm{2}$ & $174\pm{24}$ & $32$ & $43$ \\
	G 04-1 & $1.8$ & $3.1$ & $0.86\pm{0.12}$ & $0.8\pm{0.1}$ & $24$ & $264\pm{6}$ & $269\pm{22}$ & $34$ & $50$ \\
	G 08-4 & $1.1$ & $2.6$ & $0.70\pm{0.10}$ & $1.1\pm{0.1}$ & $30$ & $104\pm{3}$ & $114\pm{11}$ & $41$ & $49$ \\
	G 10-1 & $2.1$ & $2.5$ & $0.72\pm{0.11}$ & $4.3\pm{0.5}$ & $49$ & $117\pm{3}$ & $103\pm{17}$ & $52$ & $62$ \\
	G 11-1 & $1.8$ & $4.3$ & $0.67\pm{0.09}$ & $2.0\pm{0.2}$ & $23$ & $302\pm{5}$ & $314\pm{37}$ & $39$ & $47$ \\
	G 13-1 & $1.5$ & $3.2$ & $0.68\pm{0.06}$ & $0.8\pm{0.2}$ & $69$ & $103\pm{1}$ & $112\pm{2}$ & $68$ & $76$ \\
	G 14-1 & $1.1$ & $2.5$ & $0.65\pm{0.06}$ & $0.8\pm{0.2}$ & $35$ & $147\pm{3}$ & $136\pm{8}$ & $64$ & $70$ \\
	G 14-3 & $1.8$ & $1.6$ & $0.74\pm{0.12}$ & $2.7\pm{0.3}$ & $41$ & $254\pm{11}$ & $159\pm{40}$ & $52$ & $81$ \\
	G 20-2 & $0.9$ & $2.0$ & $0.69\pm{0.05}$ & $0.8\pm{0.2}$ & $34$ & $173\pm{1}$ & $166\pm{10}$ & $37$ & $45$ \\
	\hline
	\end{tabular}
	\end{minipage}
	\begin{minipage}{175mm}
	(1)~Name of the galaxy in the DYNAMO catalogue.
	(2)~The full width at half-maximum seeing used in the fitting procedure.
	(3)~Exponential disc scale radius from SDSS \textit{r}-band photometry. 
	(4)~Turn-over radius derived by our fitting method. 
	(5)~Turn-over radius of \citet{2014MNRAS.437.1070G} after applying a correction factor of $0.7$ (for details, see Section~\ref{sec:fitting-results-dynamo}). 
	(6)~Disc inclination from SDSS \textit{r}-band photometry. 
	(7)~Circular velocity at a radius of 2.2~$\times$~\textit{r}-band exponential disc scale lengths derived by our fitting method. 
	(8)~Circular velocity at a radius of 2.2~$\times$~\textit{r}-band exponential disc scale lengths as found in \citet{2014MNRAS.437.1070G}. 
	(9)~Velocity dispersion derived by our fitting method. The statistical error is 1 -- 3~km~s$^{-1}$.
	(10)~Velocity dispersion as found in \citet{2014MNRAS.437.1070G}. 
	\end{minipage}
\end{table*}

\label{lastpage}

\end{document}